\documentclass[twocolumn]{aastex631}

	\usepackage{graphicx}
	\usepackage{amsmath}
	\usepackage{url}
	\usepackage{enumerate}
	\usepackage{multirow}
	\usepackage{soul}
	\usepackage{bm}
	\usepackage{float}
	\usepackage{soul}
	\usepackage{tablefootnote}
	\hypersetup{colorlinks = true, linkcolor = green, anchorcolor = red, citecolor = blue, filecolor = red, pagecolor = red, urlcolor = red}
	\usepackage{url}
	\usepackage{colortbl}

	\RequirePackage{color}

	\newfont{\myfont}{cmmib10}

	\def\lapprox{\mathrel{\hbox{\rlap{\hbox{\lower4pt\hbox{$\sim$}}}\hbox{$<$}}}}
	\def\gapprox{\mathrel{\hbox{\rlap{\hbox{\lower3pt\hbox{$\sim$}}}\hbox{$>$}}}}
	
	\definecolor{apricot}{rgb}{0.88,0.81,0.65}

	\graphicspath{{./}{figures/}}
	
\begin{document}
	
	\title{Unveiling the emission properties of three long-period pulsars using FAST}

	\correspondingauthor{D. Li, P. Wang}
	\email{dili@tsinghua.edu.cn}
	\email{wangpei@nao.cas.cn}

	\author[0000-0002-5815-6548]{H. M. Tedila}
	\affiliation{National Astronomical Observatories, Chinese Academy of Sciences, A20 Datun Road, Chaoyang District, Beijing 100101, People's Republic of China}
	\affiliation{Arba Minch University, Arba Minch 21, Ethiopia}
	
	%\author{D. Li}
	\author[0000-0003-3010-7661]{D. Li}
	\affiliation{Department of Astronomy, Tsinghua University, Beijing 100084, People's Republic of China}
	\affiliation{National Astronomical Observatories, Chinese Academy of Sciences, A20 Datun Road, Chaoyang District, Beijing 100101, People's Republic of China}
	\affiliation{Research Center for Astronomical Computing, Zhejiang Lab, Hangzhou, Zhejiang 311121, People's Republic of China}

	%\author{P. Wang}
	\author[0000-0002-3386-7159]{P. Wang}
	\affiliation{National Astronomical Observatories, Chinese Academy of Sciences, A20 Datun Road, Chaoyang District, Beijing 100101, People's Republic of China}
	\affiliation{Institute for Frontiers in Astronomy and Astrophysics, Beijing Normal University, Beijing 102206, People's Republic of China}
	
	\author{R. Yuen}
	\affiliation{Xinjiang Astronomical Observatory, Chinese Academy of Sciences, 150 Science 1-Street, Urumqi, Xinjiang, 830011, People's Republic of China}
	\affiliation{Xinjiang Key Laboratory of Radio Astrophysics, 150 Science1-Street, Urumqi, Xinjiang, 830011, People's Republic of China}
	
	\author[0000-0002-5381-6498]{J. P. Yuan}
	\affiliation{Xinjiang Astronomical Observatory, Chinese Academy of Sciences, 150 Science 1-Street, Urumqi, Xinjiang, 830011, People's Republic of China}
	\affiliation{Xinjiang Key Laboratory of Radio Astrophysics, 150 Science1-Street, Urumqi, Xinjiang, 830011, People's Republic of China}
	
	\author[0000-0002-9786-8548]{N. Wang}
	\affiliation{Xinjiang Astronomical Observatory, Chinese Academy of Sciences, 150 Science 1-Street, Urumqi, Xinjiang, 830011, People's Republic of China}
	\affiliation{Xinjiang Key Laboratory of Radio Astrophysics, 150 Science1-Street, Urumqi, Xinjiang, 830011, People's Republic of China}
	
	\author[0000-0003-2991-7421]{Z. G. Wen}
	\affiliation{Xinjiang Astronomical Observatory, Chinese Academy of Sciences, 150 Science 1-Street, Urumqi, Xinjiang, 830011, People's Republic of China}
	\affiliation{Xinjiang Key Laboratory of Radio Astrophysics, 150 Science1-Street, Urumqi, Xinjiang, 830011, People's Republic of China}
	
	\author{S. J. Dang}
	\affiliation{School of Physics and Electronic Science, Guizhou Normal University, Guiyang, 550001, People’s Republic of China}
	
	\author{A. G. Tegegne}
	\affiliation{Xinjiang Astronomical Observatory, Chinese Academy of Sciences, 150 Science 1-Street, Urumqi, Xinjiang, 830011, People's Republic of China}
	\affiliation{University of Chinese Academy of Sciences, 19A Yuquan Road, 100049 Beijing, People's Republic of China}
	\affiliation{Arba Minch University, Arba Minch 21, Ethiopia}
	
	\author{R. Rejep}
	\affiliation{National Astronomical Observatories, Chinese Academy of Sciences, A20 Datun Road, Chaoyang District, Beijing 100101, People's Republic of China}
	
	\author{C. C. Miao}
	\affiliation{Research Center for Astronomical Computing, Zhejiang Lab, Hangzhou, Zhejiang 311121, People's Republic of China}
	%\author{...}

	\author{FAST Collaboration}
	
	\title{Unveiling the emission properties of three long-period pulsars using FAST}
	
	\begin{abstract}
		
		We detail the emission behaviors of three long-period pulsars detected using the Five-hundred-meter Aperture Spherical radio Telescope (FAST) during the CRAFTS survey. Their rotational periods range from 1.83\,s to 4.75\,s, and the null fractions measure between 28\% and 53\%. PSR J1945+1211 and PSR J2323+1214 exhibited quasi-periodic nulls, with duration of around 57 seconds. The longest null was observed in PSR J1945+1211, lasting 76 seconds. PSR J2323+1214 displayed varying null fractions between its leading and trailing components. For the first time in PSR J2323+1214, we detected five dwarf pulses, which are much weaker and narrower pulses than typical burst pulses. In addition, we investigate the microstructure of PSR J1900$-$0134 for the first time, revealing intricate pulses of up to 2.05 milliseconds and noting its complex emission characteristics. Bright pulses occur in all of these sources at different rates. These observations suggest complex magnetospheric processes, potentially related to magnetic reconnections, and provide insights into the origins of bright and microstructure pulses, as well as their distinctions from ordinary pulses.
		
	\end{abstract}
	
	
	\section{INTRODUCTION}\label{sec:intro}
	
	Pulsars are highly magnetized, rapidly rotating neutron stars, and well-known for their stable emission of radio pulses. Conventionally, the integrated pulse profile, obtained from averaging over numerous individual pulses, has been regarded as a characteristic feature of a pulsar at a given observational frequency \citep{HMT75, Edwards03, Wangz2021}. However, despite this consistency, pulsars exhibit intriguing variability that deviates from the norm, manifested as phenomena such as nulls \citep[e.g.,][]{Backer1970, Ritch76, Wang07, Anumarlapudi2023}, mode changing \citep[e.g.,][]{Wang07, Basu2018, Rahaman2021}, bright pulses \citep[e.g.,][]{Mahajan2018, Wen21, HMT2022}, and microstructures \citep[e.g.,][]{Cordes1979, Lange1998, Popov2002, Mitra2015}.
	
	Null is a phenomenon where pulsars undergo phases of diminished or entirely absent emission, first observed in PSR B0834+06, PSR B1133+16, PSR B1237+25, and PSR B1929+10 \citep{Backer1970}. This behavior has since been detected in more than 214 pulsars, representing less than 10 percent of known pulsars \citep{GYR17, Wang2020}, though the true number may be underestimated due to limitations in observational sensitivity and duration. Nulls can occur on various timescales, ranging from a single pulsar rotation to several hours and, in some cases, even months \citep{Wang07}. After these pauses, pulsars typically resume their emissions and continue as usual. The most prevalent nulls, which last for just a single period, are generally considered stochastic processes within the pulsar magnetosphere \citep[e.g., PSR B2021+51;][]{Basu17}. However, longer nulls, such as those lasting 2-5 hours in PSR B1706$-$16, may be linked to changes in plasma processes within the magnetosphere \citep{Naidu2018}. The erratic nature of nulls is thought to be caused by disruptions in the magnetospheric particle supply, possibly stemming from instabilities within the pulsar magnetosphere or alterations in the magnetospheric currents \citep{Ritch76, Backer1970, GYR17}. The `null fraction' (NF), or the fraction of time a pulsar remains in this null state, displays remarkable variability, ranging from 0\% to 95\% \citep{Wang07}, and has been shown to correlate with the characteristic age of the pulsar \citep{Ritch76} and its pulse period \citep{Bigg92}.
	Several models have been proposed to explain these mechanisms. Some magnetospheric switching models suggest that a change in emission arrangements might induce a mode change, evident in pulsars that frequently alternate between distinct emission modes \citep{Timokhin10, Melrose16}. Alternatively, other theories propose that nulls may arise from a temporary depletion in the plasma supply that hinders coherent radio emission \citep{Rankin2008, Rankin2017}, or shifts in the magnetospheric currents that drive the emissions \citep{Zhang2007}.
	
	\citet{HerfindalRankin07} conducted the first significant study of periodic nulls in a larger sample of pulsars, challenging the belief that pulse nulls are random. Recent observations, including those from the Five-hundred-meter Aperture Spherical radio Telescope (FAST), have shown that some classical null pulsars exhibit periodic null patterns, such as PSR J1727$-$2739 \citep{Wen16, Rejep22} and PSR J1909+0122 \citep{Chen23}. The correlation between null length and pulse period \citep{Bigg92} is consistent with findings that intermittency may be more common in older pulsars \citep{KAL06}, suggesting that they may share the same underlying mechanism. This underscores the importance of studying nulls for insights into intermittency. While Rotating Radio Transients (RRATs) exhibit irregular waiting times for single pulse emission \citep{KAL06, Wang07}, their connection to nulls remains tenuous.

	Bright pulses, or the more intense `giant pulses', signify the zenith of pulsar radio emission. These occasional flares originate from localized areas of increased magnetospheric activity associated with the magnetic poles of the pulsar \citep{Cognard1996, Knight2006}. However, such brilliance is not universally exhibited across all pulsars and seems to favor the young and energetic ones.
	Microstructure pulses, on the other hand, can be visualized as `ripples' or nuances in the broader canvas of pulsar emissions. These transient variations, which occasionally last for very short time scales ranging from microseconds to milliseconds \citep[e.g.,][]{Mitra2015, De2016, Liu2022}, potentially unravel the core processes that drive pulsar emissions. These microbursts might be the result of intricate wave-particle interactions or the outcomes of coherent emission processes at play within the magnetosphere, offering a microscopic lens into the very heart of pulsar emission dynamics  \citep{Hankins1975, Weatherall1998, Popov2002}. \cite{Lange1998} found that 30\%--70\% of single pulses in several pulsars exhibit microstructure at both high and low frequencies. More importantly, they did not observe significant differences in the characteristics of this microstructure between the high and low frequencies \citep{Hankins1975, Weatherall1998, Popov2002}. This indicates that microstructure is a fundamental aspect of pulsar emission and should be explained by any theory of pulsar radio emission \citep{Machabeli2001}.
	
	\begin{table*}
		\setlength{\tabcolsep}{14 pt}
		\centering
		\caption{Information of the three long-period pulsars. It shows their rotational periods ($P$), the first derivative of the rotational periods ($\dot{P})$, dispersion measures (DM), the characterstic age ($\tau_{c}$), the surface dipole magnetic fields ($B_d$), and the spin-down energy loss rates ($\dot{E}$).}
		
		\begin{tabular}{ccccccc} % four columns, alignment for each
			\hline
			PSR &$P$ &$\dot{P}$&DM& $\tau_{c}$&$B_{d}$&$\dot{E}$\\
			& (s) &($\times 10^{-15}$ \,s\,s$^{-1}$)&(cm$^{-3}$pc)&(\,Myr)&($\times 10^{12}$\,G)&($\times 10^{30}$\,erg\,s$^{-1}$) \\
			\hline
			J1945$+$1211 & 4.75 &$3.79$& 92.7 & 19.8&4.25&1.4\\
			J2323$+$1214 & 3.76 &1.85& 29 &32.2&2.64&1.37\\
			J1900$-$0134 & 1.83 & 3.049  &178.1 &0.95&7.47&196 \\
			\hline
		\end{tabular}
		\label{table:pulsarss}
	\end{table*}
	
	The exploration of pulsar phenomena has significantly advanced with the introduction of powerful instruments like the FAST telescope. This advancement is particularly evident in the Commensal Radio Astronomy FAST Survey \citep[CRAFTS;][]{Li18}, which has played a pivotal role in the endeavor. The CRAFTS survey initiated its pulsar search mode in 2017, coinciding with the commissioning and early science stages of FAST. During this period, an ultra-wideband (UWB) receiver was employed, covering a frequency range from 270 to 1620 MHz. This setup was later replaced by the more advanced FAST L-band array of 19 beams \citep[FLAN;][]{Li18}. To date, CRAFTS has successfully discovered more than 180 new pulsars\footnote{https://crafts.bao.ac.cn}, including 50 slow pulsars (with periods $\geq$ 1 s), a feat made possible by the combined use of the UWB and FLAN systems.
	
	Key discoveries include PSR J1900$-$0134, the first pulsar identified by FAST with a period of 1.83 seconds \citep{Qian2019, CLH+20} with notable features like a strong magnetic field and significant spin-down energy loss rate. Other significant findings include PSR J1945+1211, a longer period and a potentially aligned rotator, and PSR J2323+1214, a long-period pulsar with unique dual-pulse emission. Our analysis, utilizing these advanced tools for each observation and interpretation, represents significant progress in unraveling the complex nature of pulsar emission mechanisms and their magnetospheric underpinnings. Detailed characteristics of each pulsar are provided in Table \ref{table:pulsarss}. The paper is organized as follows. Our observations and data processing methods are presented in Section \ref{sec:observe}. The analysis of single-pulse sequences in relation to the different pulse emissions is presented in Section \ref{sec:analysis_result}. The detailed dwarf pulses, bright pulses, and microstructure emission properties are examined in Sections \ref{sec:dwarfPulses}, \ref{sec:brightPulses}, and \ref{sec:microstructure}, respectively. We briefly discuss the implications of our results and the main emission features of the pulsars in Section \ref{sec:discus}, and summarize our results in Section \ref{sec:summary}.
	
	\begin{figure*}
		\centering
		\includegraphics[width=2\columnwidth]{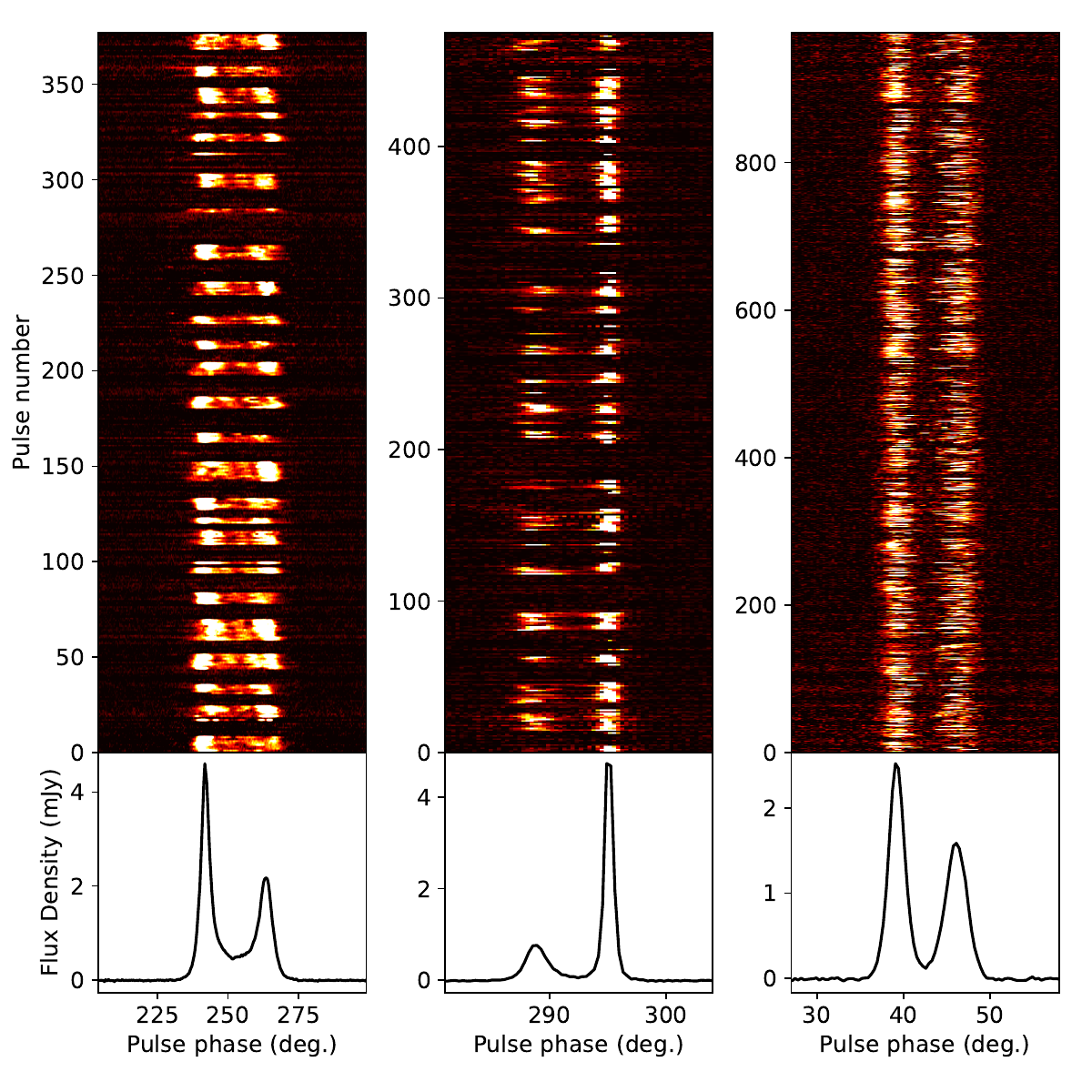}
		\caption{The single-pulse sequence (upper panel) and the associated integrated pulse profile (lower panel) are shown for PSR J1945+1211 (left), PSR J2323+1214 (middle) and PSR J1900$-$0134 (right), respectively.}
		\label{stack}
	\end{figure*}
	
	\section{OBSERVATION AND DATA PROCESSING}\label{sec:observe}
	
	In September 2016, the FAST telescope was completed with a maximum effective aperture of an impressive 300 meters in diameter \citep{Li18, Jiang19}.  Subsequently, single-pulse observations of our long-period pulsars were conducted during the pilot scans of the CRAFTS Survey using the FAST telescope.  These observations utilized the 19-beam receiver and spanned a frequency range from 1.05 to 1.45 GHz. We used the Reconfigurable Open Architecture Computing Hardware-version2 (ROACH2) signal processor in the observations \citep{Jiang19, Jiang20}. The data, with time and frequency resolutions of 49.152 µs and 0.488 MHz, respectively, were recorded on 1024 frequency channels in the search mode PSRFITS format \citep{Hotan04}. Each observation lasted for 30 minutes. Specific details for each observation can be found in Table \ref{table:observations}.

	In the offline processing, we used the {\tt\string DSPSR} software package \citep{Straten11} to disperse and produce single-pulse integrations. Spectral band edges were eliminated using the Pulsar Archive Zapper ({\tt\string PAZ}), and the radio frequency interference (RFI) was excised from the data using {\tt\string PAZI} plug-ins in the {\tt\string PSRCHIVE} software package \citep{Hotan04}. 
	Subsequently,  the total intensity was derived by averaging the single-pulse sequences across the frequencies. Following that, we removed the baseline to reduce the noise level to an approximate zero for each distinct pulse. The refined single-pulse sequences were organized into a pulse stack. This two-dimensional representation displays pulse longitude on the abscissa, with each subsequent period positioned on the ordinate. The radiometer equation, given by
	\begin{equation}\label{radiometer}
		S_{\rm av} = \frac{S_{\rm sys}}{\sqrt{n_{\rm pol} \cdot f_{\rm eff} \cdot t_{\rm smp}}},
	\end{equation}
	was employed to calibrate the observed pulses based on the off-pulse noise \citep{Dicke1982, Wen21, HMT2024}. In this equation, $n_{\rm pol}=4$ represents the number of polarizations, $f_{\rm eff} = 400\,{\rm MHz}$ is the effective frequency bandwidth, and $t_{\rm smp} = 49.152\,\mu{\rm s}$ is the sampling time. The system equivalent flux density (SEFD) of FAST, denoted as $S_{\rm sys}$, is approximately $1\,{\rm Jy}$ \citep{Yu17}. The estimated flux density ($S_{\rm av}$) is approximately 3.5 mJy. Using this calibration, flux calibration was then carried out on all pulse profiles by multiplying with $S_{\rm av}$. 
	
	\begin{table}
		\setlength{\tabcolsep}{10 pt}
		\centering
	\caption{Information of the three long-period pulsars. The symbols T$_{obs}$ and N$_{sub}$ denote the observation duration and total pulse number, respectively.}
		\begin{tabular}{cccc} % four columns, alignment for each
			\hline
			PSR & Date & T$_{obs}$& N$_{sub}$ \\
			&(yyyy$-$mm$-$dd) & (s)   \\
			\hline
			J1945$+$1211 & 2020$-$09$-$24 & 1800& 379  \\
			J2323$+$1214 & 2020$-$09$-$22 & 1800& 479  \\
			J1900$-$0134 & 2020$-$11$-$07& 1800& 983 \\
			
			\hline
		\end{tabular}
		\label{table:observations}
	\end{table}
	
	\begin{figure*}
		\centering
		\includegraphics[width=2\columnwidth]{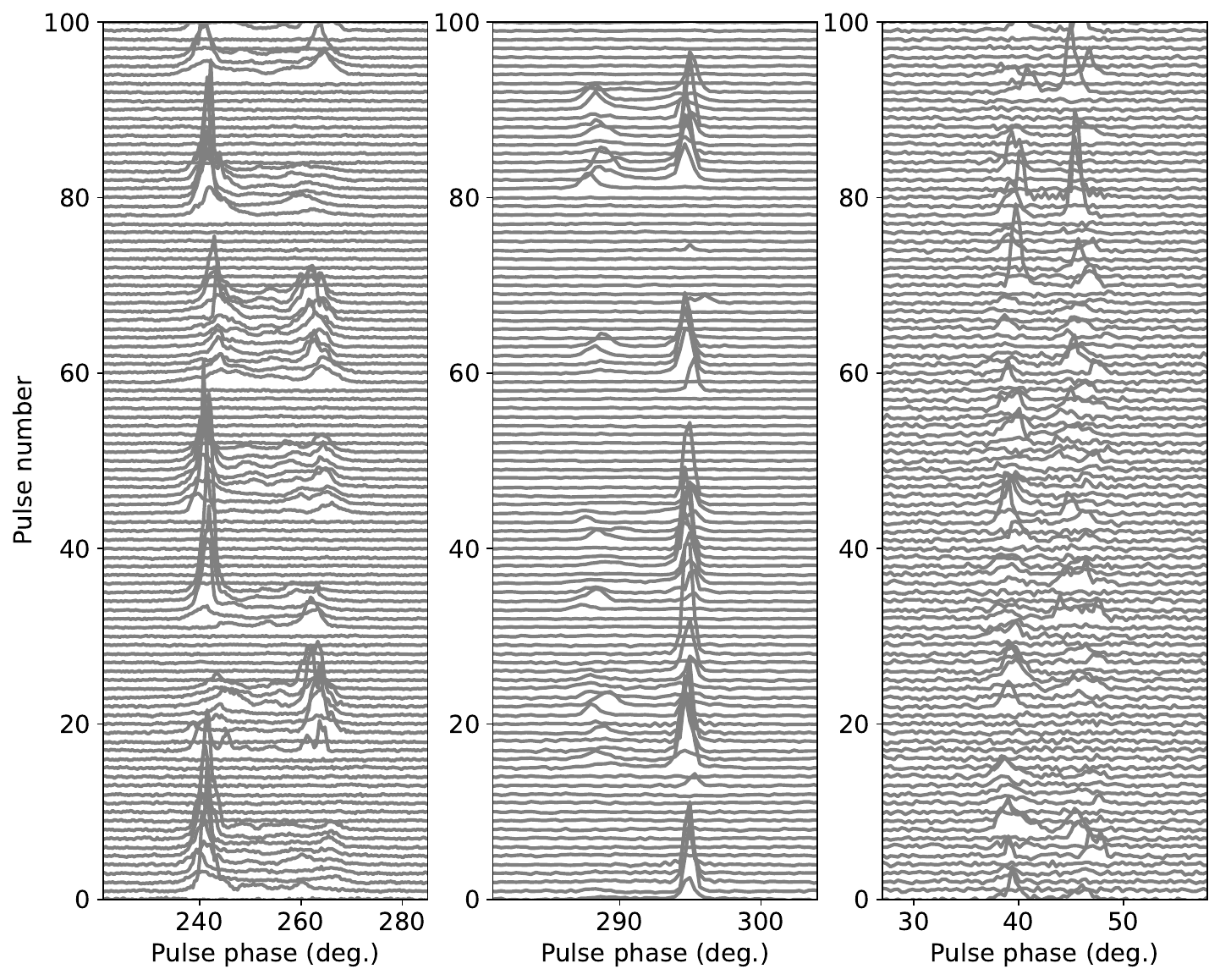}
		\caption{The first 100 single pulses from PSR J1945+1211 (left plot), PSR J2323+1214 (middle plot), and PSR J1900$-$0134 (right plot), respectively.}
		%\end{minipage}
		\label{single_pulse}
	\end{figure*}
	
	\section{Single-Pulse Emission}\label{sec:analysis_result}
	
	Figure~\ref{stack} presents the pulse sequence and integrated pulse profile of PSR J1945+1211, PSR J2323+1214, and PSR J1900$-$0134. The upper panels of the figure illustrate the observed pulse sequences over time. Each horizontal line represents a single pulse, with the vertical axis indicating the pulse number. It is apparent from Figure~\ref{single_pulse} that the pulse intensity varies and that some pulses are significantly brighter than others, which is a phenomenon often referred to as burst pulses. The lower panels of Figure~\ref{stack} show the integrated pulse profile for each pulsar.  Notably, each pulse profile comprises two primary components, commonly identified as the leading and trailing components of the profile. By visual inspection, a clear emission feature between the two profile components, known as the bridge emission, is identified in each of the three pulsars. For PSR J1945+21211, the peak intensity in the leading component is almost twice that of the trailing component, and for PSR J1900$-$0134, the leading component also exceed the trailing component by about 45\%. However, the case is significantly different for PSR J2323+1214, where the peak intensity of the trailing component is three times greater than that of the leading component.
	An intriguing aspect of this analysis is the observed variation in pulse width among the pulsars measured at 10\% ($W_{10}$) of the maximum intensity of the integrated pulse profile. Specifically, PSR J1945+1211 exhibits the widest pulse profile at 29.53$^{\circ}$, while PSR J2323+1214 has the narrowest at 7.73$^{\circ}$. This variation suggests potential differences in the emission mechanism or geometry of these pulsars. Overall, our observations reveal repeatable patterns in the single-pulse emission, encompassing both burst and null pulses, with exceptionally high brightness occasionally observed in some pulses. By analyzing these differences, we can gain insight into the factors that influence null behavior, emission properties, and potential differences in the emission mechanisms or geometries of these neutron stars.
	
	\begin{figure*}
		\centering
		\includegraphics[width=2\columnwidth]{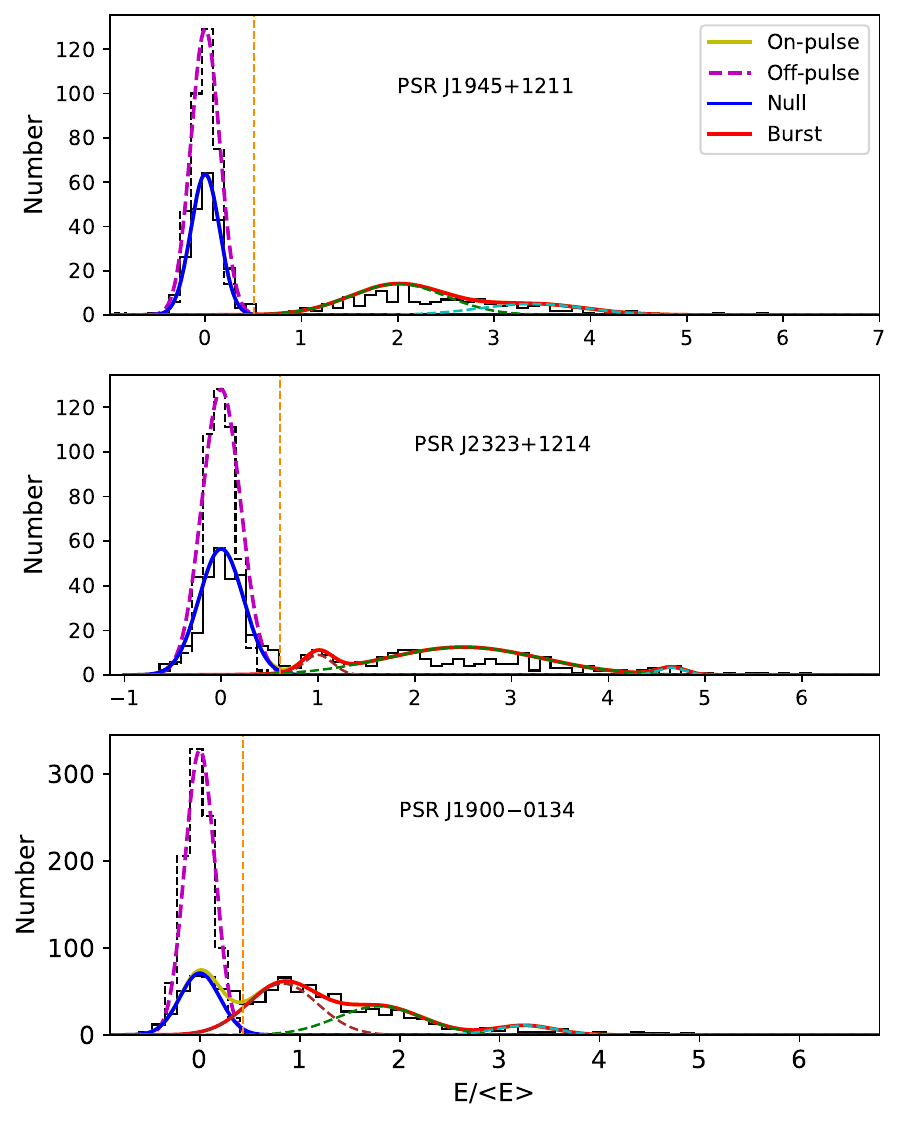}
		\caption{The pulse energy distributions for PSR J1945+1211 (upper plot), PSR J2323+1214 (middle plot) and PSR J1900$-$0134 (lower plot). All pulse energy is normalized by the respective mean on-pulse energy. The solid black lines representing the on-pulse regions and dashed red lines indicating the off-pulse regions. The vertical dotted orange dashed line represents the threshold energy value for Gaussian fitting to both null pulses and off-pulse regions. The representation of brown, green, and cyan lines is described in the main text and in Table~\ref{table:fittingx}. The x-axis is normalized by the mean on-pulse energy.}
		\label{energy_all}
	\end{figure*}

	\subsection{Pulse energy distribution}\label{sec:pulsenergy}
	
	In our effort to investigate the emission state behaviors exhibited by PSR J1945+1211, PSR J2323+1214, and PSR J1900$-$0134, we conducted an analysis of their pulse energy distributions. This initial step is important for our subsequent classifications, following the methodologies described by \citet{Ritch76}. A key metric in this framework is the on-pulse energy, carefully computed by integrating the intensity across the on-pulse region. We also determined the off-pulse energy within an equivalent width in the off-pulse region. All energy values, both on-pulse and off-pulse, were normalized using the average on-pulse energy.
	
	In the upper panel of Figure~\ref{energy_all} for PSR J1945+1211, we illustrate the energy distributions for the on-pulse and off-pulse regions. The off-pulse distribution is characterized by a histogram in dotted black and centers around zero, which can be fitted by a Gaussian curve signifying the presence of the Gaussian random noise. In contrast, the on-pulse distribution (solid black histogram) displays a bimodal pattern. The first peak at zero energy corresponds to the null state, fitted with a blue Gaussian, while the second peak at higher energies corresponds to the burst state, fitted with two Gaussians (green and cyan). The cyan component represents occasional bright pulses observed at the highest energy levels, and the green component captures the remaining burst pulses. This dual-peak structure of the on-pulse distribution, along with the use of multiple Gaussian fits for the burst state, highlights the complexity of the pulsar's emission dynamics and the statistical methods used to discern the various emission states.
	
	For PSR J2323+1214, presented in the middle plot of Figure~\ref{energy_all}, the on-pulse distribution features two predominant peaks. The first peak at zero energy corresponds to the null state (marked with blue), and the second peak, representing burst state (marked with red), fitted with three Gaussians (brown, green, and cyan). The brown component fits occasional `dwarf pulses' at lower energies, the cyan component represents bright pulses, and the green component covers the remaining burst pulses.
	
	The histograms for PSR J1900$-$0134, shown in the lower plot of Figure~\ref{energy_all}, also distinguish between off-pulse and on-pulse energies. The on-pulse distribution includes a Gaussian base at zero energy, corresponding to the null state (marked with blue), and another peak representing the burst state (marked with red), fitted with three Gaussians (brown, green, and cyan). The brown and green components fit burst pulses containing microstructure features, while the cyan component fits the bright pulses. The fitting parameters for the null and burst emission states are listed in Tables~\ref{table:fitting} and~\ref{table:fittingx}, respectively. The Gaussian distribution is defined by
	\begin{equation}
		P(E)=\frac{\alpha}{\sigma \sqrt{2\pi}}\exp\left(-\frac{(E-\mu)^2}{2\sigma^2}\right),
	\end{equation}
	where $\alpha$ represents the shape parameter, $\mu$ is the mean, and $\sigma$ delineates the standard deviation of the distribution.
	
	\begin{table}
		\centering
		\caption{List of parameters used to fit the energy distributions for the null states (marked with a blue Gaussian) and NF denotes null fraction for the three long-period pulsars in our analysis.}
		
		\setlength{\tabcolsep}{10 pt}
		%\centering
		\begin{tabular}{l|ccc|cc}
			\hline
			\multicolumn{1}{l|}{PSR name } &\multicolumn{3}{c|} {$\rm Parameters$} &\multicolumn{1}{c} {\rm NF}\\
			& $\rm \alpha$ &$ \mu $ & $ \sigma$ & $(\%)$ \\
				\hline
				\hline
			%\centering
			J1945+1211 & 63.5  & 56.5 & 71& $52.46 \pm 2.89$\\ 
			J2323+1214  & 0  & 0 & 0 & $48.48 \pm 2.51$ \\ 
			J1900$-$0134 & 0.15  & 0.23& 0.2 &$27.51 \pm 1.37$ \\ 
				\hline
		
		\end{tabular}
		%\end{adjustbox}
		\label{table:fitting}
	\end{table}
	
		\begin{table*}
		\centering
		\caption{Parameters used to fit the burst state energy distributions (red Gaussian in Figure~\ref{energy_all}) for the three long-period pulsars. Brown, Green, and Cyan correspond to different components of the red Gaussian fit.}

		\setlength{\tabcolsep}{15 pt}
		%\centering
		\begin{tabular}{l|cc|ccc|ccc}
			\hline
			\multicolumn{1}{l|}{$\rm Parameters$} &\multicolumn{2}{c|} {\rm PSR J1945+1211} & \multicolumn{3}{c|} {\rm PSR J2323+1214} &\multicolumn{3}{c} {\rm PSR J1900$-$0134}\\
			&{\rm Green} & {\rm Cyan} & {\rm Brown} & {\rm Green} &  {\rm Cyan} & {\rm  Brown}& Green& Cyan \\
			\hline
			\hline
			\centering
			$\rm \alpha$ & 14 & 5 & 9& 12.5& 3&59&33&11\\  
			$\mu$ & 2 & 3.4 & 1& 2.5& 4.65&0.83&1.8&3.25\\ 
			$\sigma$ & 0.5 & 0.55 & 0.15 & 0.8 & 0.14&0.35&0.43&0.3\\ 
			\hline
		\end{tabular}
		%\end{adjustbox}
		\label{table:fittingx}
	\end{table*}
	
	Using these data, we investigated the null behavior of the pulsars. A key aspect of our analysis was the quantification of NF. Following earlier studies \citep{Ritch76, Wang07}, we estimated the NF for each observation by subtracting a scaled version of the off-pulse histogram from the on-pulse histogram, ensuring the sum of the difference counts in bins with $E < 0$ was zero \citep{Wen16}. The uncertainty in the NF was gauged using the formula $\sqrt{n_{\rm p}}/N$, where $n_{\rm p}$ represents the number of null pulses and $N$ is the total count of observed pulses \citep{Wang07}. Our estimates place the NF of PSR J1945+1211, PSR J2323+1214, and PSR J1900$-$ 0134 at approximately 52.46\% $\pm$ 2.89\%, 48.48\% $\pm$ 2.51\% and 27.51\% $\pm$ 1.37\%, respectively.
	
	Interestingly, further analysis of PSR J2323+1214 revealed significant differences in NF across the two emission components. The leading component exhibited an NF of approximately 61.34 $\pm$ 1.82\%, indicating a greater tendency to enter a null state. In contrast, the trailing component had a lower NF of about 50.27 $\pm$ 1.73\%, which is lower than the previous observations by \citet{WWD21}. Our observations, made with the FAST telescope, which is 2.1 times more sensitive than the Arecibo telescope, detected weak emissions at higher levels, resulting in generally lower NF values. This disparity in NF between the components suggests underlying asymmetries in the pulsar's emission mechanisms or magnetic field structure.
	
	\begin{figure}
		\centering
		\includegraphics[width=1\columnwidth, angle=0]{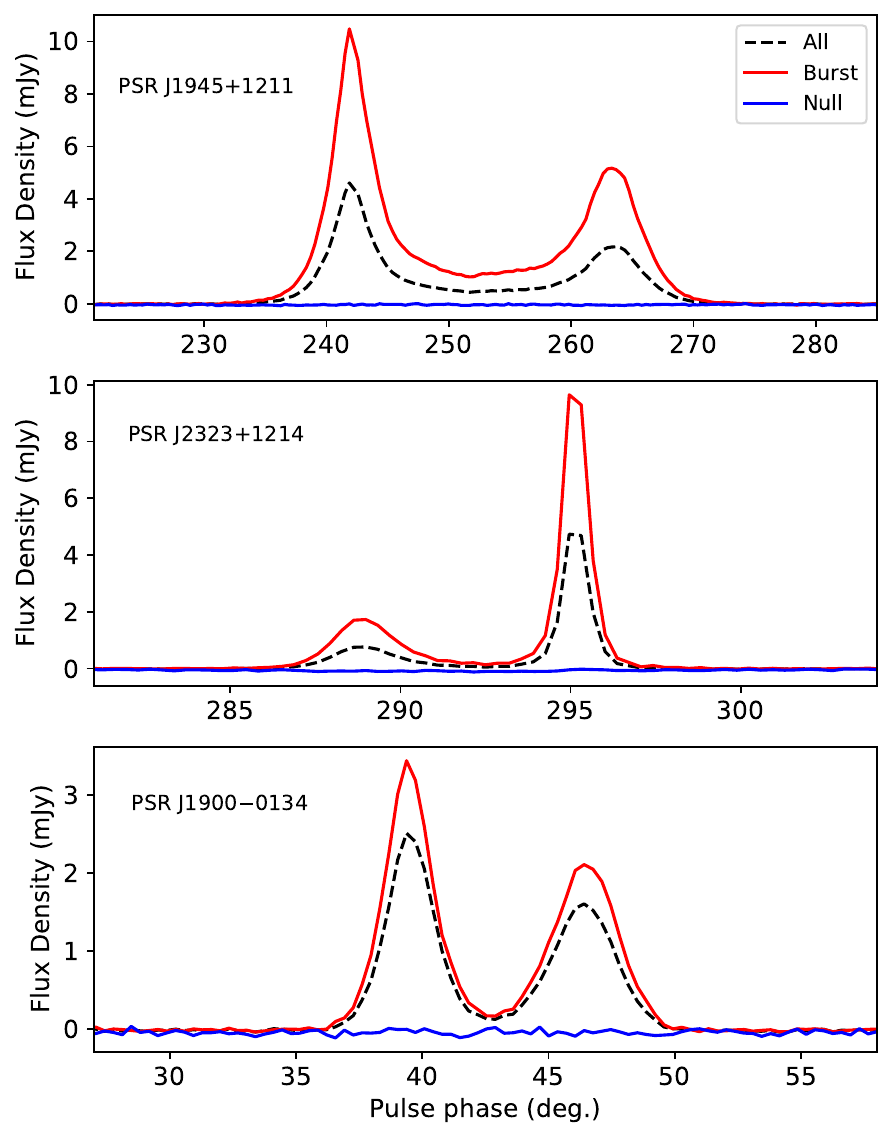}
		\caption{Integrated pulse profiles of PSR J1945+1211 (upper plot), PSR J2323+1214 (middle plot), and PSR J1900$-$0134 (lower plot), observed in different emission states: null (blue), burst (red), and overall (dotted-black).}
		\label{profile_all} 
	\end{figure}	
	
	\subsection{Emission separation}\label{sec:characterization}
	
	To distinguish between different emissions from pulsars, we adopted a methodology based on \citep{Bhat10, HMT2022}. This approach involves establishing an uncertainty threshold in the on-pulse energy for each pulse by considering overlaps in the energy distributions of different emission states. The threshold is mathematically defined as $\eta_{\rm on} = \sqrt{n_{\rm on}}\sigma_{\rm off}$, where $n_{\rm on}$ is the bin number for the on-pulse region. $\sigma_{\rm off}$ signifies the root mean square (RMS) associated with the off-pulse region, calculated from a region whose width matches that of the on-pulse window. The on-pulse range is determined by identifying the maximum intensity at 10\% of the integrated profile at the beginning and end of the leading and trailing components.
	
	\begin{figure*}
		\centering
		\includegraphics[width=1\columnwidth, angle=0]{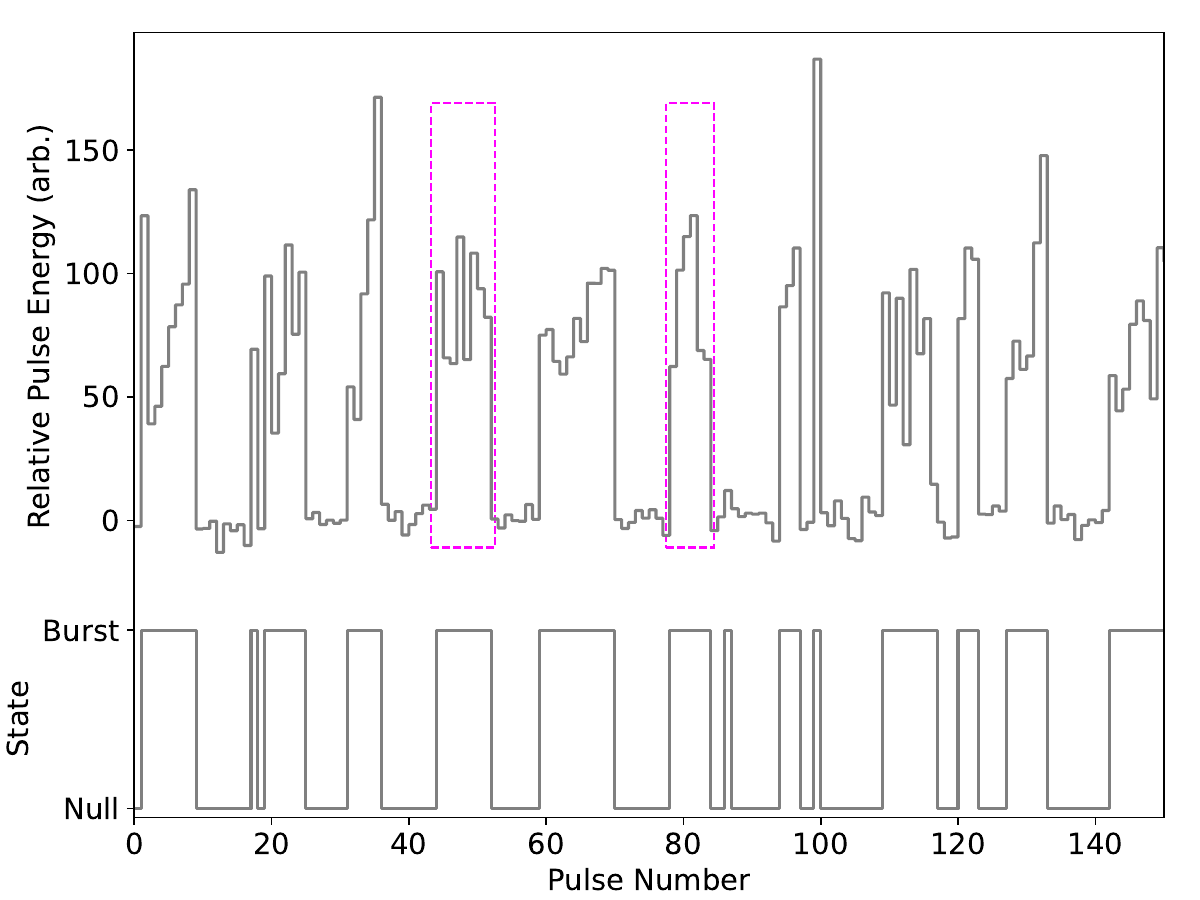}
		\includegraphics[width=1\columnwidth, angle=0]{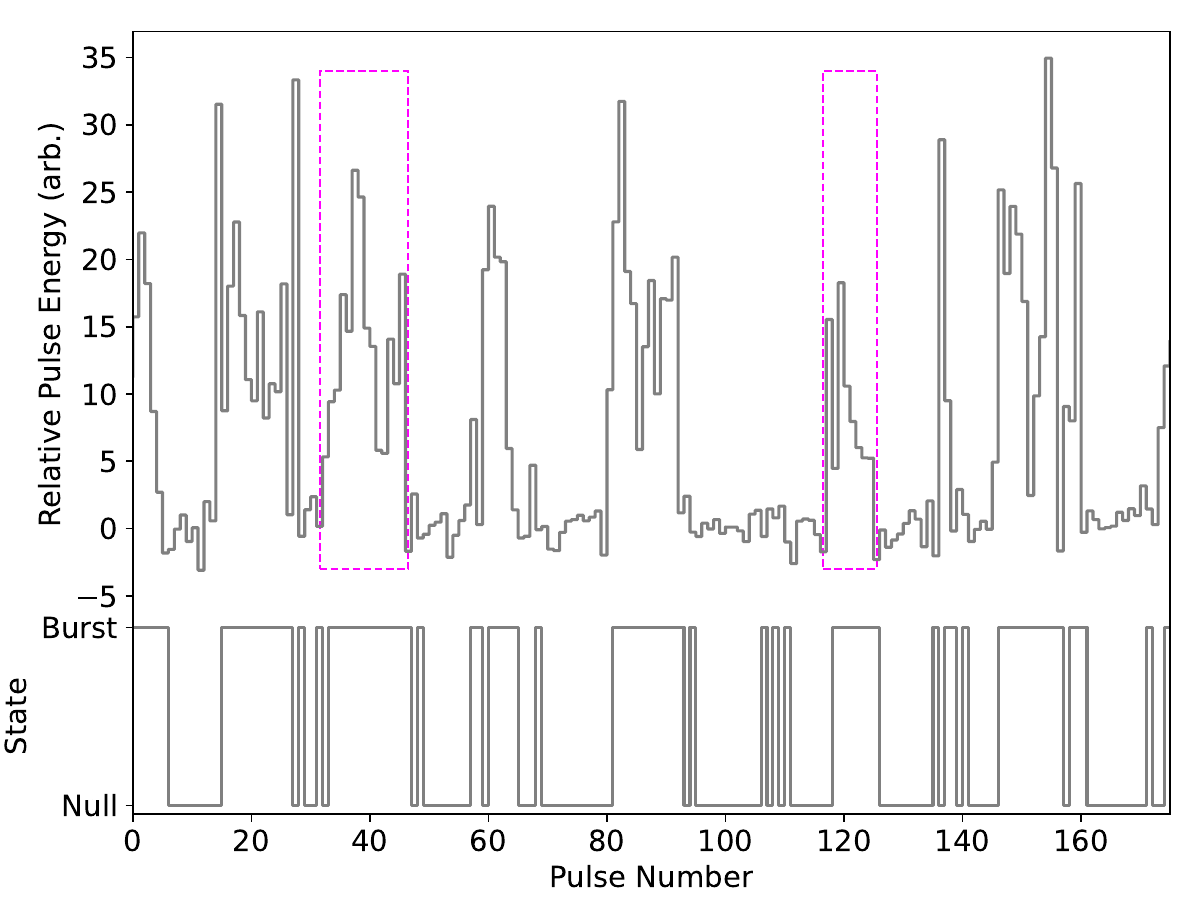}
		\caption{The upper panels display the pulse energy sequences for PSR J1945+1211 (left plot) and PSR J2323+1214 (right plot), and the lower panels indicate their corresponding null and burst states. The pink rectangular boxes indicate transitional patterns.} 
		\label{energyseries_1945} 
	\end{figure*}

	\begin{figure}
		\centering
		\includegraphics[width=1\columnwidth, angle=0]{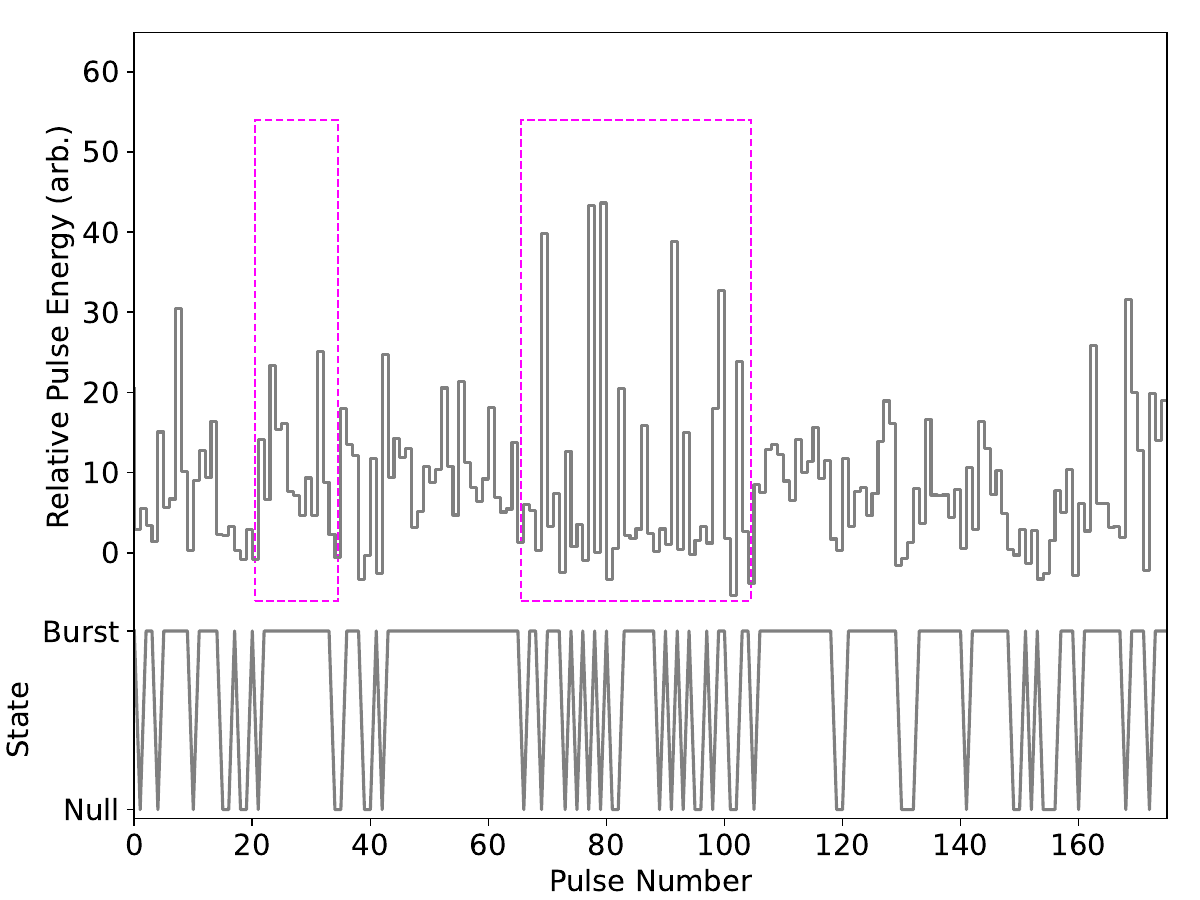}
		\caption{Similar to Figure~\ref{energyseries_1945}, but for PSR J1900$-$0134.}
		\label{energyseries_1900} 
	\end{figure}	
	
	Our observations revealed unique on-pulse energy distributions for three pulsars, each indicating different energy states. PSR J1945+1211 exhibited a bimodal distribution, with a threshold energy value for Gaussian fitting indicated by the vertical dotted orange dashed line (Figure~\ref{energy_all}), suggesting two distinct energy states. To categorize the pulses, we classified those with energy below $\gamma \times \eta_{\rm on}$ as `null emission states' and those at or above this threshold as `burst emission states'. This classification offers insights into the diverse energetic behaviors of these pulsars. The parameter $\gamma$, which encapsulates key observational metrics, aligns with the NF estimates detailed in Table \ref{table:fitting}. We estimated the $\gamma$ values for PSR J1945+1211, PSR J2323+1214, and PSR J1900$-$0134 approximately 8, 5, and 3, respectively.
	
	To further investigate the behavior of PSR J1945+1211, PSR J2323+1214, and PSR J1900$-$0134, we examined their integrated pulse profiles, as shown in Figure~\ref{profile_all}. For each pulsar, the burst emission state is depicted with the red solid line, the null emission state with blue solid line, and the entire observation period in dotted black line. In particular, burst state profiles show peak intensities exceeding those of the entire observation and have slightly wider widths at the 10\% and 50\% intensity levels. Null-state profiles exhibit the weakest emissions, often near zero.
	
	\begin{figure*}
		\centering
		\includegraphics[width=0.67\columnwidth, angle=0]{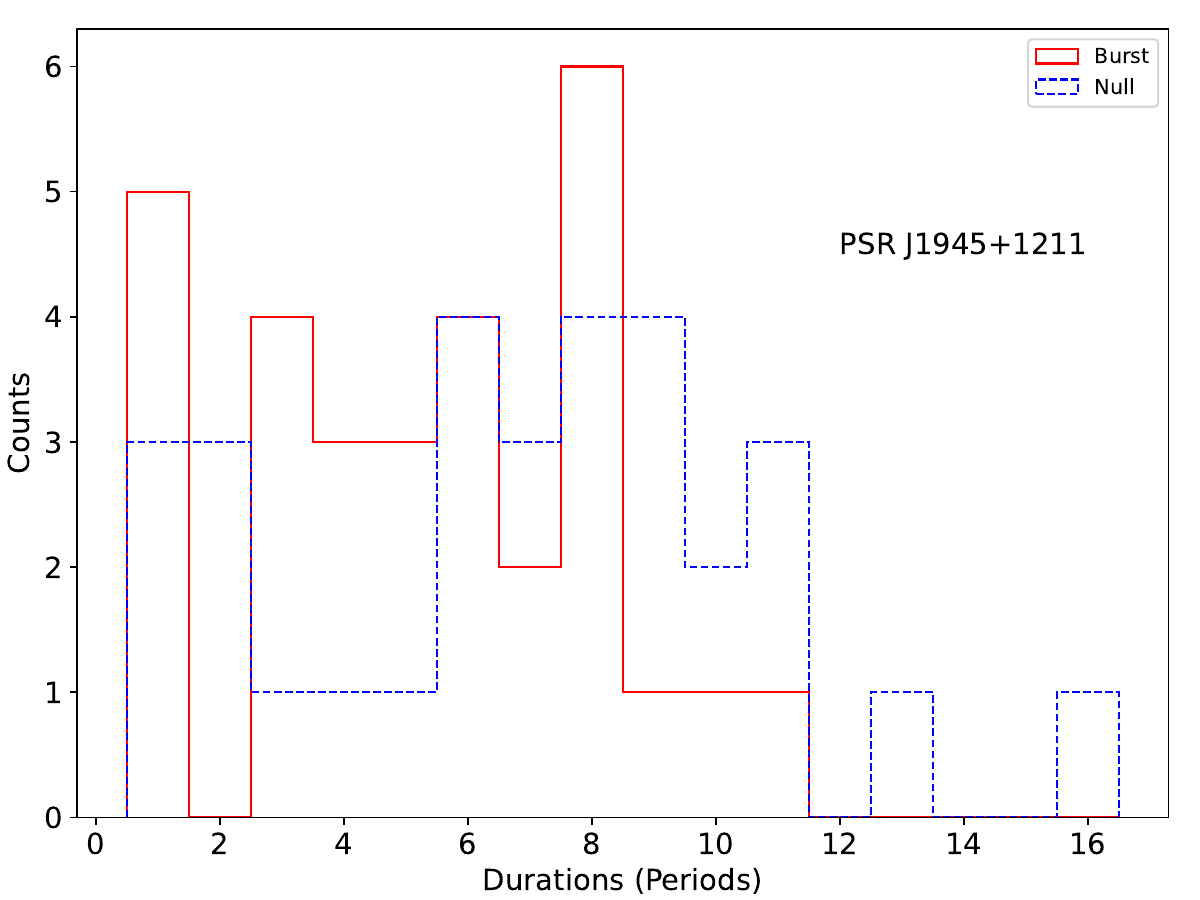}
		\includegraphics[width=0.67\columnwidth, angle=0]{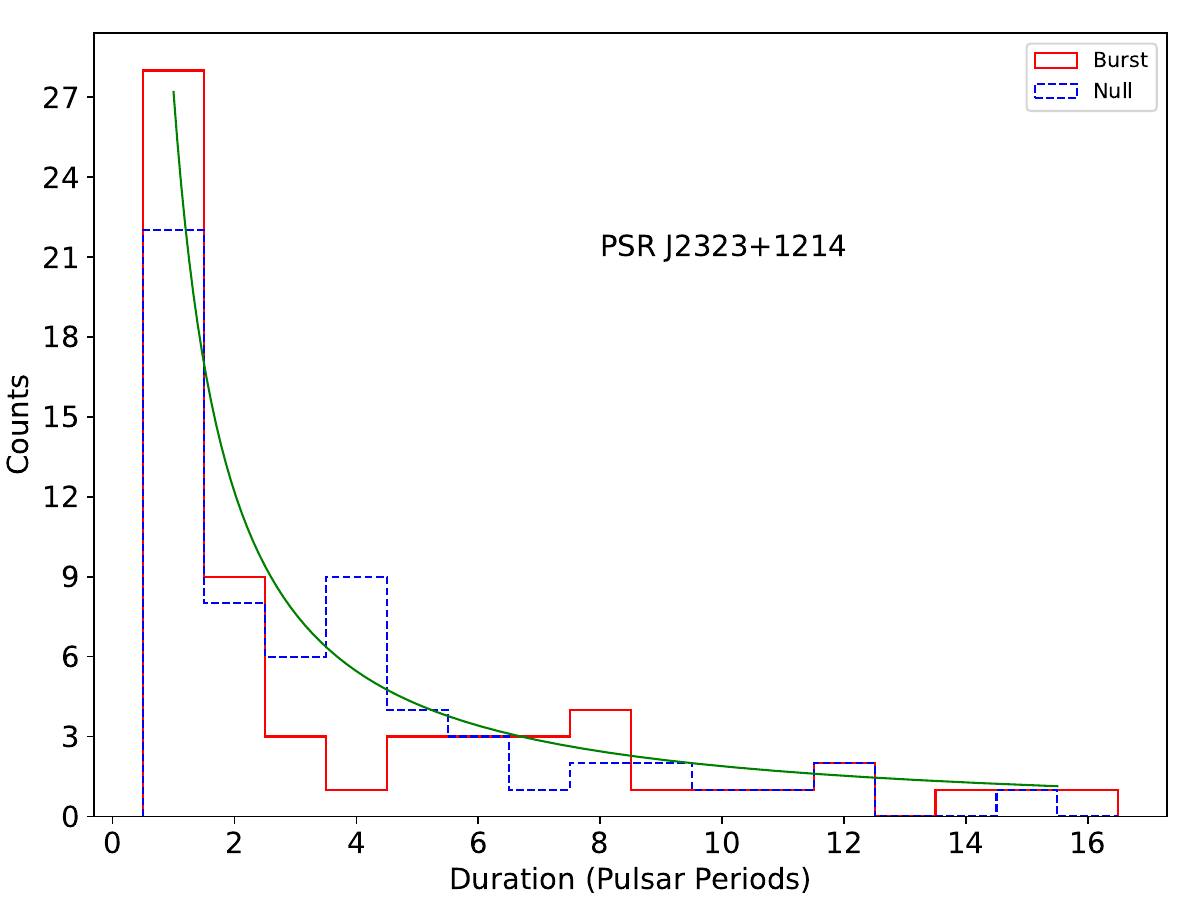}
		\includegraphics[width=0.67\columnwidth, angle=0]{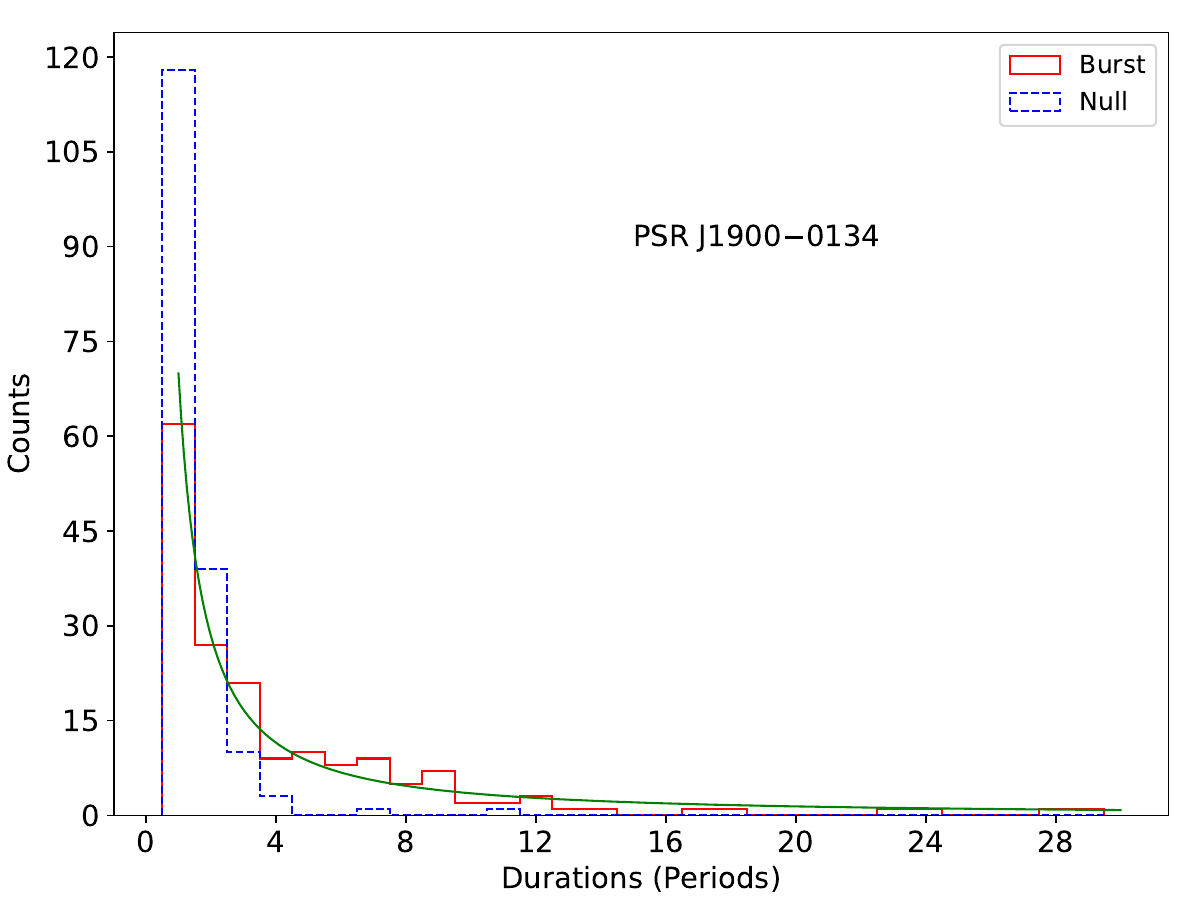}
		\caption{Histograms displaying the durations of null and burst states for PSR J1945+1211 (left plot), PSR J2323+1214 (middle plot) and PSR J1900$-$0134 (right plot), respectively. In the middle and right plots, the solid green line signifies a power law distribution fitted to the durations of both null and burst emission states of the respective pulsar.}
		\label{Duration_1945}
	\end{figure*}
	
	\begin{figure*}
		\centering
		\includegraphics[width=0.69\columnwidth]{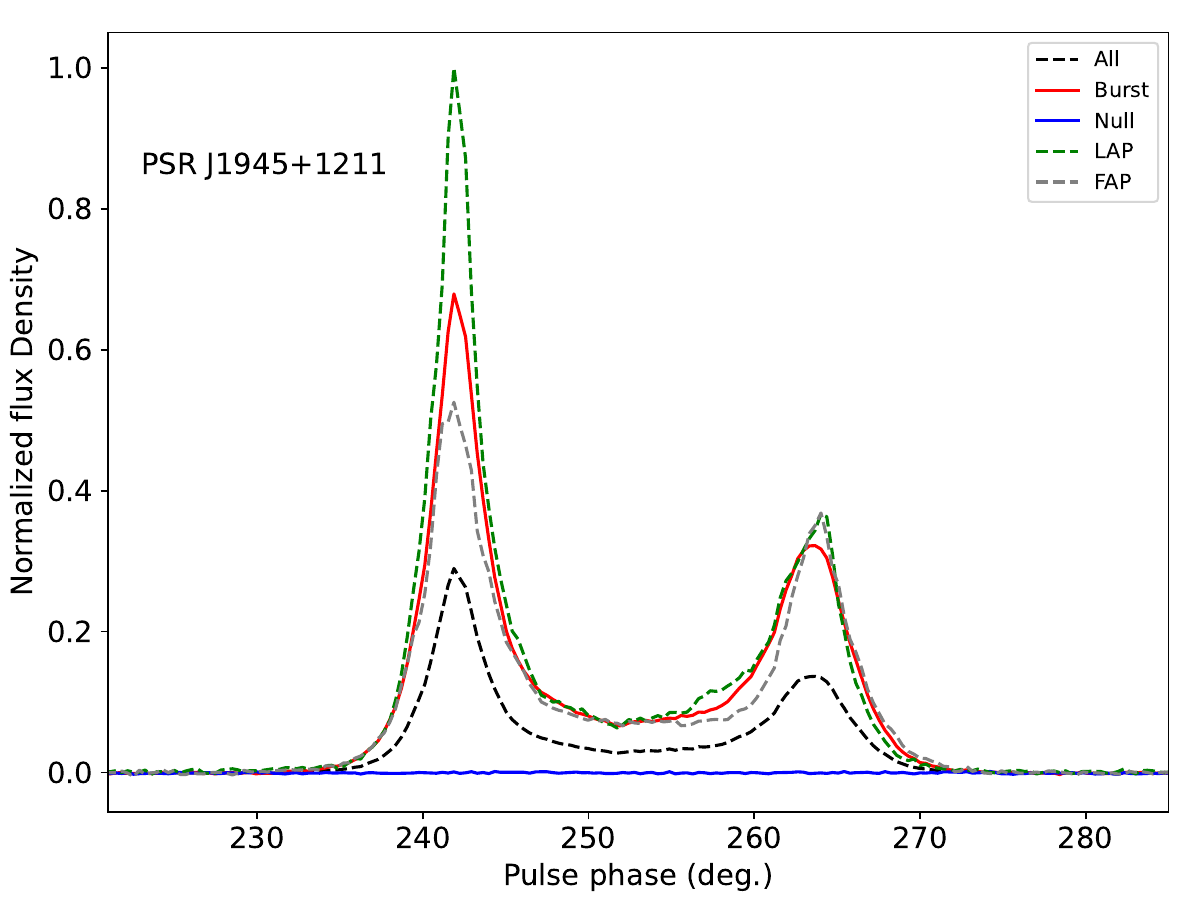}
		\includegraphics[width=0.69\columnwidth]{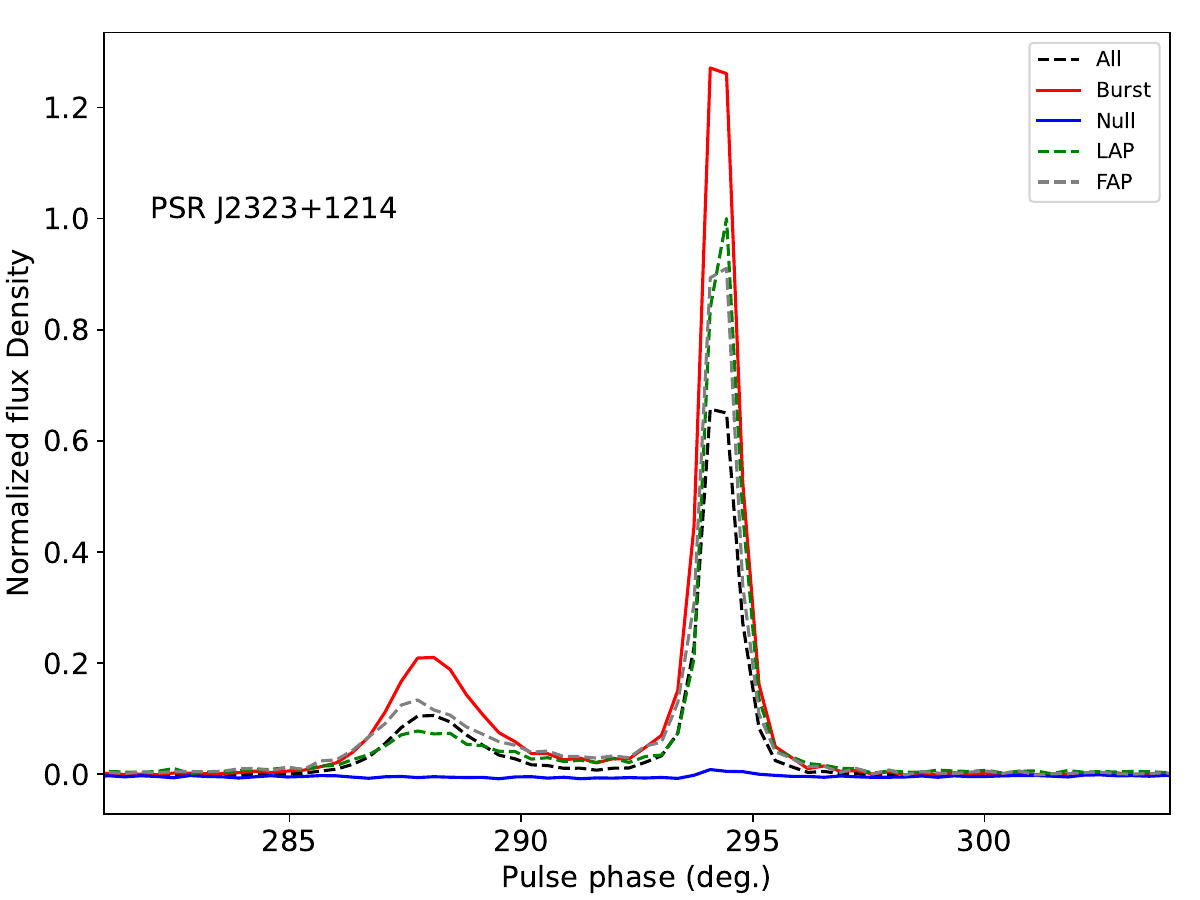}
		\includegraphics[width=0.69\columnwidth]{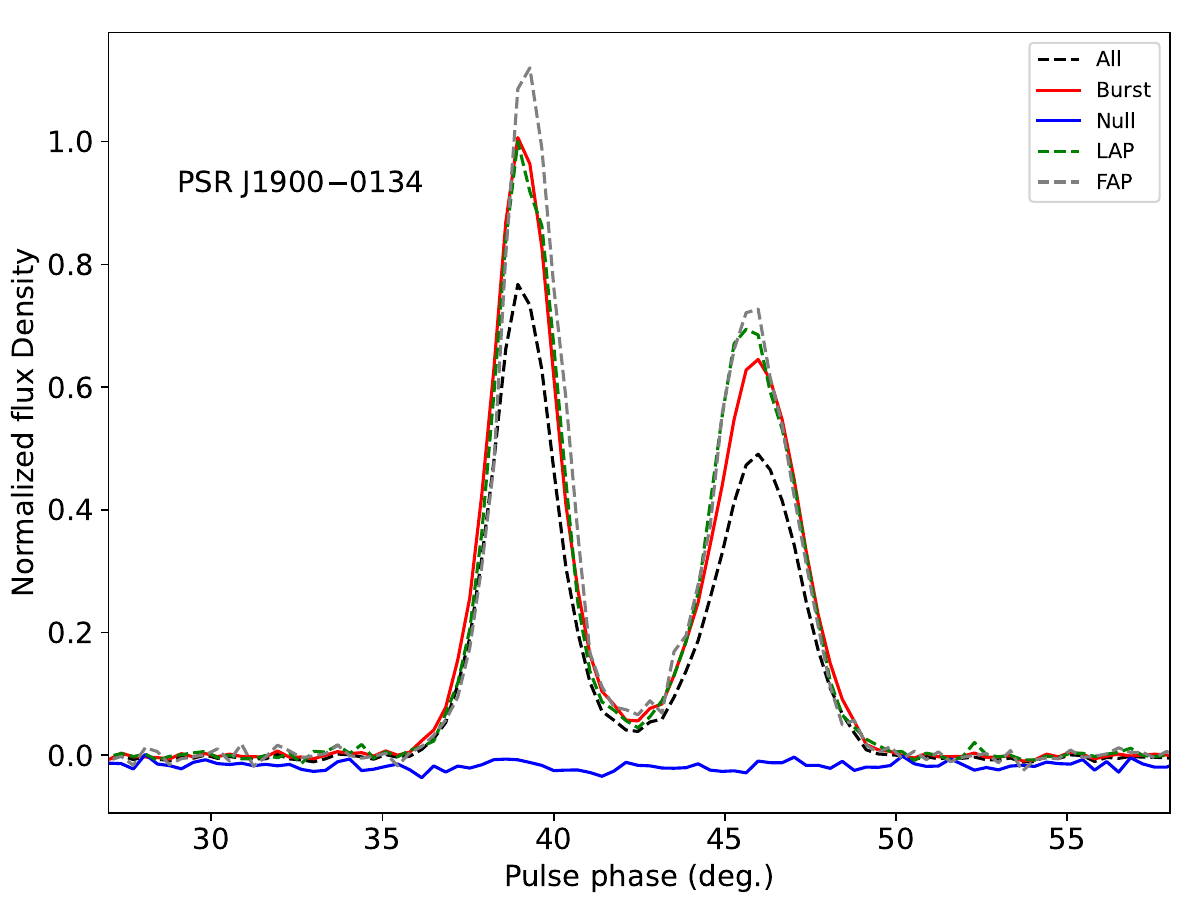}
		\caption{The integrated profiles of all burst pulses (solid red line), first active pulses (FAP, dashed gray line), last active pulses (LAP, dotted green line), and the overall observation (dashed black line) for PSR J1945+1211 (left plot), PSR J2323+1214 (middle plot), and PSR J1900$-$0134,  respectively. All profiles are normalized to the peak flux density of the LAP.}
		\label{FAP_LAP_1945}
	\end{figure*}
	\subsection{Durations of nulls and bursts}\label{sec:null_burst_1945}
	
	Some pulsars exhibit a periodic transition between null and burst emission states, providing insights into their emission mechanisms \citep{Lyne88}. These transitions can vary between either gradual or abrupt as depicted in Figure~\ref{energyseries_1945}. The upper panel categorizes individual pulses as either null or burst, while the lower panel visually represents the regularity and repeatability of the transitions, similar to previous research on the varying behavior of pulsars \citep{KAL06, Wang07}. The nature of these transition patterns can offer insights into the underlying mechanisms. PSR B0818$-$41 \citep{Bhat10}, PSR J1727$-$2739 \citep{Wen16, Rejep22}, and PSR J1909+0122 \citep{Chen23} each display unique complexities in the emission transitions. Our findings, highlighted in Figure~\ref{energyseries_1945} and illustrated by the dotted rectangles, not only emphasize the diversity in transition patterns among pulsars but also suggest that the transitions observed in PSR J1945+1211 and PSR J2323+1214 are typically abrupt. 
	
	In contrast, PSR J1900$-$0134 exhibits a different behavior, with transitions from bursts to nulls or vice versa occurring either gradually or abruptly. Figure~\ref{energyseries_1900} presents two typical examples of pulse energy variations over time, illustrating the transitional patterns.
	The first rectangle demonstrates a gradual transition from a burst to a null state, immediately followed by an abrupt return to a burst state. Specifically, the pulse energy decreases steadily over several pulses, indicating a gradual transition to the null state. Then, there is a sudden increase in pulse energy within one or two pulses, marking an abrupt return to the burst state.
	The second rectangle reveals a different pattern, characterized by frequent and erratic shifts from short bursts to null states. Here, transitions are both abrupt and gradual. Some transitions from burst to null, or vice versa, occur abruptly within one or two pulses, while others show a more gradual change in pulse energy over a few pulses before changing states.
	
	Figure~\ref{Duration_1945} presents a detailed examination of the emission durations for the three pulsars. The histogram for PSR J1945+1211 in the left panel distinguishes between burst emission and nulls using a dotted blue line and a solid red line, respectively. 
	It reveals that both burst emission and nulls have a similar median duration of six pulsar periods. While burst emission can persist for up to eleven pulsar periods, nulls can extend slightly longer, up to 16 pulsar periods. However, the majority of nulls overlap significantly with burst emission duration, ranging from 0 to 11 pulsar periods, with only a few null pulses extending beyond 11 periods. This suggests that although null pulses can potentially outlast burst emissions, their duration are generally comparable.
	The histogram in the middle panel pertains to PSR J2323+1214. The data indicates that the burst emission in this pulsar can last for a maximum of 16 periods, with a median duration of just two periods, while nulls have a maximum duration of 15 periods and a median duration of three periods. This contrasts with the longer median duration for both the burst emission and nulls in PSR J1945+1211. This points to intrinsic differences in the emission processes or perhaps in the emission environments for the two pulsars.
	The emission states for PSR J1900$-$0134, depicted in the right panel of Figure~\ref{Duration_1945}, show notable disparities. The burst state for this pulsar is capable of lasting up to 29 pulsar periods, but typically has a median duration of just 2 periods. On the other hand, the null state exhibits a duration of approximately eleven pulsar periods, with a median at a single pulsar period. The significant contrast in the duration of the burst and null states for PSR J1900$-$0134 further emphasizes the different emission properties among the three pulsars.
	
	Previous investigations of null pulsars \citep{Wen16, Guo23} showed that the duration of nulls generally adheres to a power-law distribution. Specifically, the extent of null lengths aligns with a power law pattern within the brief span of 0–15 pulsar periods. The nulls in PSR 2323+1214 seem to be consistent with this model, and the power law can also be applied to the burst state of PSR J1900$-$0134. However, PSR J1945+1211 does not exhibit this power-law behavior for the duration of either its null or burst states.
	
	\subsection{Active pulses around nulls}
	
	Figure~\ref{FAP_LAP_1945} presents a detailed analysis of the integrated profiles of all burst pulses, first active pulses (FAP), and last active pulses (LAP), highlighting the differences in the temporal behavior of active pulses around the null pulses for the three pulsars. For PSR J1945+1211, there is a marked difference in the profile intensity obtained from the FAP than that from the LAP. The LAP displays a notably higher peak intensity in the leading component than that from both the FAP and the integrated profile for all burst pulses. This suggests an increase in emission just before the pulsar transitions into a null state, potentially indicative of heightened magnetospheric activity  \citep{Rankin2008, Krishnakumar2015} as the pulsar depletes its charged particles. In contrast, the FAP shows a significant reduction in intensity, which may signify a weakened emission mechanism immediately after a null period. PSR 2323+1214 presents a contrasting profile, where the FAP's detectability is significantly pronounced, with its peak intensity resembling that of the average profile. However, the LAP in PSR 2323+1214 manifests a lower peak intensity compared to both the FAP and the average profile, highlighting the complexity of emission processes around null pulses. For PSR 1900$-$0134, the profiles indicate that both the FAP and the LAP have intensities close to the average pulse profile, with the FAP showing slightly higher intensity than the LAP. This pattern suggests a relatively stable emission process in this pulsar, with minor fluctuations around the null periods. The observed differences in the FAP and LAP intensities underscore the diverse behaviors of pulsars, contributing to a deeper understanding of the pulsar emission mechanisms, particularly around null periods.
	
	\begin{figure*}
		\centering
		\includegraphics[width=1\columnwidth]{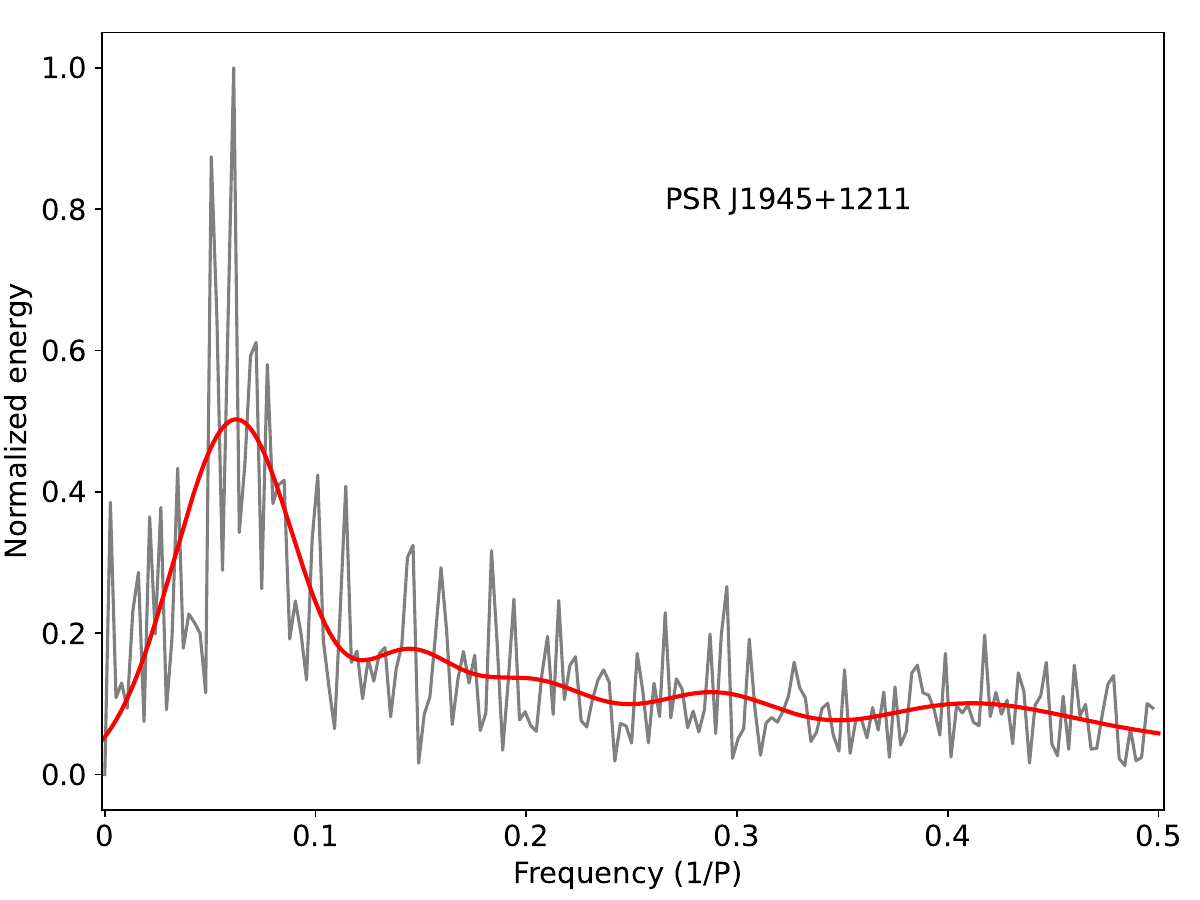}
		\includegraphics[width=1\columnwidth]{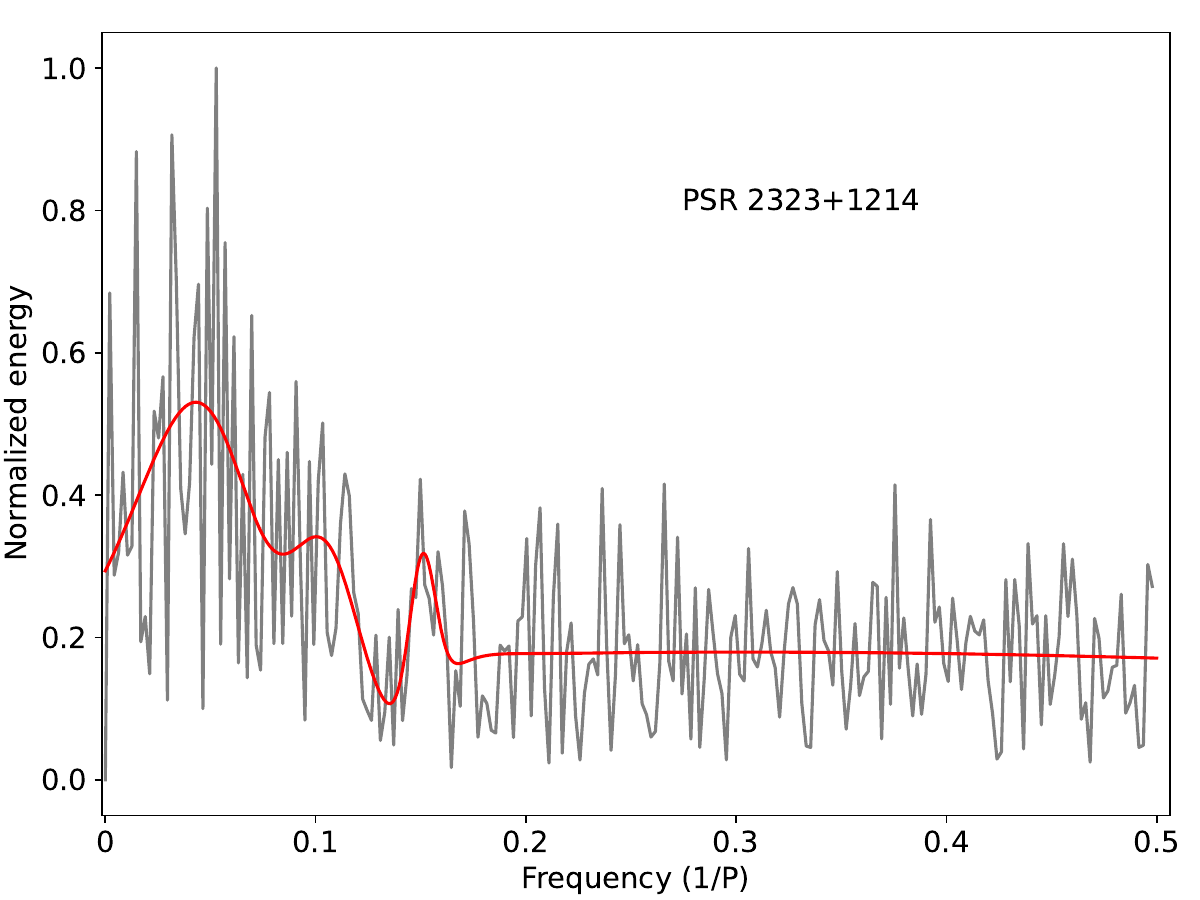}
		\caption{The Fast Fourier Transformation (FFT) fluctuation spectra analysis is presented for PSR J1945+1211 (left plot), revealing a periodicity of approximately 0.065 (1/P) equivalent to 15.38 pulsar periods. For PSR J2323+1214 (right plot), the time series analysis indicates a peak value of roughly 0.045 (1/P), corresponding to 22.22 pulsar periods in both null and burst states. The solid red line in both plots represents a smoothed curve to emphasize the dominant features in the fluctuation spectrum.}
		\label{FFT_1945}
	\end{figure*}	
	
	In addition to intensity variations, the pulse profiles of PSR J1945+1211 and PSR J2323+1214 reveal differences in the width of active pulses. For PSR J1945+1211, the FAP, measured at $W_{10}$, spans a width of approximately $31.3^\circ$. This width is greater than the $29.53^\circ$ of the average profile (represented by the dashed black line), suggesting that the FAP has a wider profile compared to the overall pulse average. LAP, however, measures about $30.04^\circ$, indicating a narrowing of the pulse width just before the pulsar enters a null state. Despite this narrowing, the LAP still has a wider profile than the average pulses obtained from the overall observation for this pulsar. These variations in pulse width, which may reflect alterations in the pulsar's beam patterns during different phases, are indicative of intrinsic properties of pulsars and are thought to be influenced by the viewing geometry \citep{Pulsars1977} and the magnetospheric physics \citep{Melrose14}. This phenomenon aligns with observations in other pulsars that exhibit similar behaviors, such as PSR J1727$-$2739 \citep{Wen16, Rejep22} and PSR J1909+0122 \citep{Chen23}. The variations can be attributed to changes in the magnetospheric conditions of the pulsar. The alignment of the LAP with the red profile in PSR 1900$-$0134, along with the subtle shift in the FAP's peak intensity away from the leading component, suggests the potential for positive drifting subsequent to null pulses. The variations in the low-level bridge emission between the LAP and FAP profiles, especially the pronounced intensity in the FAP for PSR J1945+1211 and the alignment of the LAP with the overall profile in PSR J2323+1214, reveal additional layers of complexity in pulsar emissions. 
	
	\begin{figure}
		\centering
		\includegraphics[width=1\columnwidth]{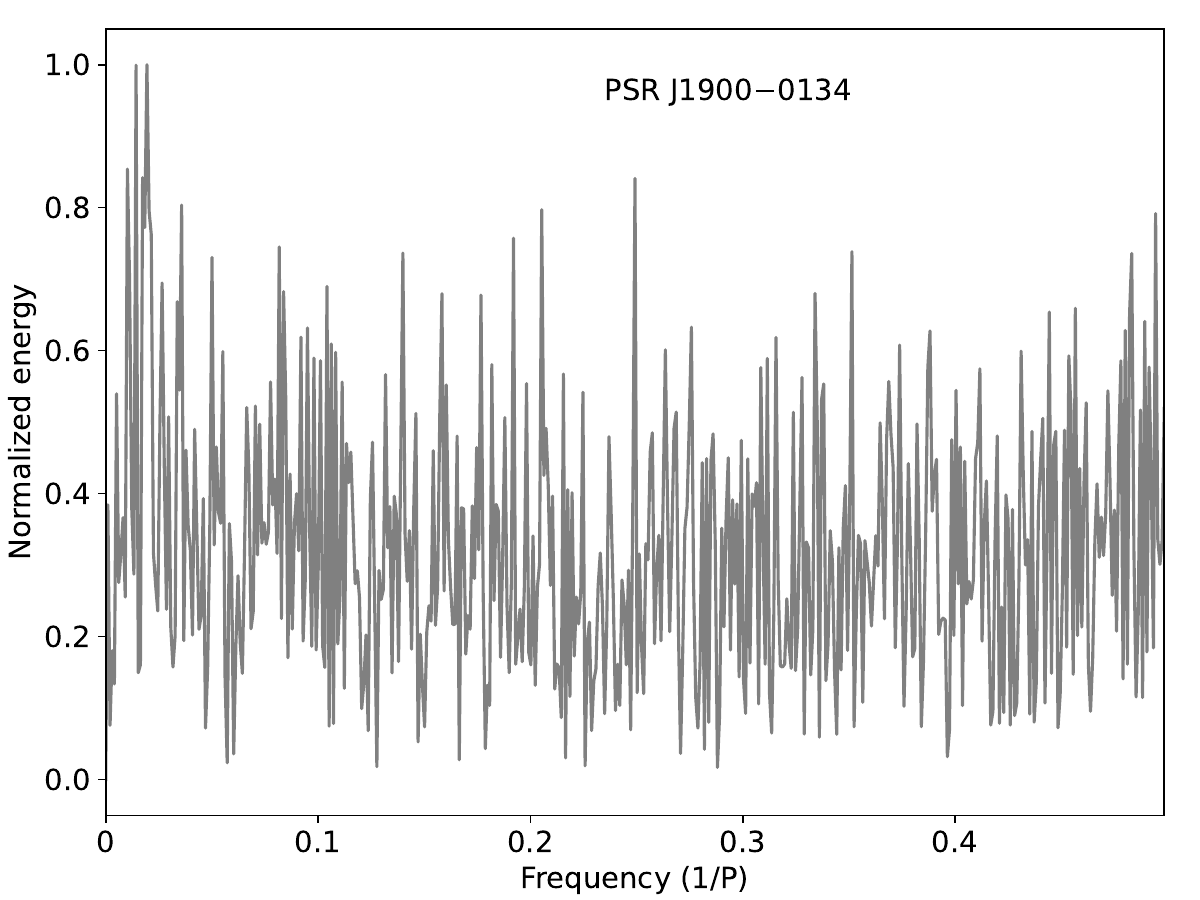}
		\caption{Similar to Figure~\ref{FFT_1945}, the FFT fluctuation spectra analysis for PSR J1900$-$0134 is presented, revealing many peaks. However, these peaks do not indicate any clear periodicity in the pulsar.}
		\label{FFT_1900}
	\end{figure}

	\subsection{Quasi-periodic fluctuations}\label{sec:periodic_null_1945}
	
	Several pulsars have been observed to display patterns of nulls that resemble quasi-periodic behavior. This quasi-periodic behavior is evident in Figure~\ref{energyseries_1945}, where the pulses from PSR J1945+1211 and PSR J2323+1214 alternate between null and burst states. Recent findings have suggested that pulsar null may follow patterns rather than being purely random \citep[e.g.,][]{RR09, Zhi23, Chen23, Wang23}. To determine whether the nulls of PSR J1945+1211, PSR J2323+1214, and PSR J1900$-$0134 are periodic, we used a method similar to that of \citet{Wang23} to process the data. In our analysis, each of the burst pulses was designated as `1' and the null pulse as `0'. We conducted a Fast Fourier Transform (FFT) analysis on one-zero sequences, covering 256 periods. This analysis, which focused on specific pulse phases, aims to identify the temporal evolution of null periodicity and transitions between burst and null states.
	
	\begin{figure*}
		\centering
		\includegraphics[width=0.67\columnwidth, angle=0]{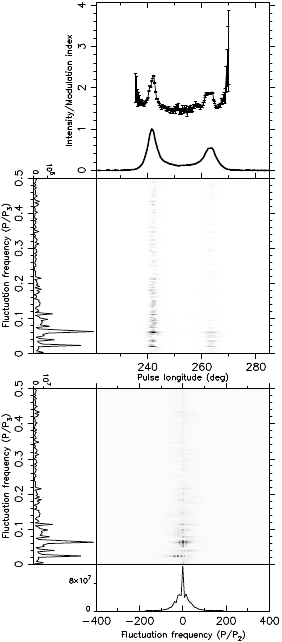}
		\includegraphics[width=0.67\columnwidth, angle=0]{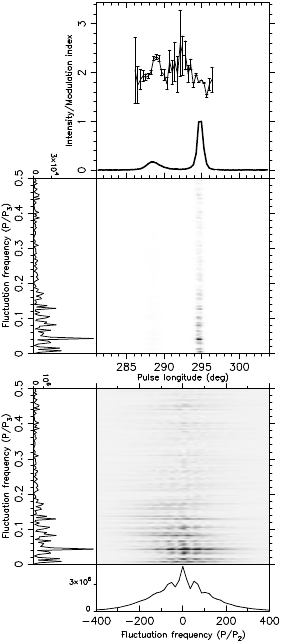}
		\includegraphics[width=0.67\columnwidth, angle=0]{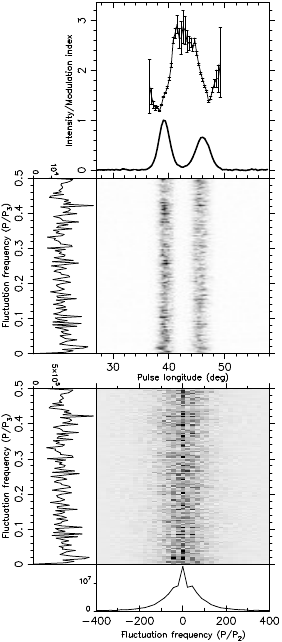}
		\caption{The fluctuation analysis for PSR J1945+1211 (left plot), PSR J2323+1214 (middle plot), and PSR J1900$-$0134 (right plot), respectively. Top: Pulse profiles with modulation index. Middle: LRFS with the side panel showing horizontally integrated power. Bottom: 2DFS for each pulsar with side panels indicating horizontal and vertical integrated powers. A 256-point Fourier transform is employed and then averaged over the blocks of the entire pulse sequence.}
		\label{Drifting_subpulse}
	\end{figure*}

	The left panel of Figure~\ref{FFT_1945} presents the results of fluctuation spectrum analysis for PSR J1945+1211. We found a peak at $0.065 \pm 0.0038$, which is equivalent to $15.38 \pm 0.90$ pulsar periods. The uncertainty values were determined using FWHM/$2\sqrt{2ln(2)}$. Here, `FWHM' stands for the full width at half maximum of the peak, while $2\sqrt{2 ln(2)}$ is a scaling factor based on the Gaussian approximation \citep{Basu16}. 
	For PSR J2323+1214, as shown in the right panel of Figure~\ref{FFT_1945}, the FFT analysis of the binary data sequence revealed the presence of a periodic component, with the dominant period calculated to be approximately $22.22 \pm 2.4$ pulsar periods. The value signifies the interval of the most prominent repeating pattern across the sequence, rather than the maximum number of consecutive 1's or 0's. The result indicates that the null-burst sequence exhibits regularity or repetitive structure. Similar analyzes were conducted for PSR J1900$-$0134, as shown in Figure~\ref{FFT_1900}. Many peaks were investigated in the sequences, which yielded no clear periodicity in the pulsar.
	
	The Kolmogorov-Smirnov (KS) test, applied to PSR 2323+1214, supports the presence of emission variations between the leading and trailing components. The analysis yielded a KS statistic of 0.2084, indicating significant divergence in the cumulative distribution functions of the two components. Notably, the p-value obtained was 1.92 $\times$ 10$^{-9}$, which is well below the conventional significance threshold of 0.05. This compels the rejection of the null hypothesis that posited the emission variations in the leading and trailing components as coming from the same distribution.

	\subsection{Subpulse drifting modulations}\label{sec:drifting}
	
	The modulation index, depicted in the upper panel of each plot in Figure~\ref{Drifting_subpulse}, represents the pulse intensity fluctuations for each pulsar. Mathematically, the modulation index is defined as $m_i = \frac{\sigma_i}{\mu_i}$ \citep{Weltevrede06}, where $\sigma_i$ is the longitude-resolved standard deviation and $\mu_i$ denotes the mean intensity for the specified longitude bin $i$. Obtained from using the longitude-resolved fluctuation power spectrum (LRFS) methodology \citep{Edwards03, Weltevrede12}, the magnitude of the modulation index at a given longitude indicates the strength of pulse intensity variation at that longitude. In addition, the modulation index varies with the phase of the average profile. Error bars for the index are determined through data bootstrapping, with each bootstrap iteration incorporating random noise equivalent to the RMS of the off-pulse region. The observation that each of the three pulsars has a modulation index exceeding one near their pulse peaks suggests a relative stability in pulse intensity during burst phases. However, there is noticeable variation in the modulation index across the leading and trailing edges, and significant fluctuations are observed between these two components, known as the bridge emission, of the pulse profile. Specifically, for PSR J1900$-$0134, there is a significant variation in pulse intensity within the bridge emission area, as indicated by a notably high modulation index.
	
	First proposed by \citet{Backer1970a}, LRFS is crucial for analyzing pulse intensity variations over time and is employed to study systematic or disordered subpulse modulation. The process begins by removing the identified nulls from the single-pulse sequence, thus creating a sequence composed only of burst pulses. This sequence is segmented into blocks, each block containing 256 pulses. The LRFS for each block is determined by performing a discrete Fourier transform on each phase bin. The final spectrum is an average of the fluctuation power spectra of these blocks, achieved by using a 256-point FFT for high frequency resolution. In addition to LRFS, the two-dimensional fluctuation spectrum (2DFS), introduced by \citet{Edwards02}, plays a pivotal role. Although LRFS is essential for characterizing the periodicity of pulse intensity modulation ($P_3$), 2DFS is used to assess whether the single-pulse drifts longitudinally ($P_2$), as shown in the middle and bottom panels of each column in Figure~\ref{Drifting_subpulse}. However, for an accurate estimation of $P_3$, LRFS needs to be supplemented with the insights gained from $P_2$ and its drift rate. This gap is effectively addressed by 2DFS. The vertical frequency axes of LRFS and 2DFS represent $P/P_{3}$, while the horizontal frequency axis of 2DFS corresponds to $P/P_{2}$, where $P$ signifies the pulsar period. For an in-depth understanding of the fluctuation power spectrum, please refer \citet{Weltevrede06}. Consequently, the 2DFS analysis for PSR J1945+1211 indicates that the modulation exhibits peak frequencies at approximately 0.025 \( P/P_{3} \), 0.065 \( P/P_{3} \), and 0.115 \( P/P_{3} \), corresponding to \( P_{3} \) values of roughly \( 40 \pm 0.24\, P \), \( 15.38 \pm 0.90 \, P \), and \( 8.45 \pm 0.11 \, P \), respectively. The periodicity at \( P_{3} = 15.38 \pm 0.90 \, P \) is associated with transitions between null and burst emission states or quasi-periodic nulling, as illustrated in Figure~\ref{FFT_1945}. Meanwhile, the periodicities at \( P_{3} = 40 \pm 0.24 \, P \) and \( 8.45 \pm 0.11 \, P \) are more probable drift periodicities for this pulsar, suggesting the presence of two distinct drift modes.
	
	For PSR J2323+1214, the peak frequency at 0.045\( P/P_{3} \) corresponds to \( P_{3} = 22.22 \pm 2.4 \,P \), also indicating periodic transitions between null and burst states (Figure~\ref{FFT_1945}). For PSR J1900$-$0134, we found that the peak frequency is at 0.43 \( P/P_{3} \), or \( P_{3} = 2.2 \pm 2.1 \, P \), suggesting rapid null-burst transitions or different modulation behavior. However, for all three pulsars, the lower panel in each column reveals that the fluctuation frequency peaks at zero. This indicates a $P_{2}$ value of zero and no sub-pulse drifting in their pulse phases, suggesting phase-stationary emission with periodic intensity modulations \citep{Yan19}.  
	
	\section{Dwarf Pulses}\label{sec:dwarfPulses}
	
	Dwarf pulses are a distinct type of weak and narrow radio emission observed from pulsars. These pulses have been particularly noted in the study of the pulsar PSR J2111+4644 \citep{Chen2023}. Dwarf pulses are characterized by their sporadic nature and occur during the pulsar's ordinary null state. They are much weaker and narrower than typical burst pulses. An example is shown in Figure~\ref{single_pulse} for PSR J2323+1214 around pulse number 74. Such a relatively weak and narrow radio emission is evident and surrounded by null pulses. Identifying dwarf pulses involves distinguishing the faint emission from the background noise and null states.

\begin{figure}
		\centering
		\includegraphics[width=1\columnwidth]{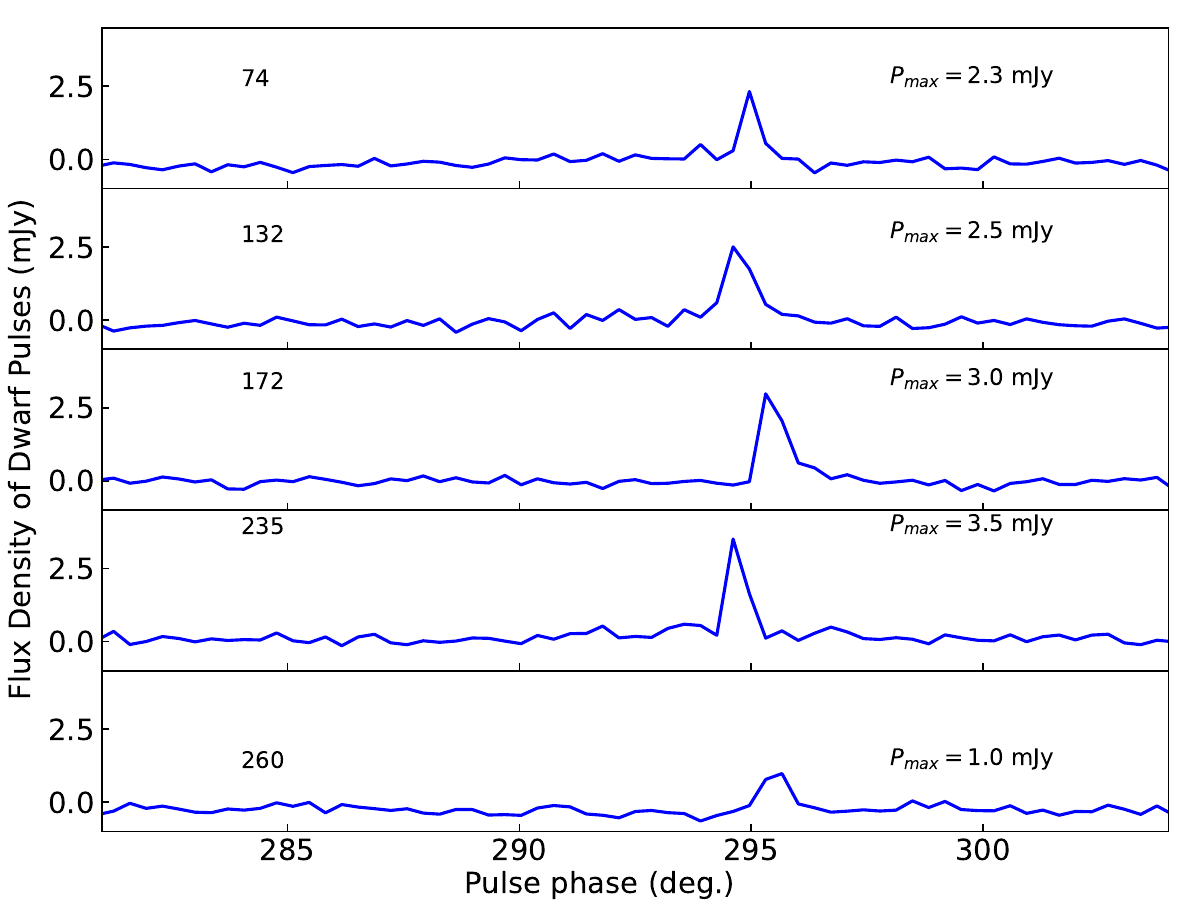}
		\caption{The pulse profiles of five dwarf pulses detected in PSR J2323+1214. The maximum peak values ($P_{\text{max}}$) of each dwarf pulse are illustrated at the right corner of each subplot. The dwarf pulses were detected at pulse numbers 74, 132, 172, 235, and 260.}
		\label{Dwarf_profile}
	\end{figure}

	To identify dwarf pulses, we followed a method that involves a two-step process. First, we identify the burst pulses as described in Section \ref{sec:characterization}. Burst pulses are those with an energy greater than $5\eta_{\rm on}$. This threshold ensures that we are selecting pulses with significantly higher energy than the average noise level, thus categorizing them as burst pulses. Second, after identifying the burst pulses, we select the pulses that meet the specific criteria indicative of dwarf pulses. From these burst pulses, we select dwarf pulses based on two criteria: A peak flux density that is less than two-third of the peak intensity in the average pulse profile, and a pulse width that is five times narrower at 10\% of the peak intensity. We identified five dwarf pulses from the 475 observed pulses of PSR J2323+1214 at pulse numbers 74, 132, 172, 235 and 260 as shown in Figure~\ref{Dwarf_profile}, which represents about 0.011\% of the total observation data. The effective width of the dwarf pulses is generally narrower, typically less than a threshold of $1.87^\circ$. Dwarf pulses occur less frequently and exhibit a more sporadic emission pattern compared to normal burst pulses. In particular, all detected dwarf pulses were found only in the trailing component of the integrated profile. 
	
	The observation of weak and narrow dwarf pulses surrounded by null pulses, as seen in PSR J2323+1214, has significant implications for the radiation process. The phenomenon suggests that the conditions necessary for radio emission are not consistently met, leading to frequent transitions between null and burst states. The presence of null pulses indicates periods where the pulsar's radio emission ceases, likely because of a deficit of outer-flowing particles, loss of coherence among the emitting particles, or the failure of gap formation within the magnetosphere. The generation of dwarf pulses suggests a `fragile' state of the pulsar's emission mechanism, where only minimal particle discharge occurs, producing weak and sporadic emissions. This sporadic activity points to a highly dynamic and unstable magnetospheric environment, particularly in older pulsars near the death line, where the magnetic field strength is insufficient to maintain continuous particle discharge.
	
\begin{figure}
		\centering
		\includegraphics[width=1\columnwidth]{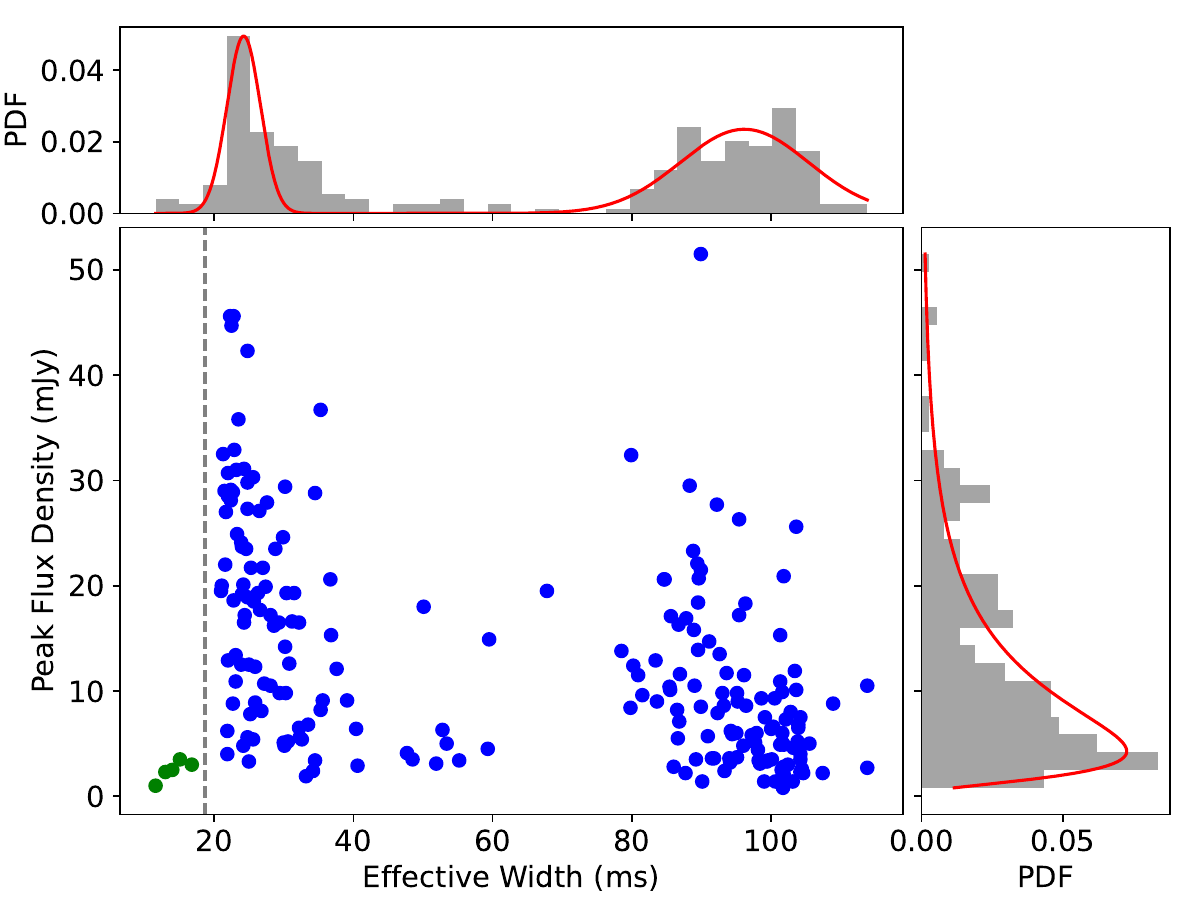}
		\caption{The population of dwarf pulses detected in PSR J2323+1214. Blue dots are burst pulses, and green dots highlight dwarf pulses. The dotted vertical gray line shows the threshold values that separate dwarf pulses from ordinary burst pulses. The red lines in the upper and right panels represent smoothed PDFs, highlighting the distribution of effective width (top) and peak flux density (right). See the main text for more descriptions.}
		\label{dwarfpulse_2323}
	\end{figure}

	Figure~\ref{dwarfpulse_2323} shows the population of dwarf pulses detected from PSR J2323+1214. The main panel shows the scatter plot of the peak flux density (in mJy) versus the effective width (in ms) of the detected pulses (see Appendix Table~\ref{table:burst2323}). The top panel shows the histogram of peak flux density, while the right panel shows the histogram of effective width. Overlaid on the histograms are fitted distribution curves: a red line in the right panel represents a log-normal fit to the pulse  probability density functions (PDFs) of peak flux density, while in the top panel, a bimodal Gaussian distribution is applied, highlighting two distinct populations of pulse widths, namely narrower and wider pulses. This figure illustrates the distinct population of dwarf pulses, characterized by their lower peak flux density and shorter effective width compared to ordinary burst pulses.

	\section{Bright pulses} \label{sec:brightPulses}
	
\begin{figure}
		\centering
		\includegraphics[width=1\columnwidth]{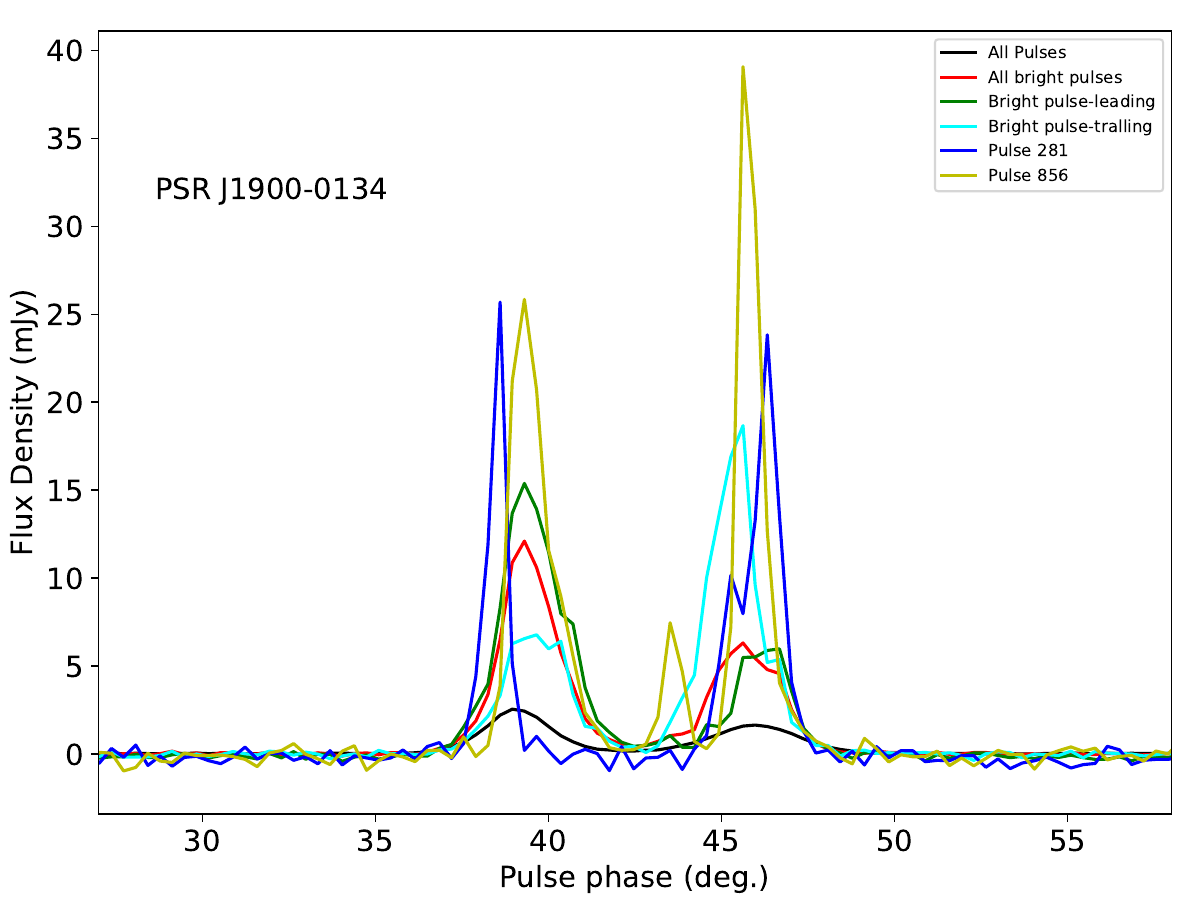}
		\caption{The average profile of bright pulses detected in PSR J1900$-$0134. The green line denotes the integrated profile of the bright emission found exclusively in the leading component, while the cyan line represents those in the trailing components. The solid black line represents the integrated pulse profile derived from the entire observation of the pulsar. See the main text for more descriptions.}
		\label{brightpulse_1900}
	\end{figure}

\begin{figure}
		\centering
		\includegraphics[width=1\columnwidth]{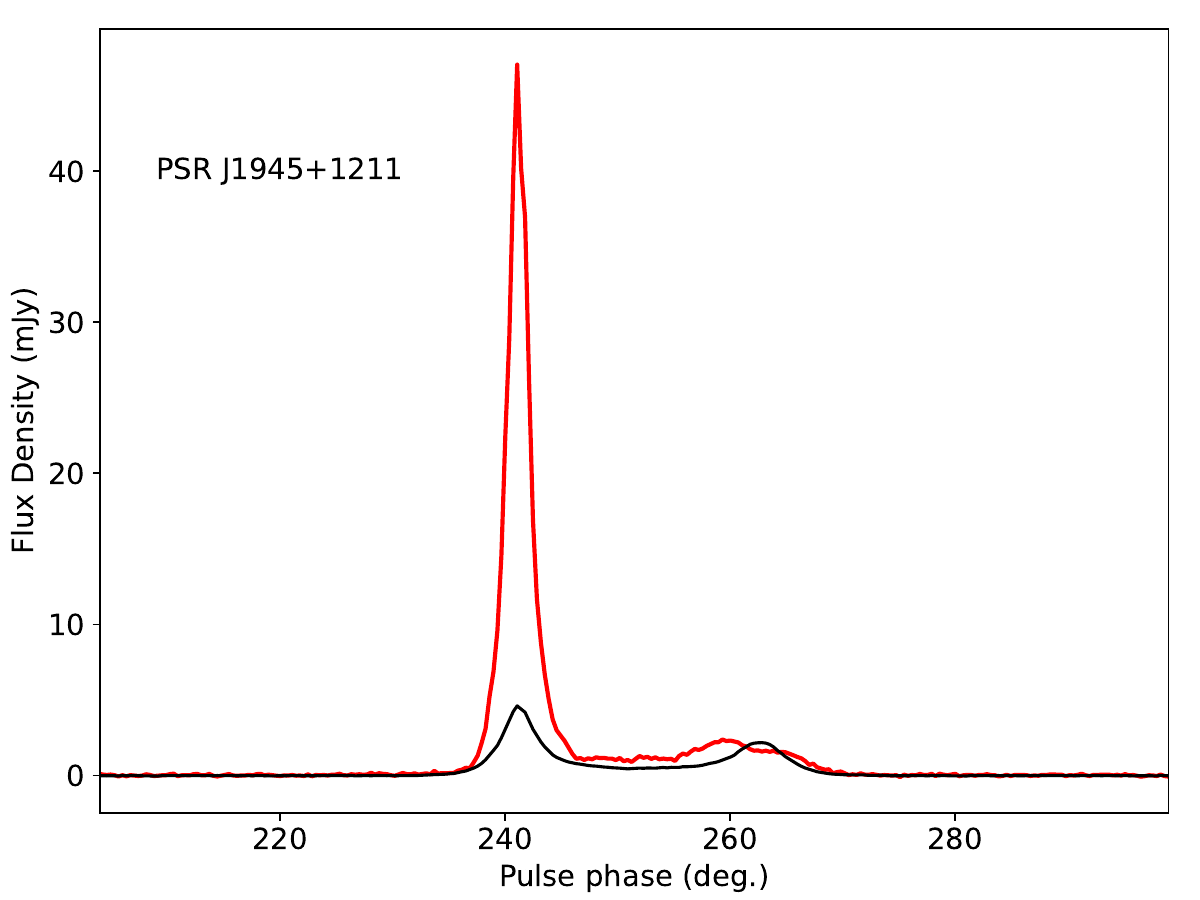}
		\caption{The red solid line represents the average pulse profile of the eleven bright pulses detected in the leading component of PSR J1945+1211. The black solid line represents the integrated pulse profile from the entire observation of the pulsar.}
		\label{brightpulse_1945}
	\end{figure}

\begin{figure}
		\centering
		\includegraphics[width=1\columnwidth]{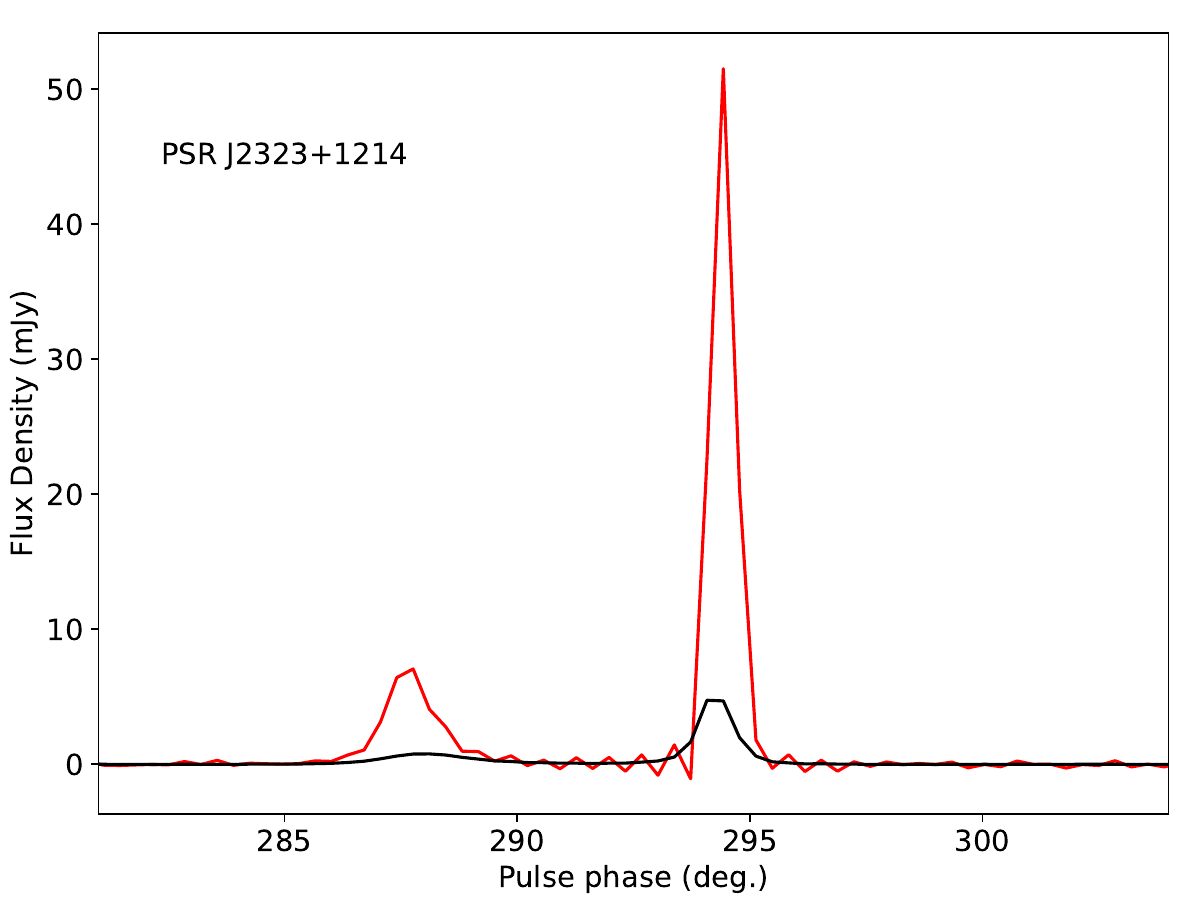}
		\caption{A single bright pulse (solid red) detected in the trailing component of PSR J2323+1214. The black solid line represents the integrated pulse profile from the entire observation of the pulsar.}
		\label{brightpulse_2323}
	\end{figure}

	We applied a stringent criterion to identify `bright pulses'  as those with peak intensity exceeding the average by a factor of ten or more. This criterion is visually demonstrated in Figure~\ref{brightpulse_1900}, Figure~\ref{brightpulse_1945}, and Figure~\ref{brightpulse_2323}, where individual bright pulse profiles are overlaid on the integrated pulse profile of the corresponding pulsar.
	
	For PSR J1900$-$0134, a nuanced analysis revealed that bright pulses made up approximately 1.3\% of the 976 single pulses. Of the 13 bright pulses detected, nine were present in the leading component and five in the trailing component. These details are illustrated in Figure~\ref{brightpulse_1900}. The integrated profile of the bright pulses in the leading and trailing components is represented by the red line. Some bright pulses were detected either in the leading or trailing component. There were also bright pulses detected simultaneously in both the leading and trailing components as those at pulse numbers 281 and 856, which are marked with blue and orange lines in the figure, respectively. 
	A subsequent investigation into the waiting times between these bright pulses uncovered an inconsistent temporal distribution. The leading component exhibited waiting times ranging from a minimum of a single pulse period to a maximum of 171 pulse periods. In contrast, the trailing component displayed waiting times from a minimum of two pulse periods to a maximum of approximately 647 pulse periods. When considering both components together, the waiting times varied from a minimum of two to a maximum of 171 pulse periods. The average waiting times were found to be around 107.5 for the leading component, 194.25 for the trailing component, and 71.67 for both components combined. These findings suggest an intricate and nonuniform pattern in the occurrence of bright pulses, which is consistent with the results from \citet{Wen21} for PSR B0031$-$07.  Figure~\ref{brightpulse_1945} shows a similar analysis for PSR J1945+1211, where the eleven bright pulses (2.92\% of 377) were all located within the leading component. These pulses, occurring at the LAP, FAP, consecutively, or in the middle of burst pulses, had an average waiting time of 35.5 pulse periods. In contrast, the bright pulses in PSR J2323+1214 stood out as single events, as shown in Figure~\ref{brightpulse_2323}, which were found exclusively in the trailing component and in the middle of burst pulses.

	\begin{figure*}
		\centering
		\includegraphics[width=2\columnwidth, angle=0]{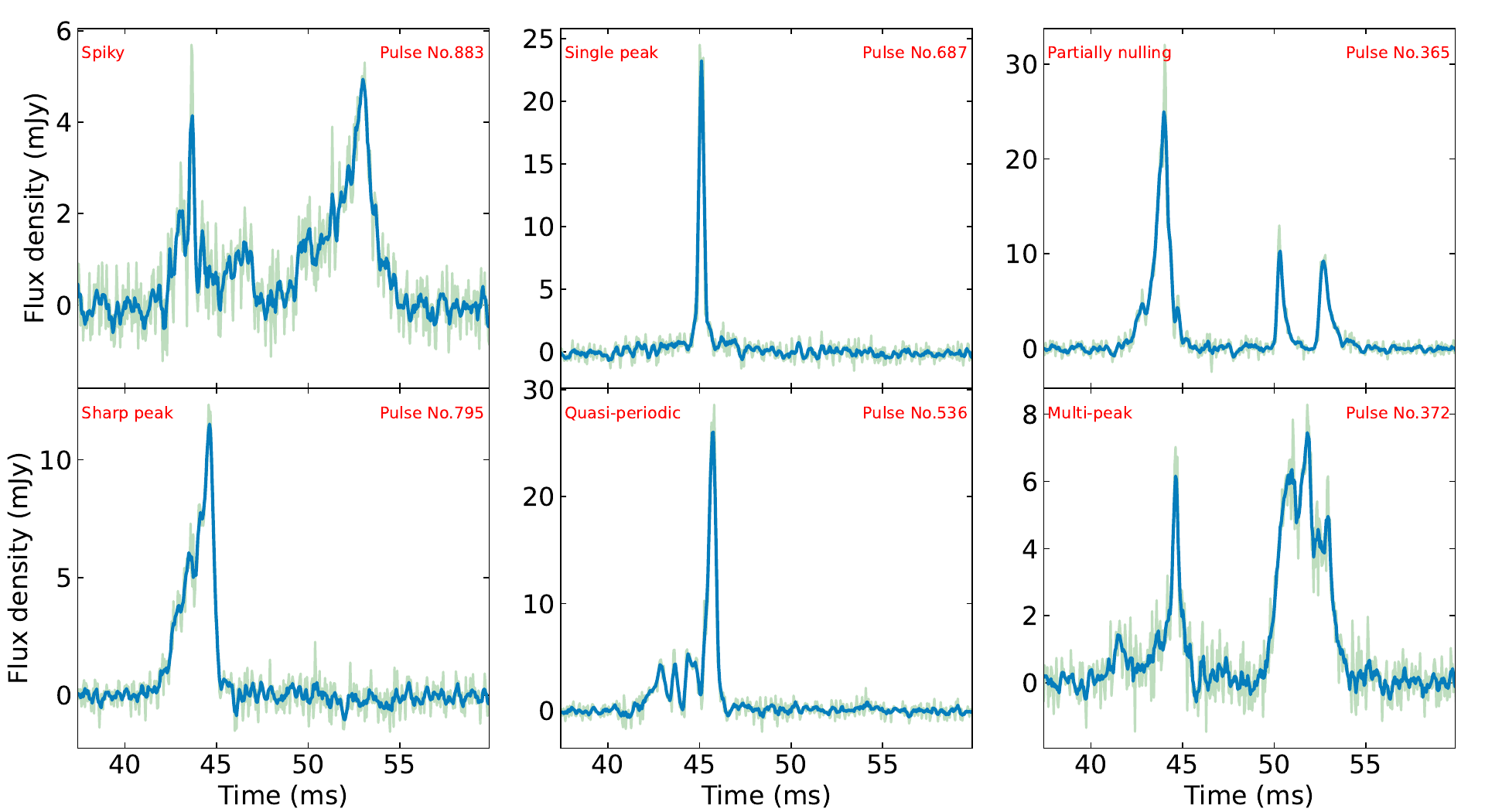}
		\caption{Example pulses from PSR J1900$-$0134 at 1.25 GHz showcasing the microstructures.}
		\label{microstructure}
	\end{figure*}
	
	\begin{figure*}
		\centering
		\includegraphics[width=2\columnwidth, angle=0]{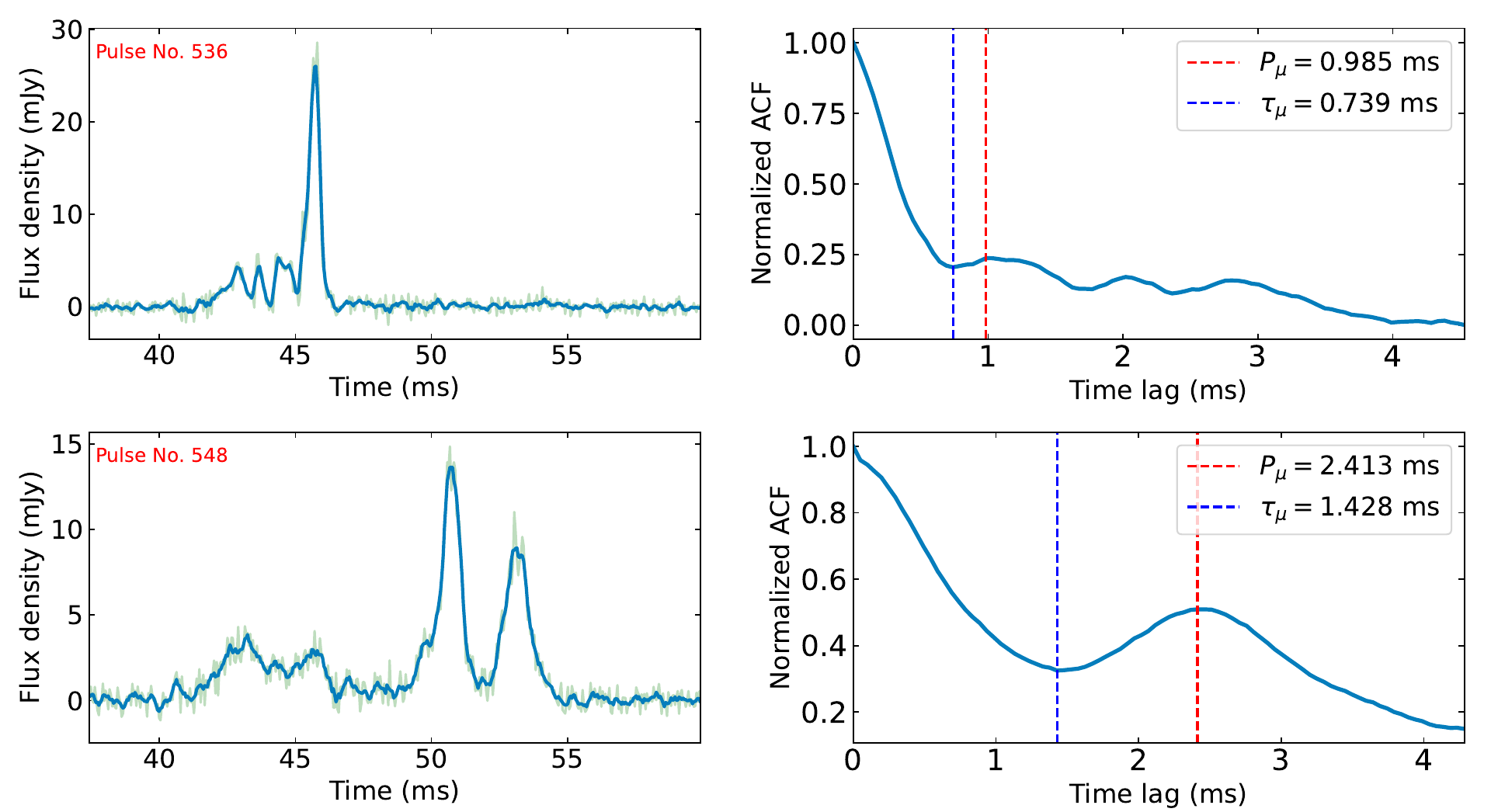}
		\caption{Examples of quasi-periodic pulses. The left column shows the pulse profiles, and the right column shows their corresponding ACFs at 49.25 $\mu$s resolution for pulse numbers 536 (top) and 548 (bottom). The blue dotted vertical line indicates the microstructure width ($t_{\mu}$), and the red dotted vertical line indicates the quasi-periodicity ($P_{\mu}$).}
		\label{Quasi-periodic}
	\end{figure*}
	
	The variability in peak intensity of the bright pulses across the three pulsars is notable, ranging from around 80 mJy for PSR J1945+1211 to about 33 mJy for PSR J1900$-$0134. This range further illustrates the natural fluctuations in pulsar emissions. These disparity could be attributed to a variety of factors, including interstellar scintillation, intrinsic variations in pulsar dynamics, or even the different orientations of each pulsar relative to our line of sight. Our results also align with and expand upon the recent findings on PSR J1900+4221 by \citet{HMT2022}, where 12 bright pulses (approximately 0.7\% of the total) were predominantly found in the leading component, with none concurrently in both components. This parallel yet distinct observation emphasizes the asymmetric and diverse nature of pulsar emissions, offering a compelling avenue for further exploration of these enigmatic celestial phenomena.
	
	\section{Microstructure pulses}\label{sec:microstructure}
	
	Our three candidate pulsars had not been previously studied for microstructure pulses. High-sensitivity telescopes like FAST are essential for detecting single pulses with satisfactory Signal-to-Noise Ratios (S/Ns) and addressing the dispersive effects of the interstellar medium across broad bandwidths. In many cases, the integrated radio profile in pulsar studies does not reveal noticeable microstructure signatures across different longitudes. This lack of visibility is primarily due to the averaging technique, which can mask the complex structures found in individual pulses. 
	Adopting methods similar to those described in \cite{Caleb22} and \citet{Dang24}, we divided each rotation period into 8,192 phase bins, achieving a time resolution of 49.152 $\mu$s. For our three pulsars, we selected burst pulses, which constituted 47\%, 51\%, and 72\% of the observational data, respectively, for investigation of the microstructures. 
	
	The auto-correlation function (ACF) is a well-established method to gauge micro-pulse timescales \citep{Cordes1990, Mitra2015}. The ACF can provide information on the temporal structure of the pulses. In the absence of microstructures, the ACF resembles a bell curve. However, when micro-pulses are present, a subtle modulation will emerge. When quasi-periodic microstructures exist, the ACF shows periodic peaks. The first peak indicates micro-pulse spacing, whereas the lag at the first minimum characterizes the width of each micro-pulse. We applied these methods to our three pulsars. Although clear microstructure periodicity was not obtained for PSR J1945+1211 and PSR J2323+1214, PSR J1900$-$0134 exhibited various microstructure pulses. 
	
	We identified 95 pulses with microstructure in PSR J1900$-$0134. They can be roughly categorized into six types based on the profile shape: spiky, single peak, multi-peak, sharp peak, partially null, and quasi-periodic. The most common types are spiky pulses (29.37\%), multi-peak pulses (25.60\%), and partially null pulses (20.82\%). Sharp peak, quasi-periodic, and single peak pulses comprise 12.07\%, 10.51\%, and 1.63\%, respectively. Figure \ref{microstructure} shows these six pulse profiles, demonstrating their distinct emission characteristics. The spiky pulse (No. 883) shows multiple sharp and narrow peaks in flux density, suggesting rapid fluctuations in emission intensity and a dynamical and variable emission process. The single peak pulse (No. 687) features a single prominent peak, representing a brief, isolated emission event. The partially null pulse (No. 365) exhibits fluctuations with significant drops in flux density, suggesting intermittent decreases or cessations in emission. The sharp peak (No. 795) presents a gradual rise in pulse intensity to its peak followed by a `sharp' drop, thus resembling a short-duration and high-intensity burst. The quasi-periodic pulse (No. 536) reveals regularly spaced peaks, implying a cyclic or periodic emission process. The multi-peak pulse (No. 372) shows several distinct maxima, indicating a complex emission process with multiple bursts occurring in quick successions.
	
	\begin{figure}
		\centering
		\includegraphics[width=\columnwidth]{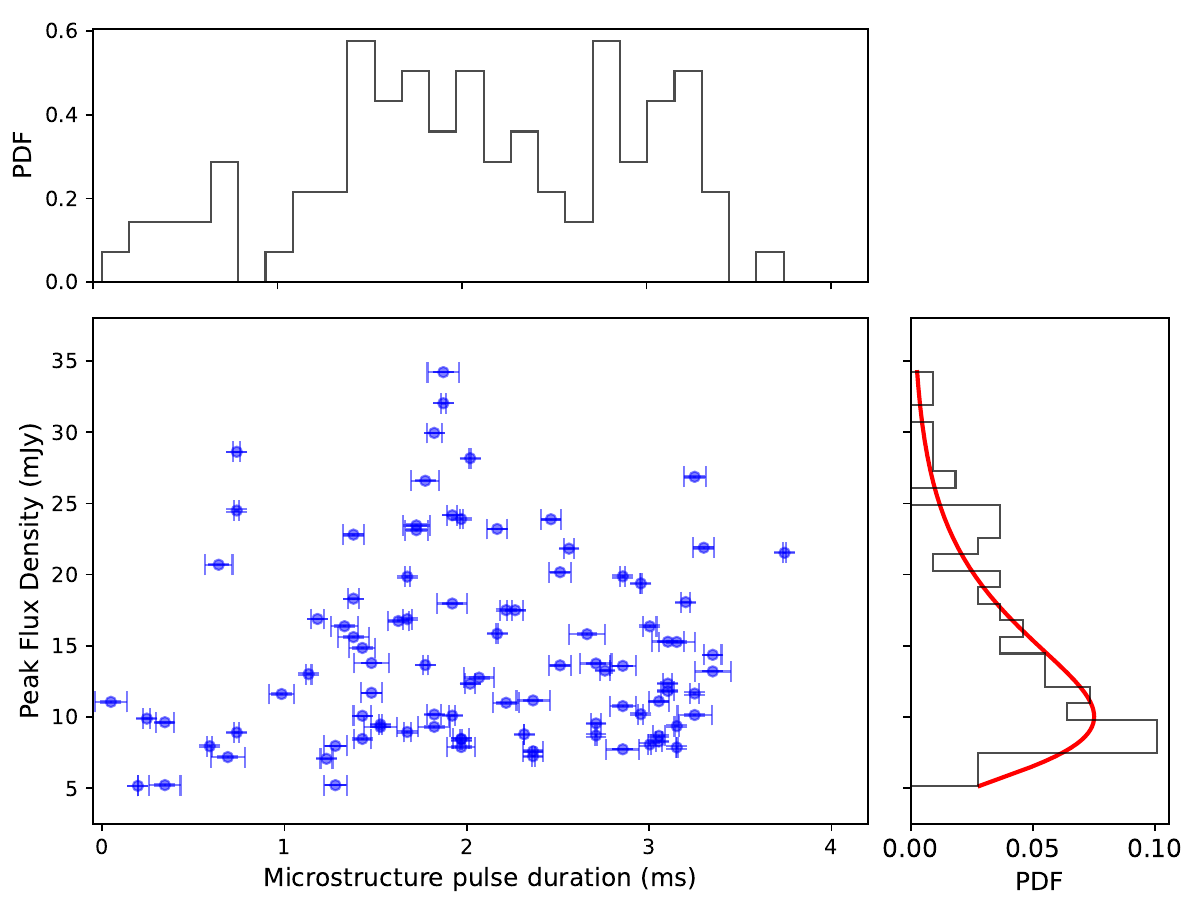}
		\caption{Main panel: A scatter plot of peak flux density and $t_{\mu}$ values for 124 microstructure pulses. Top panel: The PDFs of the $t_{\mu}$ values. Right panel: The PDFs of peak flux density, fitted with a log-normal distribution.}
		\label{distribution_microstructure}
	\end{figure}
	
	Figure \ref{Quasi-periodic} presents examples pulses with quasi-periodic microstructures. The pulse at No. 536 exhibits more than one quasi-period, indicating that some pulses exhibit multiple quasi-periods within a single pulsar rotation. However, the pulse at No. 548 demonstrates only one quasi-period. The periodicity, \( P_{\mu} \), varies between 0.985 ms and 2.413 ms, corresponding to timescales \( t_{\mu} \) of 0.739 ms and 1.428 ms, respectively. The overall periodicity varies between 0.098\,ms and 3.792\,ms, with a median of 2.13\,ms. The overall timescale is approximately 2.05\,ms. The main panel of Figure \ref{distribution_microstructure} illustrates a scatter plot showing the relationship between peak flux density and \( t_{\mu} \) for the detected microstructure pulses. This plot suggests that the pulsar's microstructure pulses may have higher peak flux density, with different ranges of pulse width and timescale. The error in the peak flux density is estimated using the RMS value of a pulse profile baseline, while the error in \( t_{\mu} \) is determined by taking the square root of their respective values. The right panel shows the PDF of the peak flux density, which is beat-fitted with a log-normal distribution centered at approximately 8.03 mJy. This indicates that the peak flux density follows a logarithmic normal distribution. The upper panel displays the PDFs of the \( t_{\mu} \) values. Moreover, the microstructure pulses from PSR J1900$-$0134 were observed to repeat in an irregular fashion.
	
	\section{Discussion} \label{sec:discus}   
	
	In this section, we discuss the findings that reveal distinct emission characteristics for PSRs J1945+1211, J2323+1214, and J1900$-$0134, and compare them with that of ordinary pulsars. This highlights the diversity in pulsar behaviors and offers new insights into the underlying mechanisms driving their emissions. Additionally, we explore how our results impact the understanding of the pulsar death line and their implications for existing pulsar theories.
	
	\subsection{Emission characteristics and comparison with ordinary pulsars}\label{Emission characteristics}
	
	PSR J1945+1211 exhibits notable null behavior, with null durations lasting around 76 seconds and a null fraction of approximately 52.46\%. Its integrated pulse profile is asymmetric with the peak intensity of the leading component almost twice that of the trailing component, as shown in Figure~\ref{profile_all}. This intensity difference suggests that the emission mechanisms in the leading and trailing components may be influenced by intrinsic factors, potentially related to the plasma dynamics within the magnetosphere. 
	The pulse profile of PSR J1945+1211 is broader compared to those of PSR J2323+1214 and PSR J1900$-$0134, with a width of 25.31$^{\circ}$ at 50\% intensity ($W_{50}$), as shown in Figure~\ref{profile_all}. The value of the inclination angle $\alpha < 60^{\circ}$ and the impact parameters $-10^{\circ} < \beta < 0^{\circ}$, as determined using the Rotating Vector Model \citep[RVM;][]{RadhakrishnanCooke1969}, suggest that the inclination angle of this pulsar is likely small \citep{CLH+20}.
	In addition, the high null fraction of PSR J1945+1211 aligns with the proposed positive correlation between the null fraction and both the pulsar age and the rotation period \citep{Ritch76, Bigg92}. The evolution of pulsar age is commonly suggested to correlate with the inclination angle $\alpha$ if assuming energy loss is through magnetic dipole radiation leading to changes in $\alpha$ from large to small \citep{Pulsars1977, LK01}. Furthermore, the \citet{RudermanSutherland1975} model indicates that the rotation period and age of a pulsar are related to $\alpha$. In addition, the pulsar has a characteristic age of 19.8 million years and a surface magnetic field of $4.25 \times 10^{12}$ G. It displays moderate energy loss, which aligns with its age and magnetic field strength. Therefore, the high null fraction and small inclination angle of PSR J1945+1211 are consistent with the correlation between null behavior and pulsar evolution.
	
	\begin{figure}
		\centering
		\includegraphics[width=\columnwidth]{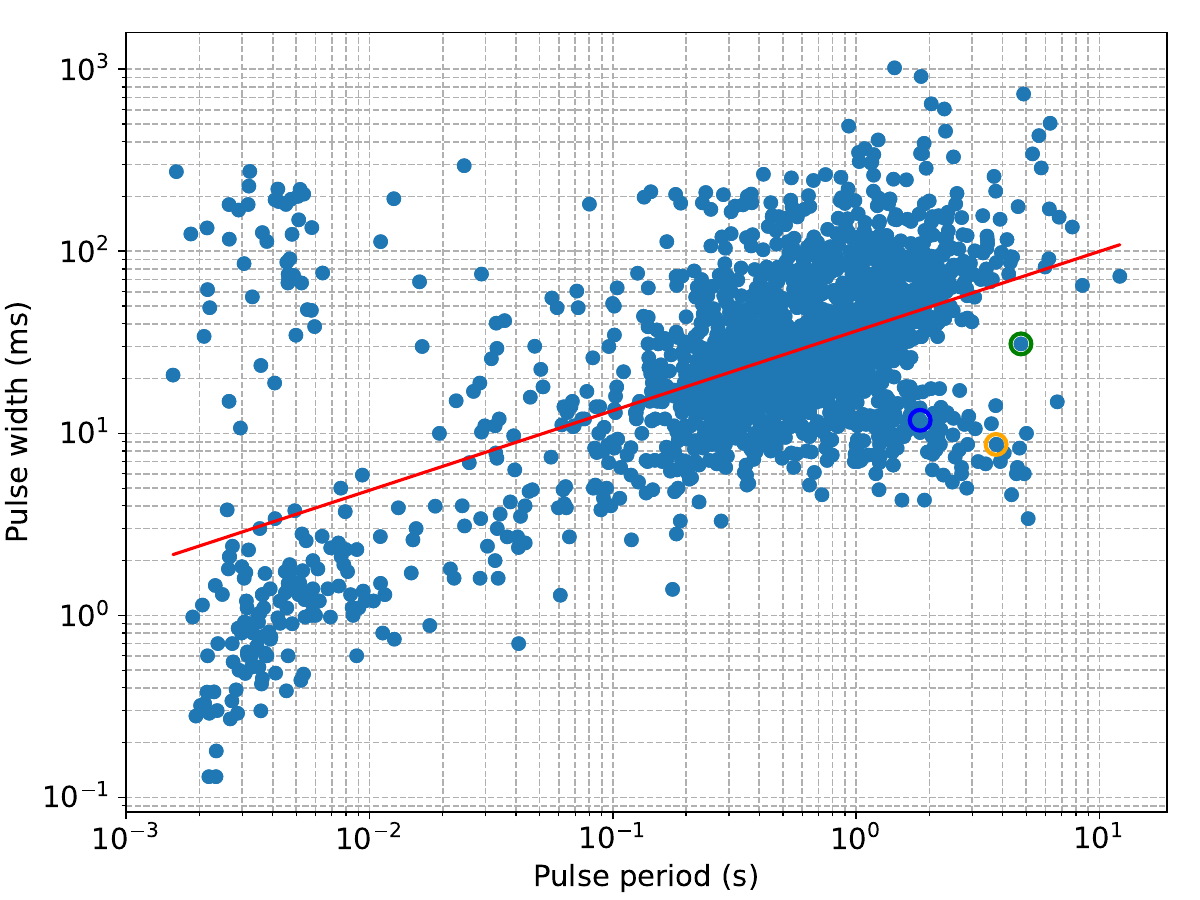}
		\caption{Log-log plot showing the relationship between pulse period and pulse width, along with a power-law fit. Data obtained from the Australia Telescope National Facility (ATNF) pulsar catalogue, http://www.atnf.csiro.au/research/pulsar/psrcat/ \citep{Manchester2005}. Our candidate pulsars are marked with colored circles: blue for (PSR J1900$-$0134), orange for (PSR J2323+1214), and green for (PSR J1945+1211).}
		\label{period-width}
	\end{figure}
	
	In contrast, PSR J2323+1214 had a null fraction of 48.48\% and a null duration of approximately 56.4 seconds, with a few dwarf pulses also exhibited in the trailing components, as shown in Figure \ref{Dwarf_profile}. This pulsar exhibited a different pattern in which the trailing component possesses a peak intensity that is three times greater than that of the leading component. The narrower pulse profile of PSR J2323+1214, measuring $7.38^\circ$ at $W_{50}$, as shown in Figure~\ref{profile_all}, suggests a more focused emission beam. This intensity distribution indicates a different emission geometry or magnetospheric configuration compared to PSR J1945+1211, as illustrated in Figure 4 of \citet{CLH+20}. Furthermore, the null behavior of PSR J2323+1214 shows periodic patterns, with a dominant period of approximately 22.22 pulsar periods, suggesting a potential underlying mechanism or a process of periodic nature. PSR J2323+1214, with a characteristic age of 32.2 million years and a surface magnetic field of $2.64 \times 10^{12}$ G, had the lowest energy loss among the three. This could imply differences in the magnetic field structure or in the size and shape of the emission region.
	
	Lastly, PSR J1900$-$0134 presented the most complex emission features among the three pulsars, including microstructure pulses with detailed structures down to 1.56 microseconds and a null fraction of 27.51\%. The emission profile of PSR J1900$-$0134 showed significant variations with complex patterns of bright and null pulses, indicating the intricate magnetospheric processes at play. The presence of microstructure pulses, as shown in Figure~\ref{microstructure}, indicates highly dynamic processes within the pulsar magnetosphere. These pulses, which are short-lived, high-intensity bursts within individual pulses, suggest mechanisms possibly related to wave-particle interactions or localized magnetic reconnections \citep{Hankins1971, Lyutikov2003}. The variability in pulse intensity and duration, with bursts lasting up to 29 pulsar periods and nulls up to 16 periods, indicates highly dynamic processes within its magnetosphere. PSR J1900$-$0134, being the youngest at 0.95 million years, had the highest surface magnetic field of 7.47 $\times 10^{12}$ G and the most significant energy loss rate. 
	
	Compared with ordinary pulsars, which typically have rotational periods ranging from milliseconds to about one second, these long-period pulsars differ significantly. Examples of ordinary pulsars include PSR B0740$-$28 \citep{Bigg92} and PSR B0329+54 \citep{Backer1970}, which have rotational periods of approximately 0.16\,s and 0.71\,s, respectively. The emission profiles of long-period pulsars often exhibit wider pulse widths (see Appendix Table~\ref{table:burst1945} and Table~\ref{table:burst2323}) and more complex substructures due to their slower rotation, allowing the emission beam to sweep more slowly across our line of sight, capturing more of the emission region, and resulting in wider profiles \citep{Weltevrede06}. In contrast, ordinary pulsars with faster rotation have more narrowly confined pulses, producing sharper profiles \citep{Rankin1993}.  The analysis presented in the log-log scatter plot (Figure~\ref{period-width}) further supports this distinction, revealing a moderate positive correlation ($r = 0.58 \pm 0.02$) between pulse period and pulse width. The power-law relationship, expressed as $y = 36.53x^{0.44}$, indicates that as the pulse period increases, the pulse width generally becomes larger. This trend aligns with the observation that long-period pulsars, with their extended periods, exhibit broader pulse profiles compared to ordinary pulsars. 
	Moreover, the null behavior differs markedly between these two classes of pulsars. Long-period pulsars, such as the PSR J1945+1211, typically exhibit higher null fractions and longer null durations, such as a null fraction of 52.46\% and null durations extending up to 16 periods. This contrasts with the shorter null durations and lower null fractions seen in ordinary pulsars like PSR J2022+5154, which has a null fraction of around 1.4\% \citep{Gajjar12}. Ordinary pulsars have relatively stable magnetospheres, resulting in less complex emission structures and lower null fractions \citep{Lorimer2005, Lyne_Graham-Smith_2012}. In contrast, the more complex emission features and higher null fractions observed in long-period pulsars suggest that their magnetospheres are more dynamic and subject to greater variability \citep{Gajjar12, Young15}. These differences in null behavior are likely due to more dynamic and variable magnetospheric environments in long-period pulsars \citep{HMT2022}, influenced by slower spin-down rates and distinct energy loss mechanisms \citep{Gao2019, Deng2020, Philippov22}.
	
	\begin{figure}
		\centering
		\includegraphics[width=1\columnwidth, angle=0]{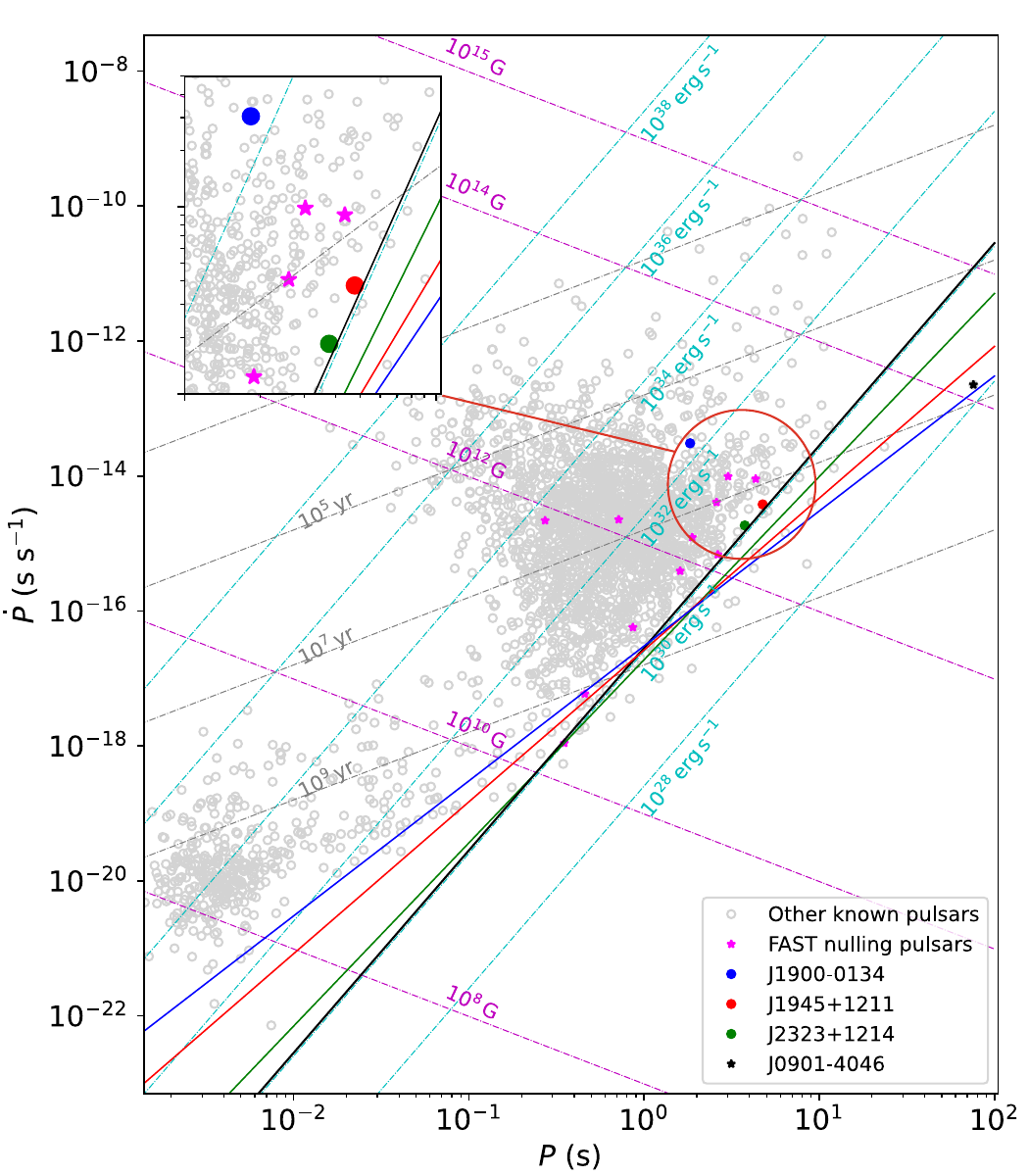}
		\caption{A $P-\dot{P}$ diagram for over 3474 pulsars. Period ($P$) is in seconds and period derivative ($\dot{P}$) in seconds per second. Light gray circles represent known pulsars, magenta stars indicate FAST null pulsars, and circles with blue highlight our candidates for PSR J1900$-$0134, red for PSR J2323+1214, and green for PSR J1945+1211. PSR J0901$-$4046, an ultra-long period pulsar \citep{Caleb22}, is marked with a black star. The inset plot provides a zoomed-in view of the region containing our candidates, while the red circle in the main plot highlights their location. Diagonal lines represent constant magnetic fields (magenta), spin-down luminosities (cyan), and characteristic ages (gray). The lower-right corner displays `death valley’ with various death lines: traditional (black), \citet{Chen1993} (red), curvature radiation from vacuum gap (green), and from the space-charge-limited flow model (blue). Data obtained from the ATNF pulsar catalogue \citep{Manchester2005}.}
		\label{p_pdot}
	\end{figure}
	
	\subsection{Impact on pulsar death line}
	
	The $P-\dot{P}$ diagram in Figure~\ref{p_pdot} illustrates the positions of our candidate pulsars relative to various death lines, which represent the boundaries where radio emission from pulsars is expected to cease due to limitations in electron-positron pair production. These death lines, derived from different models \citep{RudermanSutherland1975, Arons1979, Chen1993, Zhang2000}, are crucial to understanding the conditions required to sustain radio emission in pulsars.
	
	In particular, our observations place PSR J2323+1214 and PSR J1945+1211 near the traditional death line, as defined by the vacuum gap model \citep[RS75 model;][]{RudermanSutherland1975, Bhattacharya1991}. The proximity of these pulsars to the death line suggests that their radio emissions are likely sustained by electron-positron pair production primarily occurring at the polar cap regions, consistent with traditional models. The high NF of these pulsars, 49\% and 53\%, respectively, further support this interpretation, as increased NF is often associated with pulsars near the death line. Similarly, other FAST-discovered pulsars, such as PSR J1611$-$0114 \citep[NF = 40\%,][]{Liu2024} and PSR J0211+4235 \citep[NF = 49\%,][]{Guo23}, also lie near the death line, a critical threshold in pulsar theory where pair production becomes less efficient. These pulsars have relatively low spin-down luminosities, 9.2 $\times$ 10$^{30}$ erg/s and 1 $\times$ 10$^{30}$ erg/s, respectively, indicating that they are close to the point where their magnetospheric conditions may no longer support continuous radio emission. Additionally, PSR J1738$-$2330, discovered in the Parkes multi-beam survey \citep[NF = 69\%,][]{Gajjar12}, and PSR J1926$-$1314, discovered by the Green Bank Telescope \citep[NF = 75\%,][]{RSM13}, are also close to the death line. In contrast, PSR J1900$-$0134, which is farther from the death line, shows a much lower NF of 28\%. This difference in NF among our observed pulsars supports the idea that the NF increases as pulsars approach the death line.
	
	This observation aligns with the predictions of traditional death line models, which suggest that pulsars with lower spin-down luminosities, closer to the death line, exhibit higher nulling activity due to marginal conditions for pair production. As high-sensitivity telescopes like FAST continue to discover pulsars with low spin-down luminosities, there will be further opportunities to test and refine these theories. The diversity in null fractions and positions relative to the death line observed in our study highlights the need for ongoing research into the complexities of pulsar magnetospheres and emission mechanisms.

	\subsection{Implications for pulsar emission theories}
	
	Our study provides new insights into pulsar emission theories by examining a range of phenomena, including quasi-periodic null, dwarf pulses, bright pulses, and microstructure pulses.
	There is increasing evidence that periodic null and periodic amplitude modulations are closely linked. Periodic null can be explained in the rotating subbeam carousel model, such that null occurs when the line of sight passes through empty regions between rotating conal subbeams. Periodic amplitude modulations, characterized by periodic intensity fluctuations without systematic drift patterns, have also been observed in many pulsars \citep[eg.,][]{HerfindalRankin07, Basu17, Yan19}. \citet{Basu16} and \citet{Mitra17} reported this phenomenon in pulsars such as PSR B1946+35, PSR J1825$-$0935, and PSR B0031$-$07. In particular, these periodic modulations share similarities with periodic null, suggesting a common origin \citep{Basu17}. Both phenomena likely stem from intrinsic changes in the pulsar magnetosphere, potentially due to alterations in the magnetospheric current distribution \citep{Melrose14}.
	
	The quasi-periodic null observed in our study suggests that pulse null may not be a purely random phenomenon. Initially, \citet{Backer1970} proposed a periodic behavior for nulls, but it was later seen to be a stochastic process related to pulsar emission conditions. However, our findings are consistent with more recent observations by \citet{Wen16}, \citet{Rejep22}, and \citet{Chen23}, indicating that periodic null patterns exist in certain pulsars. Theoretical models propose various mechanisms for null in pulsars. One model suggests null results from changes in the pulsar viewing geometry, where the line of sight no longer intersects the emission beam due to changes in the magnetic field configuration \citep{Timokhin10}. Another model links null to a temporary cessation of plasma generation in the pulsar magnetosphere, disrupting coherent radio emission \citep{Filippenko82}. These theories highlight the complex interaction between geometric factors and magnetospheric processes.

	\begin{figure}
		\centering
		\includegraphics[width=\columnwidth]{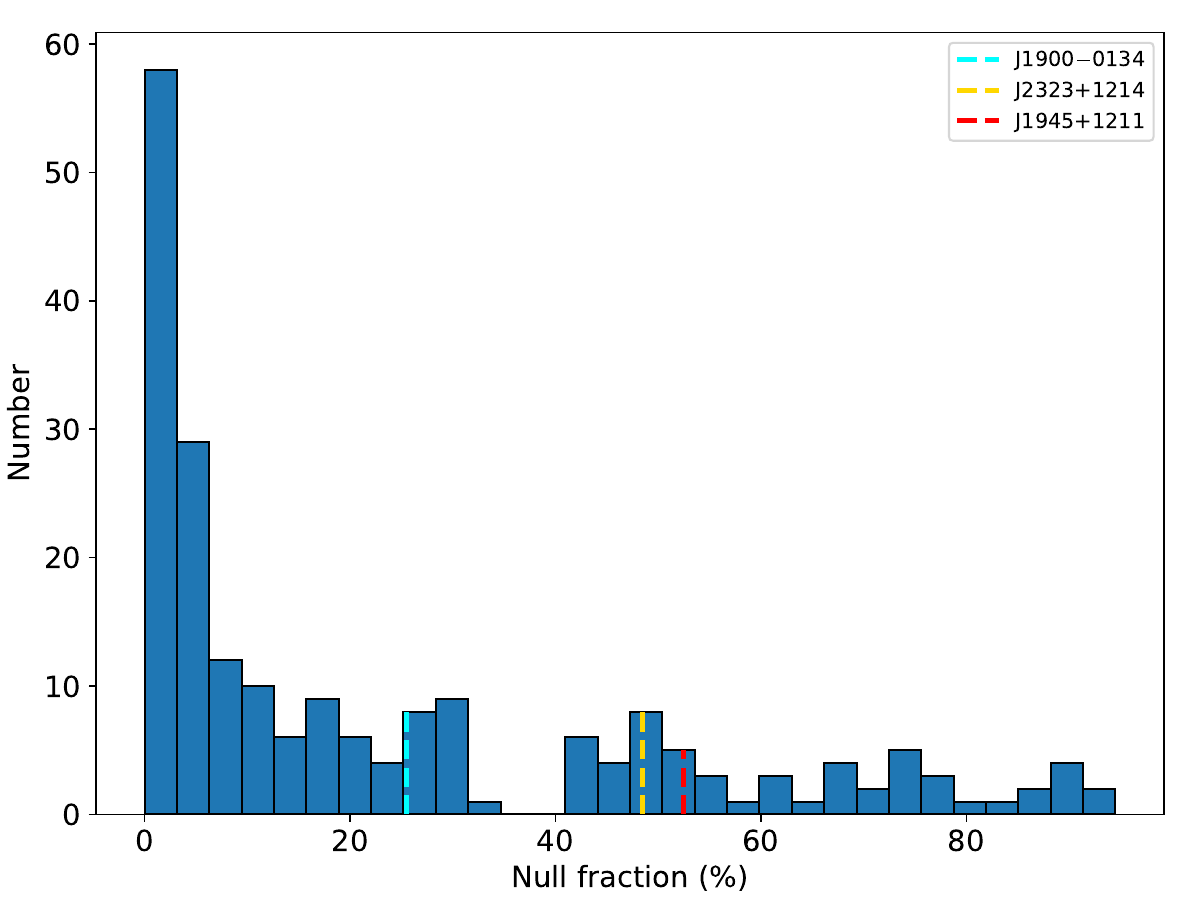}
		\caption{Histogram of null fractions observed in various pulsars. The distribution indicates a higher frequency of pulsars with low null fractions and decreases as the null fraction increases. The vertical dotted lines represent the null fractions for PSR J1900$-$0134 (cyan), PSR J2323+1214 (yellow), and PSR J1945+1211 (red), highlighting the positions of these pulsars within the distribution.}
		\label{Null_dist}
	\end{figure}
	
	The distribution of null fractions across various pulsars reveals two distinct populations \citep{Sheikh2021}, as shown in Figure~\ref{Null_dist}. The more frequent group has null fractions below 40\%, indicating that most pulsars exhibit relatively low null levels. In contrast, the less common group has null fractions above 40\%, suggesting that high levels of null are rare. These trends highlight a relationship between null behavior, pulsar age, and rotational period, supporting the idea that older, slower pulsars tend to have higher null fractions due to decreased plasma production and emission activity.
	This idea is further supported by \citet{Ritch76}, who suggested that older pulsars may null more frequently as they age, and by \citet{Bigg92}, who found a strong correlation between null and pulsar periods. Our $P-\dot{P}$ diagram, shown in Figure~\ref{p_pdot}, places our three candidate pulsars among older long-period pulsars, reinforcing the connection between null and characteristic age, likely linked to the electric potential in the acceleration gap \citep{RudermanSutherland1975}.
	Interestingly, PSR J2323+1214 shows significant differences in NF between its emission components, with the leading component having a higher NF of 61.34\% $\pm$ 1.82\% compared to the trailing component's 50.27\% $\pm$ 1.73\%. This variation aligns with \citet{WWD21}, who reported higher null fractions for both components using the Arecibo telescope. The overall lower null fractions detected by the FAST telescope likely result from its higher sensitivity, which captures more weak emissions.

	Our observations of PSR J1945+1211, PSR J2323+1214, and PSR J1900$-$0134 indicate that emissions immediately before and after a null tend to exhibit pulse energy variations, often resulting in the occurrence of a burst pulse or a bright pulse. Figure~\ref{energyseries_1945} demonstrates that the transitions in PSR J1945+1211 and PSR J2323+1214 are typically abrupt, suggesting a concurrent cessation and restarting of plasma generation across the emission region within a pulse period. Conversely, PSR J1900$-$0134 exhibits both gradual and abrupt transitions, as shown in Figure~\ref{energyseries_1900}, indicating diverse emission behaviors around nulls. This dual behavior is due to the variability in the pulsar's magnetospheric dynamics and emission processes. Gradual transitions may indicate slow changes in the magnetic field or plasma generation rates, while abrupt transitions suggest rapid reconfigurations of the magnetic field or sudden changes in plasma density \citep{Goldreich69}.
	
\begin{figure}
	\centering
	\includegraphics[width=\columnwidth]{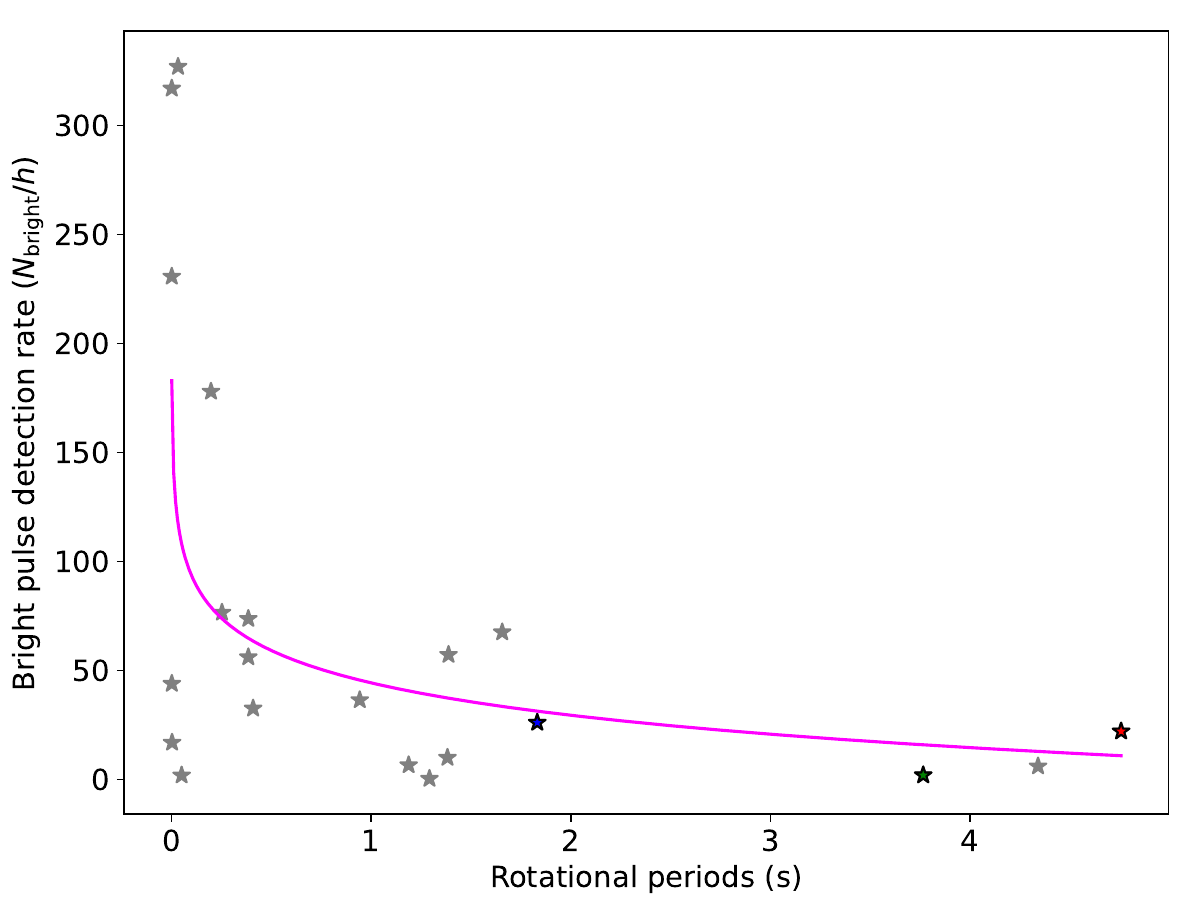}
	\caption{The distribution of the detection rate of bright pulses ($N_{\mathrm{bright}}/h$) for different pulsars. Data for PSR J1900$-$0134 (blue star), J2323+1214 (green star), and J1945+1211 (red star) are from this work. The magenta curve represents a power law fit.}
	\label{bright_population}
\end{figure}

	The bright pulses detected in PSR J1945+1211, PSR J2323+1214, and PSR J1900$-$0134 suggest high-intensity emissions from localized magnetospheric activity, likely near the magnetic poles. For instance, ordinary pulsars, such as PSR B0531+21 \citep{Lundgren95} and PSR B0656+14 \citep{Kuzmin06}, as well as millisecond pulsars like PSR B1937+21 \citep{Cognard1996} and PSR J0034$-$0721 \citep{Tuoheti11}, show much brighter pulses, with peak intensities exceeding 20 times the average profile, and up to 120 times in some cases. This supports the hypothesis that rotational period and the number of bright pulses are likely correlated (see Figure~\ref{bright_population} and Appendix Table~\ref{table:burst}), aligning with models that link shorter periods to more energetic plasma production and more frequent bright pulses \citep[eg.,][]{Philippov2020}. 
	
	Our analysis of dwarf pulses in PSR J2323+1214 indicates that short, sporadic emissions can occur during null states, suggesting that emission can persist even during these periods. This might connect null pulsars to RRATs, potentially linking them as stages in the evolution of the pulsar \citep{Burke-Bailes2010}.
	RRATs were discovered in the Parkes Multi-beam Pulsar Survey \citep{McLaughlin2006}, exhibit irregular pulses of short duration, typically between 2–30 milliseconds. A key difference between null pulsars and RRATs is the null fraction, also known as the burst rate. Most RRATs have low burst rates, meaning their null fractions are large\footnote[2]{http://astro.phys.wvu.edu/rratalog/}. The dwarf pulses in PSR J2323+1214, with a typical pulse width of $1.87^\circ$ (about 20 ms), fall within the RRAT pulse duration range. The brightest dwarf pulse was nearly 70\% as intense as the average profile, indicating that significant emissions can occur even during null states.
	Assuming radio emission is related to an electric potential in a vacuum gap, which decreases as the rotation period increases, older pulsars may exhibit more frequent and longer nulls. However, short sporadic emissions resembling RRATs can still occur. 
	
	Finally, the microstructure pulses observed, particularly in PSR J1900$-$0134, provide insight into the fine-grained processes within the pulsar magnetosphere. These microstructures exhibit a variety of shapes and periodicities, indicating that pulsar emission involves complex wave-particle interactions and coherent emission processes \citep{Hankins1975, Weatherall1998}. The diversity in microstructure shapes, from spiky to quasi-periodic, reflects the dynamic nature of these processes.
	
	Our findings offer a more nuanced understanding of pulsar emission mechanisms, showing that different emission phenomena may be interconnected and influenced by both intrinsic and extrinsic factors. Although our results largely conform to existing models, they also provide new details that refine these theories. Future studies with enhanced sensitivity, longer observation durations, and multi-frequency data will be essential to further unravel the underlying mechanisms driving these emission phenomena and to refine pulsar emission theories.

	\section{Summary} \label{sec:summary}

	We have reported a detailed analysis of the emission behaviors of three long-period pulsars, PSR J1945+1211, J2323+1214, and J1900$-$0134, detected using the FAST Telescope during the Commensal Radio Astronomy FAST Survey (CRAFTS). 
	We summarize the emission features in our three long-period pulsars below.
	\begin{enumerate}
		\item Detection of a quasi-periodic null phenomenon with durations ranging from 57 to 71.44 seconds.
		\item The null fractions for PSR J1945+1211, J2323+1214, and J1900$-$0134 are 52.46\% $\pm$ 2.89\%, 48.48\% $\pm$ 2.51\%, and 27.51\% $\pm$ 1.37\%, respectively.
		\item Complex emission features in PSR J1900$-$0134, with microstructure pulses down to 2.05 milliseconds.
		\item Asymmetric emissions, with bright pulses predominantly in the leading component for PSR J1945+1211 and PSR J1900$-$0134.
		\item Quasi-periodic microstructures in PSR J1900$-$0134, suggesting a periodic emission process in the pulsar magnetosphere. 
		\item Various microstructure pulses in PSR J1900$-$0134, indicating diverse emission characteristics.
		\item Occurrence of bright pulses at different rates in all three pulsars.
		\item Variation in intensity among pulse profiles, with burst state profiles having higher peak intensities and wider widths.
		\item Periodic transitions between null and burst states, with abrupt transitions in PSR J1945+1211 and J2323+1214, and a mix in PSR J1900$-$0134.
		\item Null and burst state durations suggesting a power-law distribution.
	\end{enumerate}

	\section*{Acknowledgments}

	This work is supported by the National Natural Science Foundation of China (NSFC)(Grant Nos. 11988101, U2031117, U1838109, 11873080, 12041301) and by the Alliance of International Science Organizations, Grant No. ANSO-VF-2024-01. We are grateful to the anonymous referee for valuable comments that have improved the presentation of this paper. H.M.T. acknowledges Arba Minch University. D.L. is a new cornerstone investigator and support by NSFC (11988101) and the 2020 project of Xinjiang Uygur Autonomous Region of China for flexibly fetching in upscale talents. P.W. acknowledges support from the NSFC Programs (11988101, 12041303), the CAS Youth Interdisciplinary Team, the Youth Innovation Promotion Association CAS (id. 2021055), and the Cultivation Project for FAST Scientific Payoff and Research Achievement of CAMS-CAS. R.Y. is supported by the National SKA Program of China No. 2020SKA0120200, the National Key Program for Science and Technology Research and Development No. 2022YFC2205201 and 2022YFC2205200, the NSFC project (12288102, 12041303, 12041304), and the Major Science and Technology Program of Xinjiang Uygur Autonomous Region No. 2022A03013-2, and the open program of the Key Laboratory of Xinjiang Uygur Autonomous Region No. 2020D04049. This research is partly supported by the Operation, Maintenance and Upgrading Fund for Astronomical Telescopes and Facility Instruments, budgeted from the Ministry of Finance of China (MOF) and administrated by the CAS. N.W. is supported by the NSFC (12041304). We also express our gratitude to all members of the FAST telescope collaboration for establishing the projects (project numbers: ZD2020-06 and PT2020-0045), which made the observations possible. 
	
	This work made use of data from the Five-hundred-meter Aperture Spherical radio Telescope (FAST). FAST is a Chinese national mega-science facility operated by the National Astronomical Observatories, Chinese Academy of Sciences.
	
	\clearpage
	% Starting the appendix section
	\appendix
	\setcounter{table}{0}
	\section{Burst Pulse Information in Longer Period Pulsars}
	
	In this appendix, we present only the burst pulses observed in the single-pulse sequences. For each pulse, we provide the arrival time in Modified Julian Date (MJD). The reference MJD for PSR J1945+1211 is 59116.597569, and for PSR J2323+1214, it is 59113.673614. Both were observed at a center frequency of 1.25 GHz. Additionally, we provide the calculated pulse width (in ms), the measured peak flux density (in mJy), and the fluence (in Jy\,ms) are provided in the table for PSR J1945+1211 and PSR J2323+1214, respectively.
	
	\setlength{\tabcolsep}{10.5 pt}  
	
	\begin{longtable}{cccccc}
		\caption{Burst pulse information for PSR J1945+1211} \label{table:burst1945} \\
		\hline
		\textbf{S. No.} & \textbf{Pulse Number} & \textbf{Pulse Arrival Time} & \textbf{Pulse Width} & \textbf{Peak Flux Density} & \textbf{Fluence} \\
		&& (MJD) & (ms) & (mJy) & (Jy\,ms) \\
		\hline
		\endfirsthead
		
		\caption[]{Burst pulse information for PSR J1945+1211} \\
		\hline
		\textbf{S. No.} & \textbf{Pulse Number} & \textbf{Pulse Arrival Time} & \textbf{Pulse Width} & \textbf{Peak Flux Density} & \textbf{Fluence} \\
		&& (MJD) & (ms) & (mJy) & (Jy\,ms) \\
		\hline
		\endhead
		
		\hline \multicolumn{6}{r}{\textit{Continued on next page}} \\
		\endfoot
		
		\hline
		\endlastfoot
		
		1  & 3   & 59116.597733931 & 466.8(6)  & 4.3(3)  & 2.021(6)  \\
		2  & 4   & 59116.597788907 & 458.5(6)  & 9.8(6)  & 4.475(7)  \\
		3  & 5   & 59116.597843884 & 467.2(7)  & 9.5(6)  & 4.420(8)  \\
		4  & 6   & 59116.597898861 & 459.4(4)  & 14.9(1) & 6.850(3)  \\
		5  & 7   & 59116.597953838 & 443.5(7)  & 21.7(4) & 9.641(9)  \\
		6  & 8   & 59116.598008815 & 444.0(7)  & 37.6(8) & 16.687(5) \\
		7  & 9   & 59116.598063792 & 434.8(6)  & 48.5(4) & 21.103(6) \\
		8  & 18  & 59116.598558583 & 388.3(5)  & 12.4(8) & 4.807(1)  \\
		9  & 20  & 59116.598668537 & 448.1(3)  & 28.0(7) & 12.533(3) \\
		10 & 21  & 59116.598723514 & 496.1(1)  & 4.0(9)  & 1.980(5)  \\
		11 & 22  & 59116.598778491 & 448.0(9)  & 112.2(3) & 50.274(6) \\
		12 & 23  & 59116.598833468 & 414.3(4)  & 24.2(8) & 10.018(3) \\
		13 & 24  & 59116.598888444 & 498.8(5)  & 8.6(5)  & 4.314(5)  \\
		14 & 25  & 59116.598943421 & 391.9(5)  & 19.3(3) & 7.575(9)  \\
		15 & 32  & 59116.599328259 & 375.9(3)  & 12.3(3) & 4.635(5)  \\
		16 & 33  & 59116.599383236 & 410.6(1)  & 688.0(0) & 282.500(4) \\
		17 & 34  & 59116.599438213 & 412.5(3)  & 28.4(8) & 11.706(9) \\
		18 & 34  & 59116.599438213 & 380.0(0)  & 28.4(8) & 10.784(4) \\
		19 & 35  & 59116.599493190 & 377.8(6)  & 38.7(7) & 14.619(3) \\
		20 & 36  & 59116.599548167 & 388.7(3)  & 79.8(7) & 31.007(5) \\
		21 & 45  & 59116.600042958 & 87.9(5) & 62.1(9) & 5.456(1) \\
		22 & 46  & 59116.600097935 & 476.8(3) & 5.2(5)  & 2.503(6) \\
		23 & 47  & 59116.600152912 & 453.1(3) & 7.3(8)  & 3.299(3) \\
		24 & 48  & 59116.600207889 & 420.6(9) & 26.9(9) & 11.311(3) \\
		25 & 49  & 59116.600262866 & 436.2(3) & 9.8(5)  & 4.253(5) \\
		26 & 50  & 59116.600317843 & 393.7(3) & 30.8(9) & 12.123(1) \\
		27 & 51  & 59116.600372819 & 410.6(1) & 18.4(7) & 7.543(1) \\
		28 & 52  & 59116.600427796 & 443.1(6) & 9.8(5)  & 4.320(3) \\
		29 & 60  & 59116.600867611 & 472.8(6) & 6.5(1)  & 3.078(6) \\
		30 & 61  & 59116.600922588 & 468.1(8) & 7.7(4)  & 3.623(5) \\
		31 & 62 & 59116.600977565 & 429.0(9) & 10.3(7) & 4.405(9) \\
		32 & 63 & 59116.601032542 & 407.8(7) & 10.3(8) & 4.193(5) \\
		33 & 64 & 59116.601087519 & 409.7(5) & 7.2(2) & 2.950(5) \\
		34 & 65 & 59116.601142495 & 403.8(5) & 10.4(5) & 4.179(5) \\
		35 & 66 & 59116.601197472 & 402.8(9) & 11.4(6) & 4.576(6) \\
		36 & 67 & 59116.601252449 & 333.0(7) & 24.3(7) & 8.083(5) \\
		37 & 68 & 59116.601307426 & 386.5(7) & 18.2(2) & 7.041(1) \\
		38 & 69 & 59116.601362403 & 379.1(9) & 13.2(2) & 5.004(4) \\
		39 & 70 & 59116.601417380 & 354.4(7) & 23.6(4) & 8.378(7) \\
		40 & 79 & 59116.601912171 & 429.0(9) & 11.4(2) & 4.899(1) \\
		41 & 80 & 59116.601967148 & 390.0(7) & 30.9(9) & 12.052(2) \\
		42 & 81 & 59116.602022125 & 91.7(9) & 49.0(9) & 4.492(5) \\
		43 & 82 & 59116.602077102 & 109.1(6) & 52.4(5) & 5.723(9) \\
		44 & 83 & 59116.602132079 & 425.7(7) & 9.5(5) & 4.044(9) \\
		45 & 84 & 59116.602187056 & 423.4(3) & 11.6(5) & 4.890(5) \\
		46 & 95 & 59116.602791801 & 460.4(7) & 9.8(1) & 4.516(5) \\
		47 & 96 & 59116.602846778 & 449.9(4) & 20.1(8) & 9.035(4) \\
		48 & 97 & 59116.602901755 & 416.5(1) & 13.9(8) & 5.782(2) \\
		49 & 100 & 59116.603066685 & 410.2(7) & 22.7(4) & 9.328(6) \\
		50 & 110 & 59116.603616454 & 396.9(9) & 18.7(1) & 7.426(2) \\
		51 & 111 & 59116.603671431 & 467.7(4) & 6.1(1) & 2.858(5) \\
		52 & 112 & 59116.603726407 & 369.6(9) & 13.6(4) & 5.041(5) \\
		53 & 113 & 59116.603781384 & 410.1(3) & 4.0(6) & 1.624(9) \\
		54 & 114 & 59116.603836361 & 343.5(9) & 28.8(1) & 9.895(9) \\
		55 & 115 & 59116.603891338 & 307.4(4) & 17.9(3) & 5.512(1) \\
		56 & 116 & 59116.603946315 & 370.4(4) & 18.7(8) & 6.918(8) \\
		57 & 117 & 59116.604001292 & 104.1(7) & 2.8(2) & 0.294(5) \\
		58 & 121 & 59116.604221199 & 392.4(7) & 10.9(7) & 4.265(5) \\
		59 & 122 & 59116.604276176 & 378.5(1) & 21.4(4) & 8.101(7) \\
		60 & 123 & 59116.604331153 & 354.4(7) & 34.4(2) & 12.198(3) \\
		61 & 128 & 59116.604606037 & 107.3(6) & 13.5(7) & 1.445(4) \\
		62 & 129 & 59116.604661014 & 415.4(1) & 9.0(8) & 3.730(8) \\
		63 & 130 & 59116.604715991 & 415.2(7) & 6.1(9) & 2.529(5) \\
		64 & 131 & 59116.604770968 & 115.6(3) & 18.7(3) & 2.165(3) \\
		65 & 132 & 59116.604825944 & 425.3(4) & 23.3(8) & 9.900(5) \\
		66 & 133 & 59116.604880921 & 399.3(9) & 38.7(3) & 15.463(6) \\
		67 & 143 & 59116.605430690 & 419.7(7) & 15.4(1) & 6.468(6) \\
		68 & 144 & 59116.605485667 & 483.7(6) & 4.5(1) & 2.182(8) \\
		69 & 145 & 59116.605540644 & 459.0(3) & 6.0(4) & 2.773(8) \\
		70 & 146 & 59116.605595620 & 421.6(5) & 15.1(3) & 6.378(5) \\
		71  & 147  & 59116.605650597  & 437.5(7)  & 16.0(0)  & 7.000(4) \\
		72  & 148  & 59116.605705574  & 411.0(6)  & 12.7(4)  & 5.236(2) \\
		73  & 149  & 59116.605760551  & 450.9(7)  & 5.8(5)   & 2.592(8) \\
		74  & 150  & 59116.605815528  & 396.5(4)  & 24.5(8)  & 9.706(5) \\
		75  & 151  & 59116.605870505  & 388.7(3)  & 19.7(2)  & 7.665(3) \\
		76  & 152  & 59116.605925481  & 345.8(9)  & 31.4(7)  & 10.849(7) \\
		77  & 163  & 59116.606530227  & 398.7(1)  & 11.2(1)  & 4.470(6) \\
		78  & 164  & 59116.606585204  & 115.1(6)  & 21.6(2)  & 2.487(4) \\
		79  & 165  & 59116.606640181  & 105.0(7)  & 50.9(9)  & 5.345(1) \\
		80  & 166  & 59116.606695157  & 374.5(3)  & 7.5(8)   & 2.801(4) \\
		81  & 167  & 59116.606750134  & 233.9(5)  & 10.9(6)  & 2.541(7) \\
		82  & 181  & 59116.607519810  & 494.4(3)  & 5.9(9)   & 2.917(6) \\
		83  & 182  & 59116.607574787  & 461.8(4)  & 6.2(3)   & 2.877(3) \\
		84  & 183  & 59116.607629764  & 467.2(7)  & 8.9(1)   & 4.163(9) \\
		85  & 184  & 59116.607684741  & 436.6(4)  & 7.6(9)   & 3.314(7) \\
		86  & 185  & 59116.607739718  & 394.1(4)  & 14.1(8)  & 5.549(2) \\
		87  & 186  & 59116.607794694  & 400.6(3)  & 18.5(5)  & 7.411(6) \\
		88  & 198  & 59116.608454417  & 492.8(5)  & 2.5(6)   & 1.212(5) \\
		89  & 199  & 59116.608509394  & 489.6(3)  & 2.8(4)   & 1.391(6) \\
		90  & 200  & 59116.608564370  & 456.3(9)  & 5.6(7)   & 2.541(4) \\
		91  & 201  & 59116.608619347  & 448.5(7)  & 6.4(2)   & 2.879(5) \\
		92  & 202  & 59116.608674324  & 407.4(6)  & 12.8(3)  & 5.228(3) \\
		93  & 203  & 59116.608729301  & 383.2(7)  & 16.2(9)  & 6.204(4) \\
		94  & 204  & 59116.608784278  & 396.5(4)  & 15.1(1)  & 5.987(9) \\
		95  & 212  & 59116.609224093  & 87.1(3)   & 58.0(6)  & 5.047(5) \\
		96  & 213  & 59116.609279069  & 106.9(5)  & 24.6(6)  & 2.625(5) \\
		97  & 214  & 59116.609334046  & 348.1(6)  & 26.8(4)  & 9.342(8) \\
		98  & 215  & 59116.609389023  & 344.0(7)  & 26.1(2)  & 8.985(4) \\
		99  & 225  & 59116.609938792  & 498.9(6)  & 4.1(1)   & 2.050(7) \\
		100 & 226  & 59116.609993769  & 463.7(7)  & 7.2(9)   & 3.334(7) \\
		101 & 227  & 59116.610048745  & 456.8(7)  & 8.4(2)   & 3.846(6) \\
		102 & 228  & 59116.610103722  & 378.4(7)  & 7.4(3)   & 2.812(3) \\
		103 & 240  & 59116.610763444  & 366.3(7)  & 17.4(5)  & 6.355(5) \\
		104 & 241  & 59116.610818421  & 452.2(2)  & 4.2(5)   & 1.877(8) \\
		105 & 242  & 59116.610873398  & 448.6(1)  & 7.2(4)   & 3.248(3) \\
		106 & 243  & 59116.610928375  & 130.6(5)  & 17.3(6)  & 2.255(5) \\
		107 & 244  & 59116.610983352  & 424.6(3)  & 13.0(5)  & 5.541(1) \\
		108 & 245  & 59116.611038329  & 391.5(7)  & 13.6(6)  & 5.324(7) \\
		109 & 246  & 59116.611093306  & 383.6(5)  & 9.3(9)   & 3.563(5) \\
		110 & 258  & 59116.611753028  & 486.9(5)  & 4.4(5)   & 2.118(5) \\
		111 & 259  & 59116.611808005  & 467.7(4)  & 5.9(6)   & 2.741(5) \\
		112 & 260  & 59116.611862981  & 316.9(4)  & 38.5(3)  & 12.211(7) \\
		113 & 261  & 59116.611917958  & 107.4(7)  & 28.9(1)  & 3.105(7) \\
		114 & 262  & 59116.611972935  & 370.5(5)  & 11.3(2)  & 4.194(6) \\
		115 & 263  & 59116.612027912  & 82.7(6)   & 52.6(6)  & 4.352(6) \\
		116 & 264  & 59116.612082889  & 421.6(5)  & 10.9(8)  & 4.587(1) \\
		117 & 265  & 59116.612137866  & 100.0(9)  & 75.6(9)  & 7.560(5) \\
		118 & 283  & 59116.613127449  & 480.9(5)  & 4.2(7)   & 2.006(5) \\
		119 & 284  & 59116.613182426  & 470.0(1)  & 4.6(5)   & 2.138(6) \\
		120 & 295  & 59116.613787171  & 350.3(5)  & 2.9(6)   & 1.002(5) \\
		121 & 296  & 59116.613842148  & 410.6(1)  & 4.3(8)   & 1.757(5) \\
		122 & 297  & 59116.613897125  & 378.7(4)  & 13.1(1)  & 4.961(7) \\
		123 & 298  & 59116.613952102  & 390.6(4)  & 12.9(4)  & 5.054(9) \\
		124 & 299  & 59116.614007079  & 454.8(5)  & 5.0(8)   & 2.265(5) \\
		125 & 300  & 59116.614062056  & 100.5(7)  & 26.9(6)  & 2.700(7) \\
		126 & 301  & 59116.614117032  & 376.3(4)  & 10.4(9)  & 3.910(4) \\
		127 & 302  & 59116.614172009  & 390.0(7)  & 14.7(7)  & 5.722(5) \\
		128 & 313  & 59116.614776755  & 292.8(3)  & 11.4(6)  & 3.326(7) \\
		129 & 320  & 59116.615161593  & 460.0(3)  & 7.3(1)   & 3.362(5) \\
		130 & 321  & 59116.615216569  & 475.1(6)  & 5.2(2)   & 2.480(5) \\
		131 & 322  & 59116.615271546  & 454.8(5)  & 7.8(6)   & 3.529(7) \\
		132 & 323  & 59116.615326523  & 387.4(9)  & 37.7(1)  & 14.608(2) \\
		133 & 332  & 59116.615821315  & 459.4(4)  & 6.1(1)   & 2.803(5) \\
		134 & 333  & 59116.615876292  & 453.6(5)  & 5.0(3)   & 2.282(3) \\
		135 & 334  & 59116.615931269  & 416.3(3)  & 7.4(6)   & 3.064(6) \\
		136 & 340  & 59116.616261130  & 357.2(1)  & 24.5(5)  & 8.751(2) \\
		137 & 341  & 59116.616316106  & 412.5(3)  & 8.0(2)   & 3.308(3) \\
		138 & 342  & 59116.616371083  & 387.8(3)  & 11.4(7)  & 4.409(6) \\
		139 & 343  & 59116.616426060  & 358.6(5)  & 22.5(6)  & 8.055(5) \\
		140 & 344  & 59116.616481037  & 383.6(5)  & 21.9(3)  & 8.412(1) \\
		141 & 345  & 59116.616536014  & 365.4(7)  & 31.9(3)  & 11.666(7) \\
		142 & 346  & 59116.616590991  & 360.3(7)  & 25.5(2)  & 9.196(8) \\
		143 & 347  & 59116.616645968  & 342.1(6)  & 12.8(6)  & 4.366(2) \\
		144 & 354  & 59116.617030806  & 416.5(1)  & 10.5(8)  & 4.365(4) \\
		145 & 355  & 59116.617085782  & 134.3(6)  & 13.9(1)  & 1.868(6) \\
		146 & 356  & 59116.617140759  & 108.7(3)  & 30.2(3)  & 3.287(8) \\
		147 & 357  & 59116.617195736  & 86.3(7)   & 59.1(3)  & 5.102(1) \\
		148 & 358  & 59116.617250713  & 337.5(6)  & 30.7(6)  & 10.348(2) \\
		149 & 368  & 59116.617800481  & 424.2(9)  & 18.3(3)  & 7.763(8) \\
		150 & 369  & 59116.617855458  & 422.9(6)  & 2.7(4)   & 1.159(9) \\
		151 & 370  & 59116.617910435  & 457.7(7)  & 11.5(2)  & 5.273(8) \\
		152 & 371  & 59116.617965412  & 421.2(7)  & 17.8(2)  & 7.505(1) \\
		153 & 372  & 59116.618020389  & 417.9(4)  & 14.0(4)  & 5.867(5) \\
		154 & 373  & 59116.618075366  & 373.1(9)  & 22.5(4)  & 8.411(6) \\
		155 & 374  & 59116.618130343  & 381.8(3)  & 20.3(1)  & 7.755(4) \\
		156 & 375  & 59116.618185319  & 407.0(1)  & 12.7(2)  & 5.178(3) \\
		
		\label{burst1945}	% Additional rows continue here...
	\end{longtable}
	
	% Long table for multi-page spanning
	\clearpage
	
	\begin{longtable}{cccccc}
		\caption{Burst pulse information for PSR J2323+1214} \label{table:burst2323} \\
		\hline
		\textbf{S. No.} & \textbf{Pulse Number} & \textbf{Pulse Arrival Time} & \textbf{Pulse Width} & \textbf{Peak Flux Density} & \textbf{Fluence} \\
		&& (MJD) & (ms) & (mJy) & (Jy\,ms) \\
		\hline
		\endfirsthead
		
		\caption[]{Burst pulse information for PSR J2323+1214} \\
		\hline
		\textbf{S. No.} & \textbf{Pulse Number} & \textbf{Pulse Arrival Time} & \textbf{Pulse Width} & \textbf{Peak Flux Density} & \textbf{Fluence} \\
		&& (MJD) & (ms) & (mJy) & (Jy\,ms) \\
		\hline
		\endhead
		
		\hline \multicolumn{6}{r}{\textit{Continued on next page}} \\
		\endfoot
		
		\hline
		\endlastfoot
		
		1  & 2   & 59113.673701037  & 22.0(7)   & 28.5(4)   & 0.629(7)  \\
		2  & 3   & 59113.673744556  & 22.4(5)   & 29.1(2)   & 0.651(9)  \\
		3  & 4   & 59113.673788074  & 21.5(4)   & 29.0(0)   & 0.623(2)  \\
		4  & 5   & 59113.673831593  & 36.8(4)   & 15.3(8)   & 0.562(1)  \\
		5  & 14  & 59113.674223259  & 24.2(8)   & 4.8(6)    & 0.115(7)  \\
		6  & 16  & 59113.674310296  & 27.6(3)   & 27.9(9)   & 0.769(6)  \\
		7  & 17  & 59113.674353815  & 94.0(0)   & 3.6(7)    & 0.336(8)  \\
		8  & 18  & 59113.674397333  & 28.8(4)   & 23.5(1)   & 0.676(9)  \\
		9  & 19  & 59113.674440852  & 21.3(7)   & 32.5(6)   & 0.692(1)  \\
		10 & 20  & 59113.674484370  & 88.8(7)   & 23.3(9)   & 2.068(3)  \\
		11 & 21  & 59113.674527889  & 24.0(2)   & 19.2(2)   & 0.462(1)  \\
		12 & 22  & 59113.674571407  & 113.8(5)  & 10.5(9)   & 1.194(2)  \\
		13 & 23  & 59113.674614926  & 95.0(5)   & 6.0(8)    & 0.568(8)  \\
		14 & 24  & 59113.674658444  & 104.4(4)  & 2.6(8)    & 0.270(7)  \\
		15 & 25  & 59113.674701963  & 113.8(5)  & 2.7(6)    & 0.303(3)  \\
		16 & 26  & 59113.674745481  & 93.0(5)   & 9.8(6)    & 0.907(2)  \\
		17 & 27  & 59113.674789000  & 37.6(6)   & 12.1(3)   & 0.456(5)  \\
		18 & 29  & 59113.674876037  & 22.5(6)   & 44.7(7)   & 1.003(7)  \\
		19 & 34  & 59113.675093630  & 87.7(3)   & 2.2(9)    & 0.192(6)  \\
		20 & 35  & 59113.675137148  & 105.5(9)  & 5.0(0)    & 0.527(5)  \\
		21 & 36  & 59113.675180667  & 83.6(6)   & 9.0(0)    & 0.752(2)  \\
		22 & 37  & 59113.675224185  & 81.5(7)   & 9.6(6)    & 0.779(5)  \\
		23 & 38  & 59113.675267704  & 24.4(8)   & 17.2(2)   & 0.421(9)  \\
		24 & 39  & 59113.675311222  & 95.4(8)   & 26.3(3)   & 2.508(6)  \\
		25 & 40  & 59113.675354741  & 35.3(2)   & 36.7(7)   & 1.294(7)  \\
		26 & 41  & 59113.675398259  & 26.5(6)   & 27.0(5)   & 0.716(8)  \\
		27 & 42  & 59113.675441778  & 101.3(2)  & 10.9(1)   & 1.105(2)  \\
		28 & 43  & 59113.675485296  & 104.2(4)  & 3.5(7)    & 0.361(5)  \\
		29 & 44  & 59113.675528815  & 104.0(9)  & 4.8(8)    & 0.497(5)  \\
		30 & 45  & 59113.675572333  & 96.1(9)   & 11.5(9)   & 1.104(3)  \\
		31 & 46  & 59113.675615852  & 25.9(4)   & 8.9(1)    & 0.231(8)  \\
		32 & 47  & 59113.675659370  & 24.8(7)   & 29.8(2)   & 0.740(7)  \\
		33 & 59  & 59113.676181593  & 23.1(4)   & 13.4(2)   & 0.310(4)  \\
		34 & 61  & 59113.676268630  & 50.1(3)   & 18.0(0)   & 0.902(4)  \\
		35 & 62  & 59113.676312148  & 34.5(7)   & 28.8(9)   & 0.992(4)  \\
		36 & 63  & 59113.676355667  & 103.6(8)  & 25.6(9)   & 2.651(7)  \\
		37 & 64  & 59113.676399185  & 101.3(2)  & 15.3(7)   & 1.547(8)  \\
		38 & 65  & 59113.676442704  & 25.8(5)   & 12.3(2)   & 0.319(3)  \\
		39 & 69  & 59113.676616778  & 25.0(3)   & 3.3(8)    & 0.082(1)  \\
		40 & 75  & 59113.676877889  & 13.0(1)   & 2.3(2)    & 0.030(4)  \\
		41 & 82  & 59113.677182519  & 40.4(2)   & 6.4(7)    & 0.258(4)  \\
		42 & 83  & 59113.677226037  & 91.1(6)   & 14.7(1)   & 1.340(6)  \\
		43 & 84  & 59113.677269556  & 84.7(4)   & 20.6(7)   & 1.741(4)  \\
		44 & 85  & 59113.677313074  & 67.8(8)   & 19.5(7)   & 1.320(4)  \\
		45 & 86  & 59113.677356593  & 22.7(9)   & 28.9(3)   & 0.656(5)  \\
		46 & 87  & 59113.677400111  & 99.5(3)   & 3.4(4)    & 0.342(8)  \\
		47 & 88  & 59113.677443630  & 101.6(5)  & 9.9(8)    & 1.004(2)  \\
		48 & 89  & 59113.677487148  & 88.3(5)   & 29.5(1)   & 2.607(7)  \\
		49 & 90  & 59113.677530667  & 95.2(1)   & 9.0(1)    & 0.858(1)  \\
		50 & 91  & 59113.677574185  & 104.2(6)  & 7.5(5)    & 0.782(8)  \\
		51 & 92  & 59113.677617704  & 103.9(2)  & 6.8(8)    & 0.705(6)  \\
		52 & 93  & 59113.677661222  & 24.3(6)   & 16.5(2)   & 0.402(8)  \\
		53 & 119 & 59113.678792704  & 86.6(4)   & 5.5(6)    & 0.473(6)  \\
		56 & 122 & 59113.678923259 & 103.8(5) & 5.2(7) & 0.537(7) \\
		57 & 123 & 59113.678966778 & 103.2(1) & 4.6(8) & 0.473(3) \\
		58 & 124 & 59113.679010296 & 32.6(8) & 5.4(9) & 0.176(2) \\
		59 & 125 & 59113.679053815 & 30.1(4) & 4.8(3) & 0.145(9) \\
		60 & 126 & 59113.679097333 & 102.1(6) & 7.3(3) & 0.749(7) \\
		61 & 133 & 59113.679401963 & 14.0(7) & 2.5(1) & 0.035(6) \\
		62 & 138 & 59113.679619556 & 92.2(1) & 27.7(6) & 2.550(4) \\
		63 & 139 & 59113.679663074 & 21.0(8) & 19.5(5) & 0.410(7) \\
		64 & 147 & 59113.680011222 & 23.1(4) & 11.0(5) & 0.253(8) \\
		65 & 148 & 59113.680054741 & 101.8(3) & 20.8(5) & 2.123(5) \\
		66 & 149 & 59113.680098259 & 22.4(2) & 28.1(3) & 0.631(1) \\
		67 & 150 & 59113.680141778 & 25.6(9) & 30.3(2) & 0.776(2) \\
		68 & 151 & 59113.680185296 & 89.5(2) & 18.4(5) & 1.642(4) \\
		69 & 152 & 59113.680228815 & 24.8(2) & 27.3(2) & 0.677(2) \\
		70 & 153 & 59113.680272333 & 25.6(5) & 5.4(8) & 0.138(2) \\
		71 & 154 & 59113.680315852 & 95.1(8) & 9.8(1) & 0.933(8) \\
		72 & 155 & 59113.680359370 & 29.3(4) & 16.5(9) & 0.483(4) \\
		73 & 156 & 59113.680402889 & 89.9(5) & 51.5(5) & 4.628(9) \\
		74 & 157 & 59113.680446407 & 24.8(7) & 42.3(9) & 1.048(6) \\
		75 & 159 & 59113.680533444 & 94.1(4) & 3.2(2) & 0.301(2) \\
		76 & 160 & 59113.680576963 & 29.4(6) & 9.8(8) & 0.287(5) \\
		77 & 161 & 59113.680620481 & 22.0(8) & 30.7(8) & 0.675(7) \\
		78 & 172 & 59113.681099185 & 30.0(2) & 5.1(7) & 0.152(4) \\
		79 & 173 & 59113.681142704 & 16.8(8) & 3.0(8) & 0.050(7) \\
		80 & 175 & 59113.681229741 & 104.2(6) & 2.3(6) & 0.236(6) \\
		81  & 176  & 59113.681273259  & 108.9(4)  & 8.8(7)   & 0.955(1)  \\
		82  & 177  & 59113.681316778  & 28.6(6)   & 16.2(2)  & 0.464(2)  \\
		83  & 179  & 59113.681403815  & 23.5(2)   & 35.8(9)  & 0.840(2)  \\
		84  & 180  & 59113.681447333  & 24.0(3)   & 23.7(4)  & 0.571(8)  \\
		85  & 205  & 59113.682535296  & 21.9(5)   & 4.0(7)   & 0.087(6)  \\
		86  & 206  & 59113.682578815  & 22.8(2)   & 45.6(4)  & 1.042(8)  \\
		87  & 208  & 59113.682665852  & 22.3(6)   & 45.6(4)  & 1.018(9)  \\
		88  & 209  & 59113.682709370  & 89.6(6)   & 20.7(7)  & 1.854(2)  \\
		89  & 210  & 59113.682752889  & 32.2(4)   & 16.5(6)  & 0.531(1)  \\
		90  & 211  & 59113.682796407  & 103.4(4)  & 11.9(8)  & 1.228(2)  \\
		91  & 212  & 59113.682839926  & 89.2(8)   & 3.5(3)   & 0.315(5)  \\
		92  & 213  & 59113.682883444  & 90.1(3)   & 1.4(8)   & 0.124(8)  \\
		93  & 218  & 59113.683101037  & 102.4(9)  & 3.0(3)   & 0.310(6)  \\
		94  & 219  & 59113.683144556  & 90.9(6)   & 5.7(7)   & 0.515(4)  \\
		95  & 224  & 59113.683362148  & 100.6(1)  & 1.4(6)   & 0.137(4)  \\
		96  & 225  & 59113.683405667  & 88.9(3)   & 15.8(5)  & 1.400(1)  \\
		97  & 226  & 59113.683449185  & 79.9(9)   & 32.4(7)  & 2.586(3)  \\
		98  & 227  & 59113.683492704  & 80.9(3)   & 11.5(3)  & 0.933(3)  \\
		99  & 228  & 59113.683536222  & 79.8(8)   & 8.4(8)   & 0.668(8)  \\
		100 & 229  & 59113.683579741  & 78.5(6)   & 13.8(1)  & 1.083(7)  \\
		101 & 230  & 59113.683623259  & 89.0(6)   & 10.5(1)  & 0.935(7)  \\
		102 & 231  & 59113.683666778  & 101.8(7)  & 2.1(9)   & 0.213(3)  \\
		103 & 236  & 59113.683884370  & 15.1(4)   & 3.5(1)   & 0.053(6)  \\
		104 & 238  & 59113.683971407  & 53.4(2)   & 5.0(8)   & 0.266(4)  \\
		105 & 245  & 59113.684276037  & 95.4(9)   & 17.2(7)  & 1.639(6)  \\
		106 & 246  & 59113.684319556  & 101.8(1)  & 4.9(3)   & 0.502(7) \\
		107 & 247  & 59113.684363074  & 98.4(2)   & 3.1(1)   & 0.305(5)  \\
		108 & 248  & 59113.684406593  & 47.7(3)   & 4.1(3)   & 0.197(6)  \\
		109 & 249  & 59113.684450111  & 52.8(6)   & 6.3(9)   & 0.332(5)  \\
		110 & 250  & 59113.684493630  & 35.3(2)   & 8.2(1)   & 0.290(7)  \\
		111 & 251  & 59113.684537148  & 30.3(3)   & 9.8(5)   & 0.295(9)  \\
		112 & 261  & 59113.684972333  & 11.6(4)   & 1.0(8)   & 0.011(1)  \\
		113 & 264  & 59113.685102889  & 89.9(6)   & 21.5(6)  & 1.930(8)  \\
		114 & 265  & 59113.685146407  & 27.0(8)   & 21.7(7)  & 0.586(5)  \\
		115 & 266  & 59113.685189926  & 99.1(1)   & 7.5(8)   & 0.742(2)  \\
		116 & 267  & 59113.685233444  & 59.5(1)   & 14.9(7)  & 0.884(2)  \\
		117 & 268  & 59113.685276963  & 86.5(3)   & 8.2(5)   & 0.705(7)  \\
		118 & 269  & 59113.685320481  & 100.0(1)  & 6.4(4)   & 0.644(2)  \\
		119 & 270  & 59113.685364000  & 33.2(4)   & 1.9(7)   & 0.062(5)  \\
		120 & 273  & 59113.685494556  & 27.2(9)   & 10.7(1)  & 0.291(1)  \\
		121 & 277  & 59113.685668630  & 104.6(7)  & 2.2(9)   & 0.229(7)  \\
		122 & 282  & 59113.685886222  & 22.9(6)   & 32.9(6)  & 0.751(8)  \\
		123 & 291  & 59113.686277889  & 97.7(1)   & 5.1(9)   & 0.497(6)  \\
		124 & 292  & 59113.686321407  & 39.1(6)   & 9.1(3)   & 0.357(7)  \\
		125 & 293 & 59113.686364926 & 23.3(4) & 24.9(6) & 0.578(5) \\
		126 & 294 & 59113.686408444 & 101.8(3) & 2.8(7) & 0.282(3) \\
		127 & 295 & 59113.686451963 & 40.6(9) & 2.9(7) & 0.117(1) \\
		128 & 302 & 59113.686756593 & 100.1(2) & 3.5(9) & 0.349(7) \\
		129 & 303 & 59113.686800111 & 32.2(7) & 6.6(5) & 0.211(8) \\
		130 & 304 & 59113.686843630 & 86.8(5) & 7.1(2) & 0.618(6) \\
		131 & 305 & 59113.686887148 & 31.5(5) & 19.3(4) & 0.610(8) \\
		132 & 306 & 59113.686930667 & 83.4(9) & 12.9(8) & 1.074(9) \\
		133 & 307 & 59113.686974185 & 85.4(5) & 10.4(1) & 0.889(3) \\
		134 & 308 & 59113.687017704 & 100.5(4) & 9.3(1) & 0.935(2) \\
		135 & 311 & 59113.687148259 & 96.0(5) & 4.8(7) & 0.458(8) \\
		136 & 312 & 59113.687191778 & 30.6(4) & 5.2(7) & 0.158(9) \\
		137 & 317 & 59113.687409370 & 21.6(8) & 22.0(2) & 0.476(3) \\
		138 & 336 & 59113.688236222 & 35.6(9) & 9.1(7) & 0.323(3) \\
		139 & 343 & 59113.688540852 & 86.0(9) & 2.8(9) & 0.240(4) \\
		140 & 344 & 59113.688584370 & 91.8(3) & 3.6(1) & 0.331(3) \\
		141 & 345 & 59113.688627889 & 89.5(2) & 13.9(1) & 1.244(7) \\
		142 & 346 & 59113.688671407 & 96.3(2) & 18.3(6) & 1.759(1) \\
		143 & 347 & 59113.688714926 & 25.3(4) & 21.7(3) & 0.549(6) \\
		144 & 349 & 59113.688801963 & 26.8(2) & 8.1(1) & 0.218(8) \\
		145 & 350 & 59113.688845481 & 48.5(7) & 3.5(1) & 0.170(6) \\
		146 & 351 & 59113.688889000 & 33.5(6) & 6.8(2) & 0.228(6) \\
		147 & 352 & 59113.688932519 & 30.8(1) & 12.6(6) & 0.387(6) \\
		148 & 353 & 59113.688976037 & 29.9(2) & 24.6(3) & 0.736(7) \\
		149 & 354 & 59113.689019556 & 30.4(3) & 19.3(1) & 0.588(2) \\
		150 & 363 & 59113.689411222 & 107.4(8) & 2.2(8) & 0.234(7) \\
		151 & 364 & 59113.689454741 & 101.1(7) & 1.4(4) & 0.145(6) \\
		152 & 365 & 59113.689498259 & 55.2(5) & 3.4(1) & 0.188(7) \\
		153 & 366 & 59113.689541778 & 94.4(6) & 5.9(7) & 0.554(5) \\
		154 & 367 & 59113.689585296 & 25.0(3) & 12.5(8) & 0.312(6) \\
		155 & 368 & 59113.689628815 & 30.2(8) & 14.2(5) & 0.428(4) \\
		156 & 369 & 59113.689672333 & 23.9(2) & 12.5(8) & 0.299(7) \\
		157 & 370 & 59113.689715852 & 25.2(3) & 7.8(9) & 0.197(3) \\
		158 & 371 & 59113.689759370 & 97.2(8) & 5.8(5) & 0.569(1) \\
		159 & 372 & 59113.689802889 & 104.2(8) & 4.0(1) & 0.418(4) \\
		160 & 375 & 59113.689933444 & 85.6(3) & 17.0(5) & 1.460(7) \\
		161 & 376 & 59113.689976963 & 101.6(5) & 5.1(7) & 0.515(4) \\
		162 & 377 & 59113.690020481 & 26.6(3) & 17.7(8) & 0.471(3) \\
		163 & 378 & 59113.690064000 & 84.6(6) & 20.6(5) & 1.738(3) \\
		164 & 379 & 59113.690107519 & 22.8(3) & 18.6(1) & 0.423(1) \\
		165 & 380 & 59113.690151037 & 21.9(2) & 6.2(3) & 0.136(2) \\
		166 & 381 & 59113.690194556 & 101.6(7) & 6.0(5) & 0.604(7) \\
		167 & 382 & 59113.690238074 & 98.2(8) & 3.4(7) & 0.331(6) \\
		168 & 383 & 59113.690281593 & 32.3(5) & 5.6(7) & 0.180(6) \\
		169 & 384 & 59113.690325111 & 89.9(3) & 8.5(7) & 0.761(3) \\
		170 & 385 & 59113.690368630 & 23.2(2) & 31.0(4) & 0.719(6) \\
		171 & 386 & 59113.690412148 & 31.2(7) & 16.6(5) & 0.516(4) \\
		172 & 387 & 59113.690455667 & 28.1(6) & 17.2(4) & 0.484(7) \\
		173 & 388 & 59113.690499185 & 86.9(4) & 11.6(9) & 1.008(8) \\
		174 & 389 & 59113.690542704 & 94.2(8) & 6.2(7) & 0.581(5) \\
		175 & 390 & 59113.690586222 & 103.9(7) & 6.5(3)  & 0.678(8) \\
		176 & 395 & 59113.690803815 & 24.8(6)  & 5.6(9)  & 0.139(4) \\
		177 & 403 & 59113.691151963 & 30.2(5)  & 29.4(4) & 0.889(5) \\
		178 & 404 & 59113.691195481 & 89.4(1)  & 22.1(1) & 1.976(8) \\
		179 & 405 & 59113.691239000 & 27.4(3)  & 20.0(5) & 0.546(4) \\
		180 & 406 & 59113.691282519 & 34.5(7)  & 3.4(4)  & 0.117(8) \\
		181 & 407 & 59113.691326037 & 95.1(9)  & 3.7(6)  & 0.348(4) \\
		182 & 412 & 59113.691543630 & 21.1(8)  & 20.0(2) & 0.422(1) \\
		183 & 413 & 59113.691587148 & 91.5(3)  & 3.6(8)  & 0.328(3) \\
		184 & 414 & 59113.691630667 & 99.4(2)  & 3.3(4)  & 0.332(2) \\
		185 & 415 & 59113.691674185 & 101.7(4) & 0.8(1)  & 0.082(1) \\
		186 & 416 & 59113.691717704 & 93.2(2)  & 8.6(7)  & 0.799(5) \\
		187 & 417 & 59113.691761222 & 103.6(8) & 10.1(4) & 1.051(3) \\
		188 & 418 & 59113.691804741 & 22.0(2)  & 12.9(9) & 0.284(7) \\
		189 & 419 & 59113.691848259 & 59.3(6)  & 4.5(8)  & 0.266(6) \\
		190 & 421 & 59113.691935296 & 99.0(4)  & 1.4(3)  & 0.142(8) \\
		191 & 422 & 59113.691978815 & 28.1(2)  & 10.5(1) & 0.295(5) \\
		192 & 423 & 59113.692022333 & 98.6(8)  & 9.3(9)  & 0.916(5) \\
		193 & 424 & 59113.692065852 & 23.9(6)  & 24.1(3) & 0.576(8) \\
		194 & 425 & 59113.692109370 & 24.2(8)  & 20.1(1) & 0.486(8) \\
		195 & 426 & 59113.692152889 & 85.5(6)  & 10.1(2) & 0.866(2) \\
		196 & 432 & 59113.692414000 & 100.3(7) & 6.6(6)  & 0.658(3) \\
		197 & 433 & 59113.692457519 & 25.7(9)  & 18.5(6) & 0.474(9) \\
		198 & 434 & 59113.692501037 & 26.3(8)  & 19.3(1) & 0.508(3) \\
		199 & 435 & 59113.692544556 & 101.3(7) & 4.9(7)  & 0.493(7) \\
		200 & 436 & 59113.692588074 & 98.1(3)  & 4.4(4)  & 0.431(5) \\
		201 & 437 & 59113.692631593 & 80.2(1)  & 12.4(4) & 0.997(2) \\
		202 & 438 & 59113.692675111 & 93.6(2)  & 11.7(1) & 1.096(1) \\
		203 & 439 & 59113.692718630 & 51.9(5)  & 3.1(2)  & 0.162(7) \\
		204 & 440 & 59113.692762148 & 92.3(2)  & 7.9(4)  & 0.733(3) \\
		205 & 441 & 59113.692805667 & 36.7(6)  & 20.6(7) & 0.754(2) \\
		206 & 442 & 59113.692849185 & 24.3(6)  & 31.1(1) & 0.757(4) \\
		207 & 443 & 59113.692892704 & 24.6(5)  & 23.5(1) & 0.579(9) \\
		208 & 444 & 59113.692936222 & 24.8(7)  & 18.9(9) & 0.468(2) \\
		209 & 845 & 59113.710387148 & 98.8(3)  & 3.3(7)  & 0.323(7) \\
		210 & 846 & 59113.710430667 & 96.4(6)  & 8.6(3)  & 0.832(5) \\
		211 & 451 & 59113.693240852 & 21.7(3)  & 27.0(1) & 0.586(5) \\
		212 & 464 & 59113.693806593 & 97.9(8)  & 6.0(5)  & 0.583(9) \\
		213 & 465 & 59113.693850111 & 101.5(1) & 2.5(2)  & 0.256(2) \\
		214 & 466 & 59113.693893630 & 93.3(6)  & 2.4(2)  & 0.226(3) \\
		215 & 467 & 59113.693937148 & 34.2(6)  & 2.4(7)  & 0.081(3) \\
		216 & 468 & 59113.693980667 & 92.6(3)  & 13.5(3) & 1.253(2) \\
		217 & 469 & 59113.694024185 & 86.7(1)  & 16.3(4) & 1.417(4) \\
		218 & 470 & 59113.694067704 & 87.8(1)  & 16.9(2) & 1.485(4) \\
		219 & 473 & 59113.694198259 & 22.7(5)  & 8.8(1)  & 0.200(8) \\			
		% Additional rows continue here...
	\end{longtable}

		\clearpage
		\section{Bright pulses}
		
	In this appendix, we provide key information for the observed pulsars, including their rotational periods ($P$), the total number of pulses ($N_{\text{sub}}$), the observation duration ($T_{\text{obs}}$) in hours (h), the number of bright pulses observed ($N_{\text{bright}}$), and the thresholds used to identify these pulses (either exceeding a set value of the peak intensity of the average profile ($X$) or a given signal-to-noise ratio (S/N)). Additionally, we list the references from which the information was obtained.

			\setlength{\tabcolsep}{10 pt}  
			\begin{longtable}{ccccccc}
			\caption{Bright pulse information for different pulsars} \label{table:burst} \\
			\hline
			PSR Name & $P$ &  $T_{obs}$ & $N_{sub}$  & $N_{bright}$ & Threshold & Ref. \\
			&(s)& (h)&& & &  \\
			\hline
			\endhead
			
			\hline \multicolumn{6}{r}{\textit{Continued on next page}} \\
			\endfoot
			
			\hline
			\endlastfoot
			
			%\endfirsthead
		PSR J1939+2134  & 0.00156 &  13.43  &1.7$\times$10$^{7}$ & 1700 & $\textgreater$ 20$X$ & \citet{Cognard1996} \\
		& & 7.36 & $3.1\times10^{7}$ &  4265 & S/N $\textgreater$  16 & \citet{McKee2019}\\
		PSR J0034$-$0721 & 0.00188  & 8.0 &  1.53$\times$10$^{7}$  & 353 & $\textgreater$ 17$X$ & \citet{Tuoheti11} \\
		PSR J1824$-$2452A & 0.00305 & 26.94 & 3.18$\times$10$^{7}$&  476 & S/N $\textgreater$ 7 & \citet{Bilous2015}\\
		PSR J0534+2200& 0.033 & 91.67 & 1$\times$10$^{7}$  & 3$\times10^{4}$ & $\textgreater$ 200$X$ & \citet{Lundgren95}\\
		PSR J0540$-$6919 & 0.051 & 71.97 &  5.08$\times$10$^{6}$ & 141 & $\textgreater$  20$X$ & \citet{Johnston2003} \\
		PSR J1047$-$6709 & 0.198 & 0.42 &  7638 &  75 & $\textgreater$  10$X$ & \citet{Sun21} \\
		PSR J0953+0755 & 0.2531 &  22.02 & 313227 & 1688 & S/N $\textgreater$ 4 &  \citet{Kazantsev2022}  \\
		PSR J0659+1414 & 0.385 & 40.67 &  56368 & 3000 & $\textgreater$  120$X$ &  \citet{Kuzmin06} \\
		 &  & 2.67 &  25,000 &  150 & $\textgreater$  116X &  \citet{Weltevrede2006}\\%, 25,000, 0.4%, 116X
		PSR J1752+2359 & 0.409 & 5.72 & 50400 & 187 & S/N $\textgreater$ 5 & \citet{Ershov06} \\
		PSR J0034$-$0721& 0.943 & 4.0 & 15272 &  146 & S/N $\textgreater$ 10 & \citet{Wen21}\\%
		PSR J1136+1551  & 1.1879 &  35.43 &  107379 & 236 & S/R $\textgreater$ 4 &    \citet{Kazantsev2022}  \\
		PSR J0814+7429 & 1.2922 & 93.72 & 261108 &  39 & S/N $\textgreater$ 4 &  \citet{Kazantsev2022} \\ 
		PSR J1239+2453   & 1.3824 & 14.40 &  37483  & 145 & S/R $\textgreater$ 4 &   \citet{Kazantsev2022}  \\
		PSR J0304+1932  & 1.3876 & 5.622 & 14588 &  322 & S/N $\textgreater$  4 &  \citet{Kazantsev2022}  \\
		PSR J1115+5030 & 1.6564 & 29.92 &  65034 & 2023 & S/N $\textgreater$ 4 &  \citet{Kazantsev2022}  \\
		PSR J1900$-$0134 & 1.832 & 0.50 &  976 &  13 & $\textgreater$ 10$X$ & This work\\
		PSR J2323+1214 & 3.765 &  0.50 & 475 &  1 & $\textgreater$ 10$X$ & This work\\
		PSR J1900+4221 & 4.341 &  1.96 &  1626 & 12 & $\textgreater$ 10$X$ & \citet{HMT2022}\\
		PSR J1945+1211 & 4.756 & 0.50 &  377 &  11 & $\textgreater$ 10$X$ & This work\\

		\end{longtable}
		
	\newpage
	\bibliography{ApJS_Tedila}

\begin{thebibliography}{}
\expandafter\ifx\csname natexlab\endcsname\relax\def\natexlab#1{#1}\fi
\providecommand{\url}[1]{\href{#1}{#1}}
\providecommand{\dodoi}[1]{doi:~\href{http://doi.org/#1}{\nolinkurl{#1}}}
\providecommand{\doeprint}[1]{\href{http://ascl.net/#1}{\nolinkurl{http://ascl.net/#1}}}
\providecommand{\doarXiv}[1]{\href{https://arxiv.org/abs/#1}{\nolinkurl{https://arxiv.org/abs/#1}}}

\bibitem[{{Anumarlapudi} {et~al.}(2023){Anumarlapudi}, {Swiggum}, {Kaplan}, \&
  {Fichtenbauer}}]{Anumarlapudi2023}
{Anumarlapudi}, A., {Swiggum}, J.~K., {Kaplan}, D.~L., \& {Fichtenbauer}, T.
  D.~J. 2023, \apj, 948, 32, \dodoi{10.3847/1538-4357/acbb68}

\bibitem[{{Arons} \& {Scharlemann}(1979)}]{Arons1979}
{Arons}, J., \& {Scharlemann}, E.~T. 1979, \apj, 231, 854,
  \dodoi{10.1086/157250}

\bibitem[{{Backer}(1970{\natexlab{a}})}]{Backer1970}
{Backer}, D.~C. 1970{\natexlab{a}}, \nat, 228, 42, \dodoi{10.1038/228042a0}

\bibitem[{{Backer}(1970{\natexlab{b}})}]{Backer1970a}
---. 1970{\natexlab{b}}, \nat, 227, 692, \dodoi{10.1038/227692a0}

\bibitem[{{Basu} \& {Mitra}(2018)}]{Basu2018}
{Basu}, R., \& {Mitra}, D. 2018, \mnras, 476, 1345,
  \dodoi{10.1093/mnras/sty297}

\bibitem[{{Basu} {et~al.}(2017){Basu}, {Mitra}, \& {Melikidze}}]{Basu17}
{Basu}, R., {Mitra}, D., \& {Melikidze}, G.~I. 2017, \apj, 846, 109,
  \dodoi{10.3847/1538-4357/aa862d}

\bibitem[{Basu {et~al.}(2016)Basu, Mitra, Melikidze, Maciesiak, Skrzypczak, \&
  Szary}]{Basu16}
Basu, R., Mitra, D., Melikidze, G.~I., {et~al.} 2016, The Astrophysical
  Journal, 833, 29, \dodoi{10.3847/1538-4357/833/1/29}

\bibitem[{{Bhattacharya} \& {van den Heuvel}(1991)}]{Bhattacharya1991}
{Bhattacharya}, D., \& {van den Heuvel}, E.~P.~J. 1991, \physrep, 203, 1,
  \dodoi{10.1016/0370-1573(91)90064-S}

\bibitem[{{Bhattacharyya} {et~al.}(2010){Bhattacharyya}, {Gupta}, \&
  {Gil}}]{Bhat10}
{Bhattacharyya}, B., {Gupta}, Y., \& {Gil}, J. 2010, \mnras, 408, 407,
  \dodoi{10.1111/j.1365-2966.2010.17116.x}

\bibitem[{{Biggs}(1992)}]{Bigg92}
{Biggs}, J.~D. 1992, \apj, 394, 574, \dodoi{10.1086/171608}

\bibitem[{{Bilous} {et~al.}(2015){Bilous}, {Pennucci}, {Demorest}, \&
  {Ransom}}]{Bilous2015}
{Bilous}, A.~V., {Pennucci}, T.~T., {Demorest}, P., \& {Ransom}, S.~M. 2015,
  \apj, 803, 83, \dodoi{10.1088/0004-637X/803/2/83}

\bibitem[{{Burke-Spolaor} \& {Bailes}(2010)}]{Burke-Bailes2010}
{Burke-Spolaor}, S., \& {Bailes}, M. 2010, \mnras, 402, 855,
  \dodoi{10.1111/j.1365-2966.2009.15965.x}

\bibitem[{{Caleb} {et~al.}(2022){Caleb}, {Heywood}, {Rajwade}, {Malenta},
  {Stappers}, {Barr}, {Chen}, {Morello}, {Sanidas}, {van den Eijnden},
  {Kramer}, {Buckley}, {Brink}, {Motta}, {Woudt}, {Weltevrede}, {Jankowski},
  {Surnis}, {Buchner}, {Bezuidenhout}, {Driessen}, \& {Fender}}]{Caleb22}
{Caleb}, M., {Heywood}, I., {Rajwade}, K., {et~al.} 2022, Nature Astronomy, 6,
  828, \dodoi{10.1038/s41550-022-01688-x}

\bibitem[{Cameron {et~al.}(2020)Cameron, Li, Hobbs, Zhang, Miao, Wang, Yuan,
  Wang, Jacobs~Corban, Cruces, Dai, Feng, Han, Kaczmarek, Niu, Pan, Qian, Tao,
  Wang, Wang, Xu, Xu, Yue, Zhang, Zhi, Zhu, Champion, Kramer, Zhou, Qiu, \&
  Zhu}]{CLH+20}
Cameron, A.~D., Li, D., Hobbs, G., {et~al.} 2020, \mnras, 495, 3515,
  \dodoi{10.1093/mnras/staa1328}

\bibitem[{{Chen} \& {Ruderman}(1993)}]{Chen1993}
{Chen}, K., \& {Ruderman}, M. 1993, \apj, 402, 264, \dodoi{10.1086/172129}

\bibitem[{{Chen} {et~al.}(2023{\natexlab{a}}){Chen}, {Yan}, {Han}, {Wang},
  {Wang}, {Jing}, {Lee}, {Zhang}, {Xu}, {Wang}, {Yang}, {Su}, {Cai}, {Wang},
  {Qiao}, {Xu}, \& {Zhou}}]{Chen2023}
{Chen}, X., {Yan}, Y., {Han}, J.~L., {et~al.} 2023{\natexlab{a}}, Nature
  Astronomy, 7, 1235, \dodoi{10.1038/s41550-023-02056-z}

\bibitem[{{Chen} {et~al.}(2023{\natexlab{b}}){Chen}, {Wang}, {Li},
  {G{\"u}gercino{\u{g}}lu}, {Zhao}, {Meng}, {Yuan}, {Niu}, {Zhu}, {Feng},
  {Miao}, {Niu}, {Wu}, {Wang}, {Wang}, {Xie}, {Xue}, {Yao}, {Yuan}, {You},
  {Yu}, {Yue}, {Zhang}, {Zhang}, {Zhang}, {Wang}, {Gan}, {Li}, {Sun}, \&
  {Wang}}]{Chen23}
{Chen}, Y., {Wang}, P., {Li}, D., {et~al.} 2023{\natexlab{b}}, Research in
  Astronomy and Astrophysics, 23, 085022, \dodoi{10.1088/1674-4527/acd89d}

\bibitem[{{Cognard} {et~al.}(1996){Cognard}, {Shrauner}, {Taylor}, \&
  {Thorsett}}]{Cognard1996}
{Cognard}, I., {Shrauner}, J.~A., {Taylor}, J.~H., \& {Thorsett}, S.~E. 1996,
  \apjl, 457, L81, \dodoi{10.1086/309894}

\bibitem[{{Cordes}(1979)}]{Cordes1979}
{Cordes}, J.~M. 1979, Australian Journal of Physics, 32, 9,
  \dodoi{10.1071/PH790009}

\bibitem[{{Cordes} {et~al.}(1990){Cordes}, {Weisberg}, \&
  {Hankins}}]{Cordes1990}
{Cordes}, J.~M., {Weisberg}, J.~M., \& {Hankins}, T.~H. 1990, \aj, 100, 1882,
  \dodoi{10.1086/115644}

\bibitem[{{Dang} {et~al.}(2024){Dang}, {Yuan}, {Shang}, {Xu}, {Tedila}, {Wen},
  {Yan}, {Bai}, {Li}, {Xiao}, {Zhao}, {Zhi}, {Dong}, {Wu}, {Rukiye}, {Zhang},
  {Yang}, {Wu}, \& {Yuen}}]{Dang24}
{Dang}, S.~J., {Yuan}, J.~P., {Shang}, L.~H., {et~al.} 2024, \mnras, 528, 1213,
  \dodoi{10.1093/mnras/stae046}

\bibitem[{{De} {et~al.}(2016){De}, {Gupta}, \& {Sharma}}]{De2016}
{De}, K., {Gupta}, Y., \& {Sharma}, P. 2016, \apjl, 833, L10,
  \dodoi{10.3847/2041-8213/833/1/L10}

\bibitem[{{Deng} {et~al.}(2020){Deng}, {Gao}, {Li}, \& {Shao}}]{Deng2020}
{Deng}, Z.-L., {Gao}, Z.-F., {Li}, X.-D., \& {Shao}, Y. 2020, \apj, 892, 4,
  \dodoi{10.3847/1538-4357/ab76c4}

\bibitem[{Dicke(1982)}]{Dicke1982}
Dicke, R.~H. 1982, The Measurement of Thermal Radiation at Microwave
  Frequencies (Dordrecht: Springer Netherlands), 106--113,
  \dodoi{10.1007/978-94-009-7752-5_11}

\bibitem[{{Edwards} \& {Stappers}(2002)}]{Edwards02}
{Edwards}, R.~T., \& {Stappers}, B.~W. 2002, \aap, 393, 733,
  \dodoi{10.1051/0004-6361:20021067}

\bibitem[{{Edwards} \& {Stappers}(2003)}]{Edwards03}
---. 2003, \aap, 407, 273, \dodoi{10.1051/0004-6361:20030716}

\bibitem[{{Ershov} \& {Kuzmin}(2006)}]{Ershov06}
{Ershov}, A.~A., \& {Kuzmin}, A.~D. 2006, arXiv e-prints, astro,
  \dodoi{10.48550/arXiv.astro-ph/0605086}

\bibitem[{{Filippenko} \& {Radhakrishnan}(1982)}]{Filippenko82}
{Filippenko}, A.~V., \& {Radhakrishnan}, V. 1982, \apj, 263, 828,
  \dodoi{10.1086/160553}

\bibitem[{{Gajjar} {et~al.}(2012){Gajjar}, {Joshi}, \& {Kramer}}]{Gajjar12}
{Gajjar}, V., {Joshi}, B.~C., \& {Kramer}, M. 2012, \mnras, 424, 1197,
  \dodoi{10.1111/j.1365-2966.2012.21296.x}

\bibitem[{{Gajjar} {et~al.}(2017){Gajjar}, {Yuan}, {Yuen}, {Wen}, {Liu}, \&
  {Wang}}]{GYR17}
{Gajjar}, V., {Yuan}, J.~P., {Yuen}, R., {et~al.} 2017, \apj, 850, 173,
  \dodoi{10.3847/1538-4357/aa96ac}

\bibitem[{{Gao} {et~al.}(2019){Gao}, {Peng}, \& {Wang}}]{Gao2019}
{Gao}, Z.-F., {Peng}, F.-K., \& {Wang}, N. 2019, Astronomische Nachrichten,
  340, 1023, \dodoi{10.1002/asna.201913710}

\bibitem[{Goldreich \& Julian(1969)}]{Goldreich69}
Goldreich, P., \& Julian, W.~H. 1969, ApJ, 157, 869

\bibitem[{{Guo} {et~al.}(2023){Guo}, {Wen}, {Yuan}, {Kou}, {Wu}, {Wang}, {Zhu},
  {Li}, {Xue}, {Wang}, {Miao}, {Zhao}, {Hu}, {Yan}, {Niu}, {Rejep}, \&
  {Huang}}]{Guo23}
{Guo}, Z., {Wen}, Z., {Yuan}, J., {et~al.} 2023, Research in Astronomy and
  Astrophysics, 23, 075008, \dodoi{10.1088/1674-4527/accdc1}

\bibitem[{{Hankins}(1971)}]{Hankins1971}
{Hankins}, T.~H. 1971, \apj, 169, 487, \dodoi{10.1086/151164}

\bibitem[{{Hankins} \& {Rickett}(1975)}]{Hankins1975}
{Hankins}, T.~H., \& {Rickett}, B.~J. 1975, Methods in Computational Physics,
  14, 55, \dodoi{10.1016/B978-0-12-460814-6.50007-3}

\bibitem[{{Helfand} {et~al.}(1975){Helfand}, {Manchester}, \& {Taylor}}]{HMT75}
{Helfand}, D.~J., {Manchester}, R.~N., \& {Taylor}, J.~H. 1975, \apj, 198, 661,
  \dodoi{10.1086/153644}

\bibitem[{{Herfindal} \& {Rankin}(2007)}]{HerfindalRankin07}
{Herfindal}, J.~L., \& {Rankin}, J.~M. 2007, \mnras, 380, 430,
  \dodoi{10.1111/j.1365-2966.2007.12089.x}

\bibitem[{{Hotan} {et~al.}(2004){Hotan}, {van Straten}, \&
  {Manchester}}]{Hotan04}
{Hotan}, A.~W., {van Straten}, W., \& {Manchester}, R.~N. 2004, \pasa, 21, 302,
  \dodoi{10.1071/AS04022}

\bibitem[{{Jiang} {et~al.}(2019){Jiang}, {Yue}, {Gan}, {Yao}, {Li}, {Pan},
  {Sun}, {Yu}, {Liu}, {Tang}, {Qian}, {Lu}, {Yan}, {Peng}, {Zhang}, {Wang},
  {Li}, \& {Li}}]{Jiang19}
{Jiang}, P., {Yue}, Y., {Gan}, H., {et~al.} 2019, Science China Physics,
  Mechanics, and Astronomy, 62, 959502, \dodoi{10.1007/s11433-018-9376-1}

\bibitem[{{Jiang} {et~al.}(2020){Jiang}, {Tang}, {Hou}, {Liu}, {Kr{\v{c}}o},
  {Qian}, {Sun}, {Ching}, {Liu}, {Duan}, {Yue}, {Gan}, {Yao}, {Li}, {Pan},
  {Yu}, {Liu}, {Li}, {Peng}, {Yan}, \& {FAST Collaboration}}]{Jiang20}
{Jiang}, P., {Tang}, N.-Y., {Hou}, L.-G., {et~al.} 2020, Research in Astronomy
  and Astrophysics, 20, 064, \dodoi{10.1088/1674-4527/20/5/64}

\bibitem[{{Johnston} \& {Romani}(2003)}]{Johnston2003}
{Johnston}, S., \& {Romani}, R.~W. 2003, \apjl, 590, L95,
  \dodoi{10.1086/376826}

\bibitem[{{Kazantsev} \& {Basalaeva}(2022)}]{Kazantsev2022}
{Kazantsev}, A.~N., \& {Basalaeva}, M.~Y. 2022, \mnras, 513, 4332,
  \dodoi{10.1093/mnras/stac473}

\bibitem[{{Knight}(2006)}]{Knight2006}
{Knight}, H.~S. 2006, Chinese Journal of Astronomy and Astrophysics Supplement,
  6, 41

\bibitem[{{Kramer} {et~al.}(2006){Kramer}, {Lyne}, {O'Brien}, {Jordan}, \&
  {Lorimer}}]{KAL06}
{Kramer}, M., {Lyne}, A.~G., {O'Brien}, J.~T., {Jordan}, C.~A., \& {Lorimer},
  D.~R. 2006, Science, 312, 549, \dodoi{10.1126/science.1124060}

\bibitem[{{Krishnakumar} {et~al.}(2015){Krishnakumar}, {Mitra}, {Naidu},
  {Joshi}, \& {Manoharan}}]{Krishnakumar2015}
{Krishnakumar}, M.~A., {Mitra}, D., {Naidu}, A., {Joshi}, B.~C., \&
  {Manoharan}, P.~K. 2015, \apj, 804, 23, \dodoi{10.1088/0004-637X/804/1/23}

\bibitem[{{Kuzmin} \& {Ershov}(2006)}]{Kuzmin06}
{Kuzmin}, A.~D., \& {Ershov}, A.~A. 2006, Astronomy Letters, 32, 583,
  \dodoi{10.1134/S1063773706090027}

\bibitem[{{Lange} {et~al.}(1998){Lange}, {Kramer}, {Wielebinski}, \&
  {Jessner}}]{Lange1998}
{Lange}, C., {Kramer}, M., {Wielebinski}, R., \& {Jessner}, A. 1998, \aap, 332,
  111

\bibitem[{{Li} {et~al.}(2018){Li}, {Wang}, {Qian}, {Krco}, {Jiang}, {Yue},
  {Jin}, {Zhu}, {Pan}, {Nan}, \& {Dunning}}]{Li18}
{Li}, D., {Wang}, P., {Qian}, L., {et~al.} 2018, IEEE Microwave Magazine, 19,
  112, \dodoi{10.1109/MMM.2018.2802178}

\bibitem[{{Liu} {et~al.}(2022){Liu}, {Antoniadis}, {Bassa}, {Chen}, {Cognard},
  {Gaikwad}, {Hu}, {Jang}, {Janssen}, {Karuppusamy}, {Kramer}, {Lee}, {Main},
  {Mall}, {McKee}, {Mickaliger}, {Perrodin}, {Sanidas}, {Stappers}, {Wang},
  {Zhu}, {Burgay}, {Concu}, {Corongiu}, {Melis}, {Pilia}, \&
  {Possenti}}]{Liu2022}
{Liu}, K., {Antoniadis}, J., {Bassa}, C.~G., {et~al.} 2022, \mnras, 513, 4037,
  \dodoi{10.1093/mnras/stac1082}

\bibitem[{{Liu} {et~al.}(2024){Liu}, {Wen}, {Wang}, {Kou}, {Yan}, {Yuan}, {Wu},
  {Zhao}, {Miao}, {Wang}, {Xue}, {Li}, {Wang}, {Sun}, {Zhu}, \&
  {Bai}}]{Liu2024}
{Liu}, S., {Wen}, Z., {Wang}, S., {et~al.} 2024, Research in Astronomy and
  Astrophysics, 24, 115001, \dodoi{10.1088/1674-4527/ad7fb6}

\bibitem[{Lorimer \& Kramer(2005)}]{Lorimer2005}
Lorimer, D., \& Kramer, M. 2005, Handbook of Pulsar Astronomy, Cambridge
  Observing Handbooks for Research Astronomers (Cambridge University Press).
\newblock \url{https://books.google.com.hk/books?id=OZ8tdN6qJcsC}

\bibitem[{{Lundgren} {et~al.}(1995){Lundgren}, {Cordes}, {Ulmer}, {Matz},
  {Lomatch}, {Foster}, \& {Hankins}}]{Lundgren95}
{Lundgren}, S.~C., {Cordes}, J.~M., {Ulmer}, M., {et~al.} 1995, \apj, 453, 433,
  \dodoi{10.1086/176404}

\bibitem[{Lyne \& Graham-Smith(2012)}]{Lyne_Graham-Smith_2012}
Lyne, A., \& Graham-Smith, F. 2012, Pulsar Astronomy, 4th edn., Cambridge
  Astrophysics (Cambridge University Press)

\bibitem[{{Lyne} \& {Manchester}(1988)}]{Lyne88}
{Lyne}, A.~G., \& {Manchester}, R.~N. 1988, \mnras, 234, 477,
  \dodoi{10.1093/mnras/234.3.477}

\bibitem[{Lyubarskii \& Kirk(2001)}]{LK01}
Lyubarskii, Y.~E., \& Kirk, J.~G. 2001, ApJ, 547, 437

\bibitem[{{Lyutikov}(2003)}]{Lyutikov2003}
{Lyutikov}, M. 2003, \mnras, 346, 540, \dodoi{10.1046/j.1365-2966.2003.07110.x}

\bibitem[{{Machabeli} {et~al.}(2001){Machabeli}, {Khechinashvili}, {Melikidze},
  \& {Shapakidze}}]{Machabeli2001}
{Machabeli}, G., {Khechinashvili}, D., {Melikidze}, G., \& {Shapakidze}, D.
  2001, \mnras, 327, 984, \dodoi{10.1046/j.1365-8711.2001.04814.x}

\bibitem[{{Mahajan} {et~al.}(2018){Mahajan}, {van Kerkwijk}, {Main}, \&
  {Pen}}]{Mahajan2018}
{Mahajan}, N., {van Kerkwijk}, M.~H., {Main}, R., \& {Pen}, U.-L. 2018, \apjl,
  867, L2, \dodoi{10.3847/2041-8213/aae713}

\bibitem[{{Manchester} {et~al.}(2005){Manchester}, {Hobbs}, {Teoh}, \&
  {Hobbs}}]{Manchester2005}
{Manchester}, R.~N., {Hobbs}, G.~B., {Teoh}, A., \& {Hobbs}, M. 2005, \aj, 129,
  1993, \dodoi{10.1086/428488}

\bibitem[{{Manchester} \& {Taylor}(1977)}]{Pulsars1977}
{Manchester}, R.~N., \& {Taylor}, J.~H. 1977, {Pulsars} (San Francisco : W. H.
  Freeman)

\bibitem[{{McKee} {et~al.}(2019){McKee}, {Stappers}, {Bassa}, {Chen},
  {Cognard}, {Gaikwad}, {Janssen}, {Karuppusamy}, {Kramer}, {Lee}, {Liu},
  {Perrodin}, {Sanidas}, {Smits}, {Wang}, \& {Zhu}}]{McKee2019}
{McKee}, J.~W., {Stappers}, B.~W., {Bassa}, C.~G., {et~al.} 2019, \mnras, 483,
  4784, \dodoi{10.1093/mnras/sty3058}

\bibitem[{{McLaughlin} {et~al.}(2006){McLaughlin}, {Lyne}, {Lorimer}, {Kramer},
  {Faulkner}, {Manchester}, {Cordes}, {Camilo}, {Possenti}, {Stairs}, {Hobbs},
  {D'Amico}, {Burgay}, \& {O'Brien}}]{McLaughlin2006}
{McLaughlin}, M.~A., {Lyne}, A.~G., {Lorimer}, D.~R., {et~al.} 2006, \nat, 439,
  817, \dodoi{10.1038/nature04440}

\bibitem[{{Melrose} \& {Yuen}(2014)}]{Melrose14}
{Melrose}, D.~B., \& {Yuen}, R. 2014, \mnras, 437, 262,
  \dodoi{10.1093/mnras/stt1876}

\bibitem[{{Melrose} \& {Yuen}(2016)}]{Melrose16}
---. 2016, Journal of Plasma Physics, 82, 635820202,
  \dodoi{10.1017/S0022377816000398}

\bibitem[{Mitra {et~al.}(2015)Mitra, Arjunwadkar, \& Rankin}]{Mitra2015}
Mitra, D., Arjunwadkar, M., \& Rankin, J.~M. 2015, The Astrophysical Journal,
  806, 236, \dodoi{10.1088/0004-637X/806/2/236}

\bibitem[{{Mitra} \& {Rankin}(2017)}]{Mitra17}
{Mitra}, D., \& {Rankin}, J. 2017, \mnras, 468, 4601,
  \dodoi{10.1093/mnras/stx814}

\bibitem[{{Naidu} {et~al.}(2018){Naidu}, {Joshi}, {Manoharan}, \&
  {Krishnakumar}}]{Naidu2018}
{Naidu}, A., {Joshi}, B.~C., {Manoharan}, P.~K., \& {Krishnakumar}, M.~A. 2018,
  \mnras, 475, 2375, \dodoi{10.1093/mnras/stx3284}

\bibitem[{{Philippov} \& {Kramer}(2022)}]{Philippov22}
{Philippov}, A., \& {Kramer}, M. 2022, \araa, 60, 495,
  \dodoi{10.1146/annurev-astro-052920-112338}

\bibitem[{{Philippov} {et~al.}(2020){Philippov}, {Timokhin}, \&
  {Spitkovsky}}]{Philippov2020}
{Philippov}, A., {Timokhin}, A., \& {Spitkovsky}, A. 2020, \prl, 124, 245101,
  \dodoi{10.1103/PhysRevLett.124.245101}

\bibitem[{{Popov} {et~al.}(2002){Popov}, {Bartel}, {Cannon}, {Novikov},
  {Kondratiev}, \& {Altunin}}]{Popov2002}
{Popov}, M.~V., {Bartel}, N., {Cannon}, W.~H., {et~al.} 2002, \aap, 396, 171,
  \dodoi{10.1051/0004-6361:20021402}

\bibitem[{{Qian} {et~al.}(2019){Qian}, {Pan}, {Li}, {Hobbs}, {Zhu}, {Wang},
  {Liu}, {Yue}, {Zhu}, {Liu}, {Yu}, {Sun}, {Jiang}, {Pan}, {Li}, {Gan}, {Yao},
  {Xie}, {Camilo}, {Cameron}, {Zhang}, \& {Wang}}]{Qian2019}
{Qian}, L., {Pan}, Z., {Li}, D., {et~al.} 2019, Science China Physics,
  Mechanics, and Astronomy, 62, 959508, \dodoi{10.1007/s11433-018-9354-y}

\bibitem[{{Radhakrishnan} \& {Cooke}(1969)}]{RadhakrishnanCooke1969}
{Radhakrishnan}, V., \& {Cooke}, D.~J. 1969, \aplett, 3, 225

\bibitem[{{Rahaman} {et~al.}(2021){Rahaman}, {Basu}, {Mitra}, \&
  {Melikidze}}]{Rahaman2021}
{Rahaman}, S. k.~M., {Basu}, R., {Mitra}, D., \& {Melikidze}, G.~I. 2021,
  \mnras, 500, 4139, \dodoi{10.1093/mnras/staa3518}

\bibitem[{{Rankin}(1993)}]{Rankin1993}
{Rankin}, J.~M. 1993, \apj, 405, 285, \dodoi{10.1086/172361}

\bibitem[{{Rankin} \& {Wright}(2008)}]{Rankin2008}
{Rankin}, J.~M., \& {Wright}, G. A.~E. 2008, \mnras, 385, 1923,
  \dodoi{10.1111/j.1365-2966.2008.13001.x}

\bibitem[{{Rankin} {et~al.}(2017){Rankin}, {Archibald}, {Hessels}, {van
  Leeuwen}, {Mitra}, {Ransom}, {Stairs}, {van Straten}, \&
  {Weisberg}}]{Rankin2017}
{Rankin}, J.~M., {Archibald}, A., {Hessels}, J., {et~al.} 2017, \apj, 845, 23,
  \dodoi{10.3847/1538-4357/aa7b73}

\bibitem[{{Redman} \& {Rankin}(2009)}]{RR09}
{Redman}, S.~L., \& {Rankin}, J.~M. 2009, \mnras, 395, 1529,
  \dodoi{10.1111/j.1365-2966.2009.14632.x}

\bibitem[{{Rejep} {et~al.}(2022){Rejep}, {Wang}, {Yan}, \& {Wen}}]{Rejep22}
{Rejep}, R., {Wang}, N., {Yan}, W.~M., \& {Wen}, Z.~G. 2022, \mnras, 509, 2507,
  \dodoi{10.1093/mnras/stab3063}

\bibitem[{{Ritchings}(1976)}]{Ritch76}
{Ritchings}, R.~T. 1976, \mnras, 176, 249, \dodoi{10.1093/mnras/176.2.249}

\bibitem[{{Rosen} {et~al.}(2013){Rosen}, {Swiggum}, {McLaughlin}, {Lorimer},
  {Yun}, {Heatherly}, {Boyles}, {Lynch}, {Kondratiev}, {Scoles}, {Ransom},
  {Moniot}, {Cottrill}, {Weaver}, {Snider}, {Thompson}, {Raycraft},
  {Dudenhoefer}, {Allphin}, {Thorley}, {Meadows}, {Marchiny}, {Liska},
  {O'Dwyer}, {Butler}, {Bloxton}, {Mabry}, {Abate}, {Boothe}, {Pritt},
  {Alberth}, {Green}, {Crowley}, {Agee}, {Nagley}, {Sargent}, {Hinson},
  {Smith}, {McNeely}, {Quigley}, {Pennington}, {Chen}, {Maynard}, {Loope},
  {Bielski}, {McGough}, {Gural}, {Colvin}, {Tso}, {Ewen}, {Zhang},
  {Ciccarella}, {Bukowski}, {Novotny}, {Gore}, {Sarver}, {Johnson},
  {Cunningham}, {Collins}, {Gardner}, {Monteleone}, {Hall}, {Schweinhagen},
  {Ayers}, {Jay}, {Uosseph}, {Dunkum}, {Pal}, {Dydiw}, {Sterling}, \&
  {Phan}}]{RSM13}
{Rosen}, R., {Swiggum}, J., {McLaughlin}, M.~A., {et~al.} 2013, \apj, 768, 85,
  \dodoi{10.1088/0004-637X/768/1/85}

\bibitem[{Ruderman \& Sutherland(1975)}]{RudermanSutherland1975}
Ruderman, M., \& Sutherland, P.~G. 1975, ApJ, 196, 51

\bibitem[{{Sheikh} \& {MacDonald}(2021)}]{Sheikh2021}
{Sheikh}, S.~Z., \& {MacDonald}, M.~G. 2021, \mnras, 502, 4669,
  \dodoi{10.1093/mnras/stab282}

\bibitem[{{Sun} {et~al.}(2021){Sun}, {Yan}, \& {Wang}}]{Sun21}
{Sun}, S.~N., {Yan}, W.~M., \& {Wang}, N. 2021, \mnras, 501, 3900,
  \dodoi{10.1093/mnras/staa3825}

\bibitem[{{Tedila} {et~al.}(2022){Tedila}, {Yuen}, {Wang}, {Yuan}, {Wen},
  {Yan}, {Wang}, {Dang}, {Li}, {Wang}, {Zhu}, {Niu}, {Miao}, {Xue}, {Zhang},
  {Tu}, {Rejep}, {Xie}, \& {FAST Collaboration}}]{HMT2022}
{Tedila}, H.~M., {Yuen}, R., {Wang}, N., {et~al.} 2022, \apj, 929, 171,
  \dodoi{10.3847/1538-4357/ac5f42}

\bibitem[{{Tedila} {et~al.}(2024){Tedila}, {Yuen}, {Wang}, {Li}, {Wen}, {Yan},
  {Yuan}, {Han}, {Wang}, {Zhu}, {Dang}, {Wang}, {Xie}, {Wu}, {Khasanov}, \&
  {FAST Collaboration}}]{HMT2024}
---. 2024, \apj, 965, 144, \dodoi{10.3847/1538-4357/ad2e06}

\bibitem[{{Timokhin}(2010)}]{Timokhin10}
{Timokhin}, A.~N. 2010, \mnras, 408, L41,
  \dodoi{10.1111/j.1745-3933.2010.00924.x}

\bibitem[{{Tuoheti} {et~al.}(2011){Tuoheti}, {Esamdin}, {Hu}, {L{\"u}}, {Wang},
  \& {Abliz}}]{Tuoheti11}
{Tuoheti}, A., {Esamdin}, A., {Hu}, H.-D., {et~al.} 2011, Research in Astronomy
  and Astrophysics, 11, 974, \dodoi{10.1088/1674-4527/11/8/009}

\bibitem[{{van Straten} \& {Bailes}(2011)}]{Straten11}
{van Straten}, W., \& {Bailes}, M. 2011, \pasa, 28, 1, \dodoi{10.1071/AS10021}

\bibitem[{{Wang} {et~al.}(2023){Wang}, {Wen}, {Duan}, {Wang}, {He}, {Wang},
  {Wang}, {Yuan}, {Yan}, {Yuen}, {Han}, {Zhi}, {Xiang}, {Dang}, {Chen}, {Lyu},
  {Wang}, \& {Ergesh}}]{Wang23}
{Wang}, H., {Wen}, Z.~G., {Duan}, X.~F., {et~al.} 2023, \apj, 950, 166,
  \dodoi{10.3847/1538-4357/acd17b}

\bibitem[{{Wang} {et~al.}(2007){Wang}, {Manchester}, \& {Johnston}}]{Wang07}
{Wang}, N., {Manchester}, R.~N., \& {Johnston}, S. 2007, \mnras, 377, 1383,
  \dodoi{10.1111/j.1365-2966.2007.11703.x}

\bibitem[{{Wang} {et~al.}(2020){Wang}, {Han}, {Han}, {Cai}, {Wang}, {Wang},
  {Chen}, {Zhou}, {Yu}, {Han}, {Xu}, {Gao}, {Hong}, {Hou}, \&
  {Dong}}]{Wang2020}
{Wang}, P.~F., {Han}, J.~L., {Han}, L., {et~al.} 2020, \aap, 644, A73,
  \dodoi{10.1051/0004-6361/202038867}

\bibitem[{Wang {et~al.}(2021)Wang, Zhu, Li, Pan, Wang, Cordes, Chatterjee, Yao,
  Qian, Yue, Zhang, Zhao, Wang, Niu, Yuan, Miao, Xie, Liu, Yu, You, Meng, \&
  Collaboration}]{WWD21}
Wang, S., Zhu, W.-W., Li, D., {et~al.} 2021, Research in Astronomy and
  Astrophysics, 21, 251, \dodoi{10.1088/1674-4527/21/10/251}

\bibitem[{{Wang} {et~al.}(2021){Wang}, {Wen}, {Yuan}, {Wang}, {Chen}, {Yan},
  {Yuen}, {Duan}, {Xiang}, \& {He}}]{Wangz2021}
{Wang}, Z., {Wen}, Z.~G., {Yuan}, J.~P., {et~al.} 2021, \apj, 923, 259,
  \dodoi{10.3847/1538-4357/ac32ba}

\bibitem[{{Weatherall}(1998)}]{Weatherall1998}
{Weatherall}, J.~C. 1998, \apj, 506, 341, \dodoi{10.1086/306218}

\bibitem[{{Weltevrede} {et~al.}(2006{\natexlab{a}}){Weltevrede}, {Edwards}, \&
  {Stappers}}]{Weltevrede06}
{Weltevrede}, P., {Edwards}, R.~T., \& {Stappers}, B.~W. 2006{\natexlab{a}},
  \aap, 445, 243, \dodoi{10.1051/0004-6361:20053088}

\bibitem[{{Weltevrede} {et~al.}(2012){Weltevrede}, {Wright}, \&
  {Johnston}}]{Weltevrede12}
{Weltevrede}, P., {Wright}, G., \& {Johnston}, S. 2012, \mnras, 424, 843,
  \dodoi{10.1111/j.1365-2966.2012.21207.x}

\bibitem[{{Weltevrede} {et~al.}(2006{\natexlab{b}}){Weltevrede}, {Wright},
  {Stappers}, \& {Rankin}}]{Weltevrede2006}
{Weltevrede}, P., {Wright}, G.~A.~E., {Stappers}, B.~W., \& {Rankin}, J.~M.
  2006{\natexlab{b}}, \aap, 458, 269, \dodoi{10.1051/0004-6361:20065572}

\bibitem[{{Wen} {et~al.}(2016){Wen}, {Wang}, {Yuan}, {Yan}, {Manchester},
  {Yuen}, \& {Gajjar}}]{Wen16}
{Wen}, Z.~G., {Wang}, N., {Yuan}, J.~P., {et~al.} 2016, \aap, 592, A127,
  \dodoi{10.1051/0004-6361/201628214}

\bibitem[{{Wen} {et~al.}(2021){Wen}, {Yuen}, {Wang}, {Tu}, {Yan}, {Yuan},
  {Yan}, {Chen}, {Wang}, {Shen}, {Wang}, {Yang}, {Duan}, {Zhang}, {Wang}, \&
  {Mao}}]{Wen21}
{Wen}, Z.~G., {Yuen}, R., {Wang}, N., {et~al.} 2021, \apj, 918, 57,
  \dodoi{10.3847/1538-4357/ac0e90}

\bibitem[{{Yan} {et~al.}(2019){Yan}, {Manchester}, {Wang}, {Yuan}, {Wen}, \&
  {Lee}}]{Yan19}
{Yan}, W.~M., {Manchester}, R.~N., {Wang}, N., {et~al.} 2019, \mnras, 485,
  3241, \dodoi{10.1093/mnras/stz650}

\bibitem[{{Young} {et~al.}(2015){Young}, {Weltevrede}, {Stappers}, {Lyne}, \&
  {Kramer}}]{Young15}
{Young}, N.~J., {Weltevrede}, P., {Stappers}, B.~W., {Lyne}, A.~G., \&
  {Kramer}, M. 2015, \mnras, 449, 1495, \dodoi{10.1093/mnras/stv392}

\bibitem[{{Yu} {et~al.}(2017){Yu}, {Pen}, {Zhang}, {Li}, \& {Chen}}]{Yu17}
{Yu}, H.~R., {Pen}, U.~L., {Zhang}, T.~J., {Li}, D., \& {Chen}, X. 2017,
  Research in Astronomy and Astrophysics, 17, 049,
  \dodoi{10.1088/1674-4527/17/6/49}

\bibitem[{{Zhang} {et~al.}(2007){Zhang}, {Gil}, \& {Dyks}}]{Zhang2007}
{Zhang}, B., {Gil}, J., \& {Dyks}, J. 2007, \mnras, 374, 1103,
  \dodoi{10.1111/j.1365-2966.2006.11226.x}

\bibitem[{{Zhang} {et~al.}(2000){Zhang}, {Harding}, \& {Muslimov}}]{Zhang2000}
{Zhang}, B., {Harding}, A.~K., \& {Muslimov}, A.~G. 2000, \apjl, 531, L135,
  \dodoi{10.1086/312542}

\bibitem[{{Zhi} {et~al.}(2023){Zhi}, {Bai}, {Shang}, {Xu}, {Dang}, {Li},
  {Zhang}, {Wang}, {Xie}, {Zhao}, {Dong}, \& {Qiao}}]{Zhi23}
{Zhi}, Q.~J., {Bai}, J.~T., {Shang}, L.~H., {et~al.} 2023, \apj, 954, 24,
  \dodoi{10.3847/1538-4357/ace362}

\end{thebibliography}
	\bibliographystyle{aasjournal}
	
\end{document}